\documentclass[11pt,a4paper]{article}
\usepackage{times}
\usepackage{euler}
\usepackage{a4}
\usepackage{amssymb}
\usepackage{amsmath}
\usepackage{amsthm} 
\usepackage{epsfig}
\usepackage{rotating}
\usepackage[hang,stable]{footmisc}
\theoremstyle{plain}
\newtheorem{theorem}{Theorem}
\newtheorem{corollary}[theorem]{Corollary}
\newtheorem{proposition}[theorem]{Proposition}
\newtheorem{lemma}[theorem]{Lemma}
\theoremstyle{definition}
\newtheorem{definition}{Definition}

\newcommand{\reals}{\mathbb{R}}    
\newcommand{\complex}{\mathbb{C}}  
\newcommand{\field}{\mathbb{F}} 
\newcommand{\End}{\mbox{End}} 
\newcommand{\Frames}[1]{\mathcal{F}_{#1}}     
\newcommand{\Stab}{\mathrm{Stab}}  
\newcommand{\Span}{\mbox{span}} 
\newcommand{\Proj}{\Pi} 
\newcommand{\trace}{\mathrm{trace}}        
                 
\newcommand{\Aff}{\mathrm{Aff}}
\newcommand{\Orb}{\mathrm{Orb}}
\newcommand{\group}{\mathsf}
\newcommand{\Lie}{\mathrm{Lie}}
\newcommand{\Lor}{\mathsf{Lor}} 
\newcommand{\ILor}{\mathsf{ILor}}
\newcommand{\Gal}{\mathsf{Gal}}
\newcommand{\IGal}{\mathsf{IGal}}         
\newcommand{\omegaup}{\omega^\uparrow}
\newcommand{\omegadown}{\omega^\downarrow}
\newcommand{\mat}{\mathbf}
\begin{document}
\ifx\href\undefined\else\hypersetup{linktocpage=true}\fi

\title{Algebraic and geometric structures of\\ 
       Special Relativity}
\author{Domenico Giulini            \\
        University of Freiburg      \\
        Institute of Physics        \\
        Hermann-Herder-Strasse 3    \\
        79104 Freiburg, Germany}
\date{}

\maketitle

\begin{abstract}
\noindent
I review, on an advanced level, some of the algebraic and geometric 
structures that underlie the theory of Special Relativity.  
This includes a discussion of relativity as a symmetry 
principle, derivations of the Lorentz group, its composition 
law, its Lie algebra, comparison with the Galilei group, 
Einstein synchronization, the lattice of causally and 
chronologically complete regions in Minkowski space, 
rigid motion (the Noether-Herglotz theorem), and the geometry of 
rotating reference frames. Representation-theoretic aspects of the 
Lorentz group are not included. A series of appendices present 
some related mathematical material. 

This paper is a contribution to the proceedings of the 339th 
WE-Heraeus-Seminar \emph{Special Relativity: Will it Survive the Next 100 Years?}, 
edited by J.\,Ehlers and C.\,L\"ammerzahl, which will be 
published by Springer Verlag in 2006.   

\end{abstract}

\newpage
\setcounter{tocdepth}{2}
\tableofcontents
\newpage

\section{Introduction}
\label{sec:Introduction}
In this contribution I wish to discuss some structural aspects of 
Special Relativity (henceforth abbreviated SR) which are, 
technically speaking, of a more advanced nature. Most of what 
follows is well known, though generally not included in 
standard text-book presentations. Against my original intention,
I decided to not include those parts that relate to the 
representation- and field-theoretic aspects of the homogeneous 
and inhomogeneous Lorentz group, but rather to be more explicit 
on those topics now covered. Some of the abandoned material is 
(rather informally) discussed in \cite{GiuliniStraumann:2006}. 
All of it will appear in \cite{Giulini:SR2}. For a comprehensive 
discussion of many field-theoretic aspects, see 
e.g.~\cite{SexlUrbantke:RGP}. 

I always felt that Special Relativity deserves more attention than 
what is usually granted to it in courses on mechanics or electrodynamics. 
There is a fair amount of interesting algebraic structure that 
characterizes the transition between the Galilei and Lorentz group,
and likewise there is some interesting geometry involved in the 
transition between Newtonian (or Galilean) spacetime and Minkowski 
space. The latter has a rich geometric structure, notwithstanding the 
fact that, from a general relativistic viewpoint, it is `just' flat 
spacetime. I hope that my contribution will substantiate these claims.
For the convenience of some interested readers I have included several 
mathematical appendices with background material that, according to my 
experience, is considered helpful being spelled out in some detail.

\section{Some remarks on `symmetry' and `covariance'}
\label{sec:PreliminaryRemarks}
For the purpose of this presentation I regard SR as the 
(mathematical) theory of how to correctly implement the \emph{Galilean 
Relativity Principle}---henceforth simply abbreviated by RP. 
The RP is a physical statement concerning a subclass of 
phenomena---those not involving gravity---which translates into 
a mathematical symmetry requirement for the laws describing them. 
But there is no unique way to proceed; several choices need to 
be made whose correctness cannot be decided by mere logic. 

Given that the symmetry requirement is implemented by a group 
action (which may be relaxed; compare e.g. supersymmetry, which is 
not based on a group), the most fundamental question is: 
\emph{what group?} In this regard there is quite a convincing  
string of arguments that, given certain mild technical assumptions, 
the RP selects either the Galilei or the Lorentz group (the latter 
for some yet undetermined velocity parameter $c$). This will be 
discussed in detail in Section\,\ref{sec:ImpactRP}.

Almost as important as the selection of a group is the question 
of how it should act on physical entities in question, like particles 
and fields. The importance and subtlety of this question is usually 
underestimated. Let us therefore dwell a little on it.

As an example we consider vacuum electrodynamics. Here the 
mathematical objects that represent physical reality are two 
spacetime dependent fields, $\vec E(\vec x,t)$ and $\vec B(\vec x,t)$,
which take values in a vector space isomorphic to $\reals^3$.
There will be certain technical requirements on these fields, 
e.g. concerning differentiability and fall-off at spatial infinity, 
which we do not need to spell out here. For simplicity we shall 
assume that the set of all fields obeying these conditions forms 
an infinite-dimensional linear space $\mathcal{K}$, which is 
sometimes called the space of `kinematical' (or `kinematically 
possible') fields. Those fields in $\mathcal{K}$ which satisfy 
Maxwell's (vacuum) equations form a proper subset, 
$\mathcal{S}\subset\mathcal{K}$, which, due to the linearity of 
the equations and the boundary conditions (fall-off to zero value, say), 
is a linear subspace. It is called the space of `physical' 
(or `dynamically possible') fields. Clearly these notions of the 
spaces of kinematically and dynamically possible fields apply to all 
sorts of situations in physics where one considers `equations of motion',
though in general neither of these sets will be a vector space. 
This terminology was introduced in~\cite{Anderson:RelativityPhysics}.    

In general, we say that a group $G$ is a \emph{symmetry group} of 
a given dynamical theory if the following two conditions are satisfied:  
\begin{itemize}
\item[1.]
There exists an (say left-) effective action 
$G\times\mathcal{K}\rightarrow\mathcal{K}$,
$(g,k)\mapsto g\cdot k$, of $G$ on $\mathcal{K}$ 
(cf. Sect.\,\ref{sec:GroupActions}). Posing effectiveness just 
means that we do not wish to allow trivial enlargements of the 
group by elements that do not move anything. It also means no loss 
of generality, since any action of a group $G$ on a set $\mathcal{K}$ 
factors through an effective action of $G/G'$ on $\mathcal{K}$,
where $G'$ is the normal subgroup of trivially acting elements 
(cf. Sect.\,\ref{sec:GroupActions}). Such an action of the group on 
the kinematical space of physical fields is also called an 
\emph{implementation} of the group into the physical theory.  
\item[2.]
The action of $G$ on $\mathcal{K}$ leaves $\mathcal{S}$
invariant (as a set, not necessarily pointwise), i.e. if 
$s\in\mathcal{S}$ then $g\cdot s\in\mathcal{S}$ for all $g$. 
This merely says that the group action restricts from 
$\mathcal{K}$ to $\mathcal{S}$. Note that, from an abstract 
point of view, this is the precise statement of the phrase 
`leaving the field equations invariant', since the field 
equations are nothing but a characterization of the subset 
$\mathcal{S}\subset\mathcal{K}$.  
\end{itemize}
If this were all there is to require for a group to count as 
a \emph{symmetry group}, then we would probably be surprised 
by the wealth of symmetries in Nature. For example, in the 
specific case at hand, vacuum electrodynamics, we often hear or 
read the statement that the Lorentz group leaves Maxwell's 
equations invariant, whereas the Galilei group does not. 
Is this really true? Has anyone really shown in this context that 
the Galilei group cannot effectively act on $\mathcal{K}$ so 
as to leave $\mathcal{S}$ invariant? Certainly not, because such 
an action is actually  known to exist; see e.g. Chap.\,5.9 in 
\cite{Fushchich:SymmDGL}. Hence, in the general sense above, 
the Galilei group \emph{is} a symmetry group of Maxwell's equations!

The folklore statement just alluded to can, however, be turned 
into a true statement if a decisive restriction for the action
is added, namely that it be \emph{local}. This means that the action 
on the space of fields is such that the value of the transformed field 
at the transformed spacetime point depends only on the value of the 
untransformed field at the untransformed point and \emph{not}, 
in addition, on its derivatives.\footnote{Here one should actually 
distinguish between `ultralocality', meaning not involving any 
derivatives, and `locality', meaning just depending on 
derivatives of at most finite order.} This is the crucial 
assumption that is implicit in all proofs of Galilean-non-invariance 
of Maxwell's equations and that is also made regarding the Lorentz 
group in classical and quantum field theories. The action of the 
Galilei group that makes it a symmetry group for Maxwell's equations 
is, in fact, such that the transformed field depends linearly on 
the original field and its derivatives to \emph{all} orders. That is,
it is highly non local.  

Returning to the general discussion, we now consider a classical 
field, that is, a map $\Psi:M\rightarrow V$ from spacetime $M$ into 
a vector space $V$. A spacetime symmetry-group has an action on $M$, 
denoted by $T:(g,x)\mapsto T_g(x)$,  as well as an action on $V$, 
which in most cases of interest is a linear representation 
$g\mapsto D(g)$ of $G$. A local action of $G$ on field space is 
then given by 
\begin{equation}
\label{eq:G-ActionFields}
(g,\Psi)\mapsto g\cdot\Psi:= D(g)\circ\Psi\circ T_g^{-1}\,,
\end{equation}       
where here and below the symbol $\circ$ denotes composition of maps. 
This is the form of the action one usually assumes. Existing generalizations
concerning possible non-linear target spaces for $\Psi$ and/or
making $\Psi$ a section in a bundle, rather than just a global 
function, do not influence the locality aspect emphasized here and 
will be ignored.    

Next to fields one also considers particles, at least in the classical 
theory. Structureless (e.g. no spin) particles in spacetime are 
mathematically idealized by maps $\gamma:\reals\rightarrow M$, 
where $\reals$ (or a subinterval thereof) represents parameter 
space. The parameterization usually does not matter, except for time 
orientation, so any reparameterization 
$f:\reals\rightarrow\reals$ with $f'>0$ gives a 
reparameterized curve $\gamma':=\gamma\circ f$ which is just 
as good. On the space of particles, the group $G$ acts as follows: 
\begin{equation}
\label{eq:G-ActionParticles}
(g,\gamma)\mapsto g\cdot\gamma:= T_g\circ\gamma\,.
\end{equation}       
Together (\ref{eq:G-ActionFields}) and (\ref{eq:G-ActionParticles})
define an action on all the dynamical entities, that is particles 
and fields, which we collectively denote by the symbol $\Phi$. 
The given action of $G$ on that space is simply denoted by 
$(g,\Phi)\mapsto g\cdot\Phi$. 

Now, the set of equations of motion for the whole system can be 
written in the general form 
\begin{equation}
\label{eq:EqMotion}
\mathcal{E}[\Sigma,\Phi]=0\,, 
\end{equation}
where this should be read as a multi-component equation (with $0$ 
being the zero `vector' in target space). $\Sigma$ stands collectively for 
non-dynamical entities (background structures) whose values are 
fixed by means independent of equation (\ref{eq:EqMotion}). 
It could, for example, be the Minkowski metric in Maxwell's equations 
and also external currents. The meaning of (\ref{eq:EqMotion}) is
to determine $\Phi$, \emph{given} $\Sigma$ (and the boundary 
conditions for $\Phi$). We stress that $\Sigma$ is a constitutive 
part of the equations of motion. We now make the following 
\begin{definition}
An action of the group $G$ on the space of dynamical entities 
$\Phi$ is said to correspond to a \emph{symmetry} of the 
equations of motion iff\footnote{Throughout we write `iff' as 
abbreviation for `if and only if'.} for all $g\in G$ we have 
\begin{equation}
\label{eq:Def:Symmetries}
\mathcal{E}[\Sigma,\Phi]=0\ \Longleftrightarrow\ 
\mathcal{E}[\Sigma,g\cdot\Phi]=0\,.  
\end{equation}
\end{definition}

Different form that is mere `covariance', which is a far more 
trivial requirement. It arises if the space of background 
structures, $\Sigma$, also carries an action of $G$ (as it 
naturally does if the $\Sigma$ are tensor fields). Then we have  

\begin{definition}
An action of the group $G$ on the space of dynamical and 
non-dynamical entities $\Phi$ and $\Sigma$ is said to 
correspond to a \emph{covariance} of the equations of motion
iff for all $g\in G$ we have
\begin{equation}
\label{eq:Def:Invariance}
\mathcal{E}[\Sigma,\Phi]=0\ \Longleftrightarrow\ 
\mathcal{E}[g\cdot\Sigma,g\cdot\Phi]=0\,.  
\end{equation}
\end{definition}
The difference to symmetries is that the background 
structures---and in that sense the equations of motion themselves---are 
changed too. Equation~(\ref{eq:Def:Symmetries}) says that if $\Phi$ 
solves the equations of motion then $g\cdot\Phi$ solves the 
\emph{very same} equations. In contrast, (\ref{eq:Def:Invariance}) 
merely tells us that if $\Phi$ solves the equations of motion
then $g\cdot\Phi$ solves the appropriately transformed equations. 
Trivially, a symmetry is also a covariance but the converse it not 
true. Rather, a covariance is a symmetry iff it stabilizes the background 
structures, i.e. if for all $g\in G$ we have $g\cdot\Sigma=\Sigma$.   

Usually one has a good idea of what the dynamical entities $\Phi$ 
in ones theory should be, whereas the choice of $\Sigma$ is more
a matter of presentation and therefore conventional. After all, 
the only task of equations of motion is to characterize the set 
$\mathcal{S}$ of dynamically possible fields (and particles) 
amongst the set $\mathcal{K}$ of all kinematically possible ones. 
Whether this is done by using auxiliary structures $\Sigma_1$ 
or $\Sigma_2$ should not affect the physics. It is for this reason 
that one has to regard the requirement of mere covariance as,
physically speaking, rather empty. This is because one can always
achieve covariance by suitably adding non-dynamical structures 
$\Sigma$. Let us give an example for this~\cite{Anderson:RelativityPhysics}. 

Consider the familiar heat equation, 
\begin{equation}
\label{eq:HeatEq}
\partial_t T-\kappa\Delta T=0\,.
\end{equation}
Here the dynamical field $\Phi= T$ is the temperature function. The 
background structure is the 3-dimensional Euclidean metric, $\delta$, 
of space which enters the Laplacian,
$\Delta:=\delta^{ab}\partial_a\partial_b$; hence $\Sigma=\delta$. 
This equation possesses time translations and Euclidean motions in 
space (we neglect space reflections for simplicity) as symmetries. 
These form the group $E_3\cong\reals^3\rtimes SO(3)$,
the semi-direct product of spatial translations and rotations. 
Clearly $E_3$ stabilizes $\delta$. 

But without changing the physics we can rewrite (\ref{eq:HeatEq})
in the following spacetime form: Let $(x^0,x^1,x^2,x^3)=(ct,x,y,z)$
be inertial coordinates in Minkowski space and 
$n=\partial_t=n^\mu\partial_\mu$ (i.e. $n^\mu=(c,0,0,0)$) the 
(covariant) constant vector field describing the motion of the 
inertial observer. The components of the Minkowski metric in these 
coordinates are denoted by $g_{\mu\nu}$. In our conventions 
(`mostly minus') $\{g_{\mu\nu}\}=\text{diag(1,-1,-1,-1)}$.
Then (\ref{eq:HeatEq}) is clearly just the same as 
\begin{equation}
\label{eq:HeatEqCov}
n^\mu\partial_\mu T-\kappa\left(c^{-2}\,n^\mu n^\nu-g^{\mu\nu}\right)\,
\partial_\mu\partial_\nu\,T=0\,.
\end{equation}
Here the dynamical variable is still $\Phi=T$ but the background 
variables are now given by $\Sigma=(n,g)$. This equations is now 
manifestly covariant under the Lorentz group if $n^\mu$ and 
$g^{\mu\nu}$ are acted upon as indicated by their indices. Hence 
we were able to enlarge the covariance group by enlarging the 
space of $\Sigma$s. In fact, we could even make the equation 
covariant under general diffeomorphisms by replacing partial 
with covariant derivatives. But note that the symmetry 
group would still be that subgroup that stabilizes (leaves invariant) 
the (flat) metric $g$ and the (covariant constant) vector field 
$n$, which again results in the same symmetry group as for the 
original equation~(\ref{eq:HeatEq}). 
For more discussion concerning also the problematic aspects of 
of the notion of `general covariance', see e.g. 
\cite{Giulini:2006}.

\section{The impact of the relativity principle on the\\  
automorphism group  of spacetime}
\label{sec:ImpactRP}
In the history of SR it has often been asked what the most general 
transformations of spacetime were that implemented the relativity 
principle (RP), without making use of the requirement of the constancy 
of the speed of light. This question was first addressed by 
Ignatowsky~\cite{Ignatowsky:1910}, who showed that under a certain 
set of technical assumptions (not consistently spelled out by him)
the RP alone suffices to arrive at a spacetime symmetry group which 
is either the Galilei or the Lorentz group, the latter for some 
yet undetermined limiting velocity $c$. More precisely, what is
actually shown in this fashion is, as we will see, that the 
spacetime symmetry group must contain either the proper orthochronous 
Galilei or Lorentz group, if the group is required to comprise at 
least spacetime translations, spatial rotations, and boosts (velocity
transformations). What we hence gain is the group-theoretic insight 
of how these transformations must combine into a common group, given 
that they form a group at all. We do not learn anything about 
other transformations, like spacetime reflections or dilations,
whose existence we neither required nor ruled out on this 
theoretical level. 
 
The work of Ignatowsky was put into a logically more coherent form
by Franck \& Rothe~\cite{FrankRothe:1911}\cite{FrankRothe:1912}, 
who showed that some of the technical assumptions could be dropped. 
Further formal simplifications were achieved by Berzi \& Gorini 
\cite{BerziGorini:1969}. Below we shall basically follow their line 
of reasoning, except that we do not impose the continuity of the 
transformations as a requirement, but conclude it from their 
preservation of the inertial structure plus bijectivity. See also 
\cite{BacryLevy-Leblond:1968} for an alternative discussion on 
the level of Lie algebras.    

The principles of SR are mathematically most concisely expressed in 
terms of few simple structures put onto spacetime. In SR these 
structures are absolute in the sense of not being subject to any 
dynamical change. From a fundamental point of view, it seems rather 
a matter of convention whether one thinks of these structures as 
primarily algebraic or geometric. According to the idea advocated 
by Felix Klein in his `Erlanger Programm'~\cite{Klein:ErlangerProgramm}, 
a geometric structure can be characterized by its 
\emph{automorphism group}\footnote{Klein calls it `Hauptgruppe'.}. 
The latter is generally defined by the subgroup of bijections of the 
set in question which leaves the geometric structure---e.g. thought 
of as being given in terms of relations---invariant. Conversely, any 
transformation group (i.e. subgroup of group of bijections) can be 
considered as the automorphism group of some `geometry' which is 
defined via the invariant relations.

The geometric structure of spacetime is not a priori given to us. 
It depends on the physical means on which we agree to measure spatial 
distances and time durations. These means refer to physical systems, 
like `rods' and `clocks', which are themselves subject to dynamical laws 
in spacetime. For example, at a fundamental physical level, the spatial 
transportation of a rod or a clock from one place to another is 
certainly a complicated \emph{dynamical} process. It is only due to the 
special definition of `rod' and `clock' that the result of such a 
process can be summarized by simple \emph{kinematical} rules. Most 
importantly, their dynamical behavior must be `stable' in the sense 
of being essentially independent of their dynamical environment.
Hence there is always an implicit consistency hypothesis underlying 
operational definitions of spatio-temporal measurements, which in case 
of SR amount to the assumption that rods and clocks are themselves 
governed by Lorentz invariant dynamical laws. 

A basic physical law is the law of inertia. It states the preference 
of certain types of motions for force-free, uncharged, zero-spin 
test-particles: the `uniform' and `rectilinear' ones. In the spacetime 
picture this 
corresponds to the preference of certain curves corresponding to the 
inertial worldlines of the force-free test particles. In the 
gravity-free case, we model these world lines by straight lines of 
the affine space $\Aff(\reals^4)$ over the vector space 
$\reals^4$. This closely corresponds to our intuitive notion of 
homogeneity of space and time, that is, that there exists an effective 
and transitive (and hence simply transitive) action of the Abelian 
group $\reals^4$ of translations (cf. Sects.\,\ref{sec:GroupActions}
and \ref{sec:AffineSpaces}). A lot could (and perhaps should) be said 
at this point about the proper statement of the law of inertia and 
precisely how it endows spacetimes with certain geometric structures. 
Instead we will simply refer the interested reader to the literature; 
see e.g. \cite{Giulini:2002b} and references therein. 

Note that we do not conversely assume any straight line to correspond 
to some inertial world-line. Hence the first geometric structure on 
spacetime, which can be thought of as imposed by the law of inertia, 
is that of a subset of straight lines.
If all straight lines were involved, the automorphism group of 
spacetime would necessarily have to map any straight line to a straight 
line and therefore be a subgroup of the affine group 
$\reals^4\rtimes\group{GL}(4,\reals)$. This is just the 
content of the main theorem in affine geometry; see e.g. 
\cite{Berger:Geometry1}. However, we can only argue that it must map 
the subset of inertial world-lines onto itself. We take this subset 
to consist of all straight lines in $\Aff(\reals^4)$ whose slope
with respect to some reference direction is smaller than a certain 
finite value $\beta$. This corresponds to all worldlines not exceeding 
a certain limiting speed with reference to some inertial frame. It is 
then still true that any bijection\footnote{If one drops the 
assumption of bijectivity, then there exist in addition the 
fractional linear transformations which map straight lines to 
straight lines, except for those points that are mapped to 
`infinity'; see e.g. the discussion in Fock's book \cite{Fock:STG},
in particular his Appendix\,A, and also \cite{FrankRothe:1912}. 
One might argue that since physics 
takes place in the finite we cannot sensibly argue for global 
bijectivity and hence have to consider those more general 
transformations. However, the group they generate does not have 
an invariant bounded domain in spacetime and hence cannot be 
considered as the automorphism group of any fixed set of physical 
events.} of $\Aff(\reals^4)$ preserving that subset must be a 
subgroup of the affine group~\cite{Hegerfeld:1972}. Also, it is 
not necessary to assume that lines map surjectively onto 
lines~\cite{ChubarevPinelis:2000}.

For further determination of the automorphism group of spacetime 
we invoke the following principles:
\begin{itemize}
\item[ST1:]\hspace{0.6cm}
Homogeneity of spacetime.
\item[ST2:]\hspace{0.6cm}
Isotropy of space.
\item[ST3:]\hspace{0.6cm}
Galilean principle of relativity.
\end{itemize}
We take ST1 to mean that the sought-for group should include all 
translations and hence be of the form $\reals^4\rtimes\group{G}$, 
where $\group{G}$ is a subgroup of $\group{GL}(4,\reals)$. 
ST2 is interpreted as saying that $G$ should include the set of all 
spatial rotations. If, with respect to some frame, we write the 
general element $A\in\group{GL}(4,\reals)$ in a $1+3$ split form 
(thinking of the first coordinate as time, the other three as space), 
we want $\group{G}$ to include all       
\begin{equation}
\label{eq:SO3Imbedding}
R(\mat{D})=
\begin{pmatrix}
1&\vec 0^\top\\
\vec 0&\mat{D}
\end{pmatrix}\,,\qquad
\text{where}\quad
\mat{D}\in\group{SO}(3)\,.
\end{equation}
Finally, ST3 says that velocity transformations, henceforth 
called `boosts', are also contained in $\group{G}$. However, at this 
stage we do not know how boosts are to be represented mathematically. 
Let us make the following assumptions: 
\begin{itemize}
\item[B1:]\hspace{0.2cm}
Boosts $B(\vec v)$ are labeled by a vector 
$\vec v\in B_c(\reals^3)$, where $B_c(\reals^3)$ is the 
open ball in $\reals^3$ of radius $c$. The physical interpretation 
of $\vec v$ shall be that of the boost velocity, as measured in the system 
from which the transformation is carried out. We allow $c$ to be 
finite or infinite ($B_\infty(\reals^3)=\reals^3$). 
$\vec v=\vec 0$ corresponds to the identity transformation, 
i.e. $B(\vec 0)=\text{id}_{\reals^4}$. We also assume that 
$\vec v$, considered as coordinate function on the group, 
is continuous. 
\item[B2:]\hspace{0.2cm}
As part of ST2 we require equivariance of boosts under rotations:
\begin{equation}
\label{eq:EquivBoostsRot}
R(\mat{D})\cdot B(\vec v)\cdot R(\mat{D}^{-1})=
B(\mat{D}\cdot\vec v)\,.
\end{equation}
\end{itemize}
The latter assumption allows us to restrict attention to boost 
in a fixed direction, say  that of the positive $x$-axis. 
Once their analytical form is determined
as function of $v$, where $\vec v=v\vec e_x$, we deduce the 
general expression for boosts using (\ref{eq:EquivBoostsRot}) and 
(\ref{eq:SO3Imbedding}). We make no assumptions involving space 
reflections.\footnote{Some derivations in the literature of the 
Lorentz group do not state the equivariance property 
(\ref{eq:EquivBoostsRot}) explicitly, though they all use it
(implicitly), usually in statements to the effect that it is 
sufficient to consider boosts in one fixed direction. Once this 
restriction is effected, a one-dimensional spatial reflection 
transformation is considered to relate a boost transformation to 
that with opposite velocity. This then gives the impression that 
reflection equivariance is also invoked, though this is not 
necessary in spacetime dimensions greater than two, for 
(\ref{eq:EquivBoostsRot}) allows to invert one axis through a 
180-degree rotation about a perpendicular one.} 
We now restrict attention to $\vec v=v\vec e_x$. We wish to determine
the most general form of $B(\vec v)$ compatible with all requirements 
put so far. We proceed in several steps: 
\begin{enumerate}
\item
Using an arbitrary rotation $\mat{D}$ around the $x$-axis, 
so that $\mat{D}\cdot\vec v=\vec v$, equation (\ref{eq:EquivBoostsRot}) 
allows to prove that
\begin{equation}
\label{eq:FormBoost1}
B(v\vec e_x)=
\begin{pmatrix}
\mat{A}(v)&0\\
0&\alpha(v)\mat{1}_2
\end{pmatrix}\,,
\end{equation}
where here we wrote the $4\times 4$ matrix in a $2+2$ decomposed form. 
(i.e. $\mat{A}(v)$ is a $2\times 2$ matrix and $\mat{1}_2$ is the 
 $2\times 2$ unit-matrix). Applying (\ref{eq:EquivBoostsRot}) once 
more, this time using a $\pi$-rotation about the $y$-axis, we learn 
that $\alpha$ is an even function, i.e. 
\begin{equation}
\label{eq:FormBoosts2}
\alpha(v)=\alpha(-v)\,.
\end{equation}
Below we will see that $\alpha(v)\equiv 1$. 

\item
Let us now focus on $\mat{A }(v)$, which defines the action of 
the boost in the $t-x$ plane. We write 
\begin{equation}
\label{eq:FormBoost3}
\begin{pmatrix}t\\x\end{pmatrix}
\mapsto
\begin{pmatrix}t'\\x'\end{pmatrix}
=\mat{A}(v)\cdot\begin{pmatrix}t\\x\end{pmatrix}=
\begin{pmatrix}a(v)&b(v)\\c(v)&d(v)\end{pmatrix}
\cdot\begin{pmatrix}t\\x\end{pmatrix}\,.
\end{equation}
We refer to the system with coordinates $(t,x)$ as $K$ and that 
with coordinates $(t',x')$ as $K'$. From (\ref{eq:FormBoost3})
and the inverse (which is elementary to compute) one infers that 
the velocity $v$ of $K'$ with respect to $K$ and the velocity $v'$ 
of $K$ with respect to $K'$ are given by 
\begin{subequations}
\label{eq:FormBoost4}
\begin{alignat}{3}
\label{eq:FormBoost4a}     
& v\,&&=\,&&-\,c(v)/d(v)\,,\\
\label{eq:FormBoost4b} 
& v'\,&&=\,&&-\,v\,d(v)/a(v)\,=:\,\varphi(v)\,.
\end{alignat}
\end{subequations}
Since the transformation $K'\rightarrow K$ is the inverse of 
$K\rightarrow K'$, the function $\varphi:(-c,c)\rightarrow (-c,c)$ 
obeys 
\begin{equation}
\label{eq:FormBoost5}
\mat{A}(\varphi(v))=(\mat{A}(v))^{-1}\,.
\end{equation}
Hence $\varphi$ is a bijection of the open interval $(-c,c)$ 
onto itself and obeys 
\begin{equation}
\label{eq:FormBoost6}
\varphi\circ\varphi=\text{id}_{(-c,c)}\,.
\end{equation}

\item
Next we determine $\varphi$. Once more using (\ref{eq:EquivBoostsRot}),
where $\mat{D}$ is a $\pi$-rotation about the $y$-axis, shows that 
the functions $a$ and $d$ in (\ref{eq:FormBoost1}) are even and the 
functions $b$ and $c$ are odd. The definition (\ref{eq:FormBoost4b}) 
of $\varphi$ then implies that $\varphi$ is odd. Since we assumed 
$\vec v$ to be a continuous coordinatization of a topological group,
the map $\varphi$ must also be continuous (since the inversion 
map, $g\mapsto g^{-1}$, is continuous in a topological group). 
A standard theorem now states that a continuous bijection of an 
interval of $\reals$ onto itself must be strictly monotonic. 
Together with (\ref{eq:FormBoost6}) this implies that $\varphi$ is 
either the identity or minus the identity map.\footnote{The simple 
proof is as follows, where we write $v':=\varphi(v)$ to save notation, 
so that (\ref{eq:FormBoost6}) now reads $v''=v$. First assume that 
$\varphi$ is strictly monotonically increasing, then $v'>v$ implies 
$v=v''>v'$, a contradiction, and $v'<v$ implies $v=v''<v'$, likewise 
a contradiction. Hence $\varphi=\text{id}$ in this case. Next assume 
$\varphi$ is strictly monotonically decreasing. Then 
$\tilde\varphi:=-\varphi$ is a strictly monotonically increasing map 
of the interval $(-c,c)$ to itself that obeys (\ref{eq:FormBoost6}). 
Hence, as just seen, $\tilde\varphi=\text{id}$, i.e. $\varphi=-\text{id}$.} If it is the 
identity map, evaluation of (\ref{eq:FormBoost5}) shows that either 
the determinant of $\mat{A}(v)$ must equals $-1$, or that $\mat{A}(v)$ 
is the identity for all $\vec v$. We exclude the second possibility 
straightaway and the first one on the grounds that we required 
$\mat{A}(v)$ be the identity for $v=0$. Also, in that case, 
(\ref{eq:FormBoost5}) implies $A^2(v)=\text{id}$ for all $v\in(-c,c)$. 
We conclude that $\varphi=-\text{id}$, which implies that the relative 
velocity of $K$ with respect to $K'$ is minus the relative velocity 
of $K'$ with respect to $K$. Plausible as it might seem, there is 
no a priori reason why this should be so.\footnote{Note that $v$ and $v'$ 
are measured with different sets of rods and clocks.}. On the face of it,
the RP only implies (\ref{eq:FormBoost6}), not the stronger relation 
$\varphi(v)=-v$. This was first pointed out in~\cite{BerziGorini:1969}. 

\item
We briefly revisit (\ref{eq:FormBoosts2}). Since we have seen 
that $B(-v\vec e_x)$ is the inverse of $B(v\vec e_x)$, we must have   
$\alpha(-v)=1/\alpha(v)$, so that (\ref{eq:FormBoosts2})
implies $\alpha(v)\equiv\pm 1$. But only $\alpha(v)\equiv +1$ is 
compatible with our requirement that $B(\vec 0)$ be the identity. 

\item   
Now we return to the determination of $\mat{A}(v)$. 
Using (\ref{eq:FormBoost4}) and $\varphi=-\text{id}$, we write 
\begin{equation}
\label{eq:FormBoost6b}
\mat{A}(v)=
\begin{pmatrix}
a(v)&b(v)\\
-va(v)&a(v)
\end{pmatrix}
\end{equation}
and 
\begin{equation}
\label{eq:FormBoost7}
\Delta(v):=\det\bigl(\mat{A}(v)\bigr)
=a(v)\bigl[a(v)+vb(v)\bigr]\,.
\end{equation}
Equation $\mat{A}(-v)=(\mat{A}(v))^{-1}$ is now equivalent to 
\begin{subequations}
\label{eq:FormBoost8}
\begin{alignat}{2}
\label{eq:FormBoost8a}
& a(-v)&&\,=\,a(v)/\Delta(v)\,,\\
\label{eq:FormBoost8b}
& b(-v)&&\,=\,-\,b(v)/\Delta(v)\,.
\end{alignat}
\end{subequations}
Since, as already seen, $a$ is an even and $b$ is an odd function,
(\ref{eq:FormBoost8}) is equivalent to $\Delta(v)\equiv 1$, i.e. 
the unimodularity of $B(\vec v)$. Equation (\ref{eq:FormBoost7})
then allows to express $b$ in terms of $a$:
\begin{equation}
\label{eq:FormBoost9}
b(v)=\frac{a(v)}{v}\left[\frac{1}{a^2(v)}-1\right]\,.
\end{equation}

\item
Our problem is now reduced to the determination of the single 
function $a$. This we achieve by employing the requirement that 
the composition of two boosts in the same direction results again 
in a boost in that direction, i.e. 
\begin{equation}
\label{eq:FormBoost10}
\mat{A}(v)\cdot\mat{A}(v')=\mat{A}(v'')\,.
\end{equation}
According to (\ref{eq:FormBoost6b}) each matrix $\mat{A}(v)$ has equal 
diagonal entries. Applied to the product matrix  on the left hand side 
of (\ref{eq:FormBoost10}) this implies that $v^{-2}(a^{-2}(v)-1)$ is 
independent of $v$, i.e. equal to some constant $k$ whose physical 
dimension is that of an inverse velocity squared. Hence we have
\begin{equation}
\label{eq:FormBoost11}
a(v)=\frac{1}{\sqrt{1+kv^2}}\,,
\end{equation}
where we have chosen the positive square root since we require 
$a(0)=1$. The other implications of (\ref{eq:FormBoost10}) are 
\begin{subequations}
\label{eq:FormBoost12}
\begin{alignat}{2}
\label{eq:FormBoost12a}
& a(v)a(v')(1-kvv')&&\,=\,a(v'')\,,\\
\label{eq:FormBoost12b}
& a(v)a(v')(1+vv')&&\,=\,v''a(v'')\,,
\end{alignat}
\end{subequations}
from which we deduce 
\begin{equation}
\label{eq:FormBoost13}
v''=\frac{v+v'}{1-kvv'}\,.
\end{equation}
Conversely, (\ref{eq:FormBoost11}) and (\ref{eq:FormBoost13})
imply (\ref{eq:FormBoost12}). We conclude that 
(\ref{eq:FormBoost10}) is equivalent to (\ref{eq:FormBoost11}) 
and (\ref{eq:FormBoost13}).  

\item
So far a boost in $x$ direction has been shown to act non-trivially 
only in the $t-x$ plane, where its action is given by the matrix that 
results from inserting (\ref{eq:FormBoost9}) and (\ref{eq:FormBoost11})
into (\ref{eq:FormBoost6b}): 
\begin{equation}
\label{eq:FormBoost14}
\mat{A}(v)=\begin{pmatrix}
a(v)&kv\,a(v)\\
-v\,a(v)& a(v)\,,
\end{pmatrix}
\qquad\text{where}\quad
a(v)=1/\sqrt{1+kv^2}\,.
\end{equation}
\begin{itemize}
\item
If $k>0$ we rescale $t\mapsto \tau:= t/\sqrt{k}$ and set 
$\sqrt{k}\,v:=\tan\alpha$. Then (\ref{eq:FormBoost14}) is seen to 
be a Euclidean rotation with angle $\alpha$ in the $\tau-x$ 
plane. The velocity spectrum is the whole real line plus 
infinity, i.e. a circle, corresponding to $\alpha\in[0,2\pi]$, 
where $0$ and $2\pi$ are identified. Accordingly, the composition 
law (\ref{eq:FormBoost13}) is just ordinary addition for the angle 
$\alpha$. This causes several paradoxa when $v$ is interpreted 
as velocity. For example, composing two finite 
velocities $v,v'$ which satisfy $vv'=1/k$ results in  
$v''=\infty$, and composing two finite and positive velocities, 
each of which is greater than $1/\sqrt{k}$, results in a finite 
but negative velocity. In this way the successive composition 
of finite positive velocities could also result in zero velocity. 
The group $\group{G}\subset\group{GL}(n,\reals)$ obtained in 
this fashion is, in fact, $\group{SO}(4)$. This group may be 
uniquely characterized as the largest connected group of bijections 
of $\reals^4$ that preserves the Euclidean distance measure. 
In particular, it treats time symmetrically with all space directions,
so that no invariant notion of time-orientability can be given in 
this case.  
\item
For $k=0$ the transformations are just the ordinary boosts of 
the Galilei group. The velocity spectrum is the whole real 
line (i.e. $v$ is unbounded but finite) and $\group{G}$ is the 
Galilei group. The law for composing velocities is just 
ordinary vector addition. 
\item
Finally, for $k<0$, one infers from (\ref{eq:FormBoost13})
that $c:=1/\sqrt{-k}$ is an upper bound for all velocities, 
in the sense that composing two velocities taken from the
interval $(-c,c)$ always results in a velocity from within 
that interval. Writing $\tau:=ct$, $v/c=:\beta=:\tanh\rho$, 
and $\gamma=1/\sqrt{1-\beta^2}$, the matrix (\ref{eq:FormBoost14}) 
is seen to be a \emph{Lorentz boost} or \emph{hyperbolic motion} in the 
$\tau-x$ plane:
\begin{equation}
\label{eq:FormBoost15}
\begin{pmatrix}\tau\\ x\end{pmatrix}
\mapsto
\begin{pmatrix}
\gamma&-\beta\gamma\\
-\beta\gamma&\gamma
\end{pmatrix}\cdot
\begin{pmatrix}\tau\\ x\end{pmatrix}=
\begin{pmatrix}
\cosh\rho&-\sinh\rho\\
-\sinh\rho&\cosh\rho
\end{pmatrix}\cdot
\begin{pmatrix}\tau\\ x\end{pmatrix}\,.
\end{equation}
The quantity 
\begin{equation}
\label{eq:def:Rapidity}
\rho:=\tanh^{-1}(v/c)=\tanh^{-1}(\beta)
\end{equation}
is called \emph{rapidity}\footnote{This term was coined by 
Robb~\cite{Robb:OptGeom}, but the quantity was used before by 
others; compare \cite{Varicak:1912}.}. If rewritten in terms of 
the corresponding rapidities the composition law 
(\ref{eq:FormBoost13}) reduces to ordinary addition: 
$\rho''=\rho+\rho'$.  
\end{itemize} 
\end{enumerate}

This shows that only the Galilei and the Lorentz group 
survive as candidates for any symmetry group implementing the 
RP. Once the Lorentz group for velocity parameter $c$ is 
chosen, one may prove that it is fully characterized by its 
property to leave a certain symmetric bilinear form invariant
(cf. Sect\,\ref{sec:IsometriesNonDegForms}).  
Endowing spacetime with that structure plus the affine 
structure from the law of inertia, we can characterize the 
Lorentz group as automorphism group of some geometric 
structure. This is often the starting point of more axiomatic 
approaches. Here we preferred to start with the opposite strategy,
which stresses that the geometry of spacetime is a contingent 
physical property, emerging through its automorphism group,
which in turn relates to the actual dynamical laws of nature. 
Having said that, we may now follow the convenient axiomatic 
line of presentation. 

\newpage
\section{Algebraic structures of Minkowski space}
\label{sec:AlgStructMink}
\begin{definition}
\emph{Minkowski space} is the affine space $\Aff(\reals^4)$
over the four-dimensional real vector space $\reals^4$, where 
the latter is endowed with a symmetric non-degenerate bilinear form 
$g$ of signature $(+,-,-,-)=(1{,}3)$. 
We write $\mathbb{M}^4=(\Aff(\reals^4),g)$. We shall usually 
restrict to bases $\{e_\mu\}_{\mu=0\cdots 3}$ of $\reals^4$ for 
which $g(e_\mu,e_\nu)=:g_{\mu\nu}=\text{diag}(1,-1,-1,-1)$. 
\end{definition}

\subsection{The Lorentz and the Galilei group}
\label{sec:LorentzGalileiGroup}
\begin{definition}
The \emph{(homogeneous) Lorentz group} is the linear group 
(subgroup of $\group{GL}(4,\reals)$) of orthogonal 
transformations of $\mathbb{M}^4$, also called $\group{O}(1{,}3)$. 
Hence $\{L^\mu_\nu\}\in\group{O}(1{,}3)$ iff 
\begin{equation}
\label{eq:DefRelLor}
g_{\mu\nu}L^\mu_\alpha L^\nu_\beta=g_{\alpha\beta}\,.
\end{equation}
\end{definition}
Note that according to Proposition\,\ref{prop:IsometriesV}
orthogonal transformations are necessarily linear.

As topological space $\group{O}(1{,}3)$ decomposes into the disjoint 
union of four connected components. Here $+/-$ stands for 
positive/negative determinant and $\uparrow/\downarrow$ for 
time-orientation preserving/reversing respectively:
\begin{equation}
\label{eq:LorentzGroup}
\group{O}(1{,}3)=
\underbrace{
\group{O}^{\uparrow}_+(1{,}3)\  \cup\
\group{O}^{\downarrow}_+(1{,}3)
}_{\displaystyle \group{SO(1{,}3)}}\ \cup\
\group{O}^{\uparrow}_-(1{,}3)\ \cup\
\group{O}^{\downarrow}_-(1{,}3)\,.
\end{equation}
Of these four components only $\group{O}^{\uparrow}_+(1{,}3)$,
the component containing the group identity, is a 
subgroup, called the group of proper orthochronous Lorentz 
transformations. Elementwise composition with space/time 
reflections gives
$\group{O}^{\uparrow}_-(1{,}3)/\group{O}^{\downarrow}_-(1{,}3)$
respectively. In the sequel we shall also write $\Lor$ for 
$\group{O}(1{,}3)$ and  $\Lor_+^{\uparrow}$ for 
$\group{O}^{\uparrow}_+(1{,}3)$. 

For any group $\group{G}\subset\group{GL}(n,\reals)$, there 
is a corresponding \emph{inhomogeneous group}, $\group{IG}$,
given by the semi-direct product 
\begin{equation}
\label{eq:SemiDirprod1}
\group{IG}=\{(a,A)\mid a\in\reals^n\,,\ A\in\group{G}\}\,,
\end{equation}
where
\begin{equation}
\label{eq:SemiDirprod2}
(a,A)(a',A')=(a+A\cdot a'\,,\, A\cdot A')\,.
\end{equation}
It can again be thought of as subgroup of $\group{GL}(n+1\,,\,\reals)$
via the embedding 
\begin{equation}
\label{eq:InhGrEmb}
(a,A)\longmapsto
\begin{pmatrix}
1&0^\top\\
a&A
\end{pmatrix}\,.
\end{equation}
In this fashion we get the inhomogeneous Lorentz groups 
$\ILor$ and $\ILor_+^{\uparrow}$ also called Poincar\'e groups. 

Let us recall the structure of the proper orthochronous homogeneous 
Galilei group, which we denote by $\Gal_+^\uparrow$. It is generated 
by spatial rotations $\vec x\mapsto\vec x'=\mat{D}\cdot\vec x$
and boosts $\vec x\mapsto \vec x'=\vec x+\vec vt$ ($t'=t$ in both cases).
Hence, if we agree to let rotations act first and then act with the 
boosts, the general form of a matrix in 
$\Gal_+^\uparrow\subset\group{GL}(4,\reals)$ will be (written in a 
$1+3$ decomposition):
\begin{equation}
\label{eq:1plus3HomGal}
G(\vec v,\mat{D})):=
\begin{pmatrix}
1&\vec 0^\top\\
\vec v&\mat{1}_3
\end{pmatrix}
\begin{pmatrix}
1&\vec 0^\top\\
\vec 0&\mat{D}
\end{pmatrix}=
\begin{pmatrix}
1&\vec 0^\top\\
\vec v&\mat{D}\,.
\end{pmatrix}
\end{equation}
Hence, given any pair $(\vec v,\mat{D})$, this tells us how to uniquely 
construct the matrix $G(\vec v,\mat{D})\in\Gal_+^\uparrow$. Conversely, 
given a matrix $G\in\Gal_+^\uparrow$, we can immediately tell 
$\vec v\in\reals^3$ and $\mat{D}\in\group{SO}(3)$ by comparison 
with the general form (\ref{eq:1plus3HomGal}). Hence there is a bijection
of sets $G:\reals^3\times\group{SO}(3)\rightarrow\Gal_+^\uparrow$. 
The group structure on $\reals^3\times\group{SO}$ that makes this 
into an isomorphism of groups is a semi-direct product: 
\begin{equation}
\label{eq:GenCompLawGalilei}
G(\vec v_1,\mat{D}_1)\cdot
G(\vec v_2,\mat{D}_2)=G(\vec v_1+\mat{D}_1\cdot\vec v_2\,,\,
\mat{D}_1\cdot\mat{D}_2)\,.
\end{equation}
Hence we have an isomorphism 
$\Gal_+^\uparrow\cong\reals^3\rtimes\group{SO}(3)$.
This also follows straightaway from comparing (\ref{eq:InhGrEmb}) 
with (\ref{eq:1plus3HomGal}). From (\ref{eq:GenCompLawGalilei}) the 
law for taking the inverse is easily deduced: 
\begin{equation}
\label{eq:GenInvLawGalilei}
\bigl(G(\vec v,\mat{D})\bigr)^{-1}=
G(-\mat{D}^{-1}\cdot v\,,\,\mat{D}^{-1})
\end{equation}

The inhomogeneous Galilei group is now isomorphic to an
iterated semi direct product: 
\begin{equation}
\label{eq:InGalGroup}
\IGal_+^\uparrow:=\reals^4\rtimes\Gal_+^\uparrow
\cong\reals^4\rtimes(\reals^3\rtimes\group{SO}(3))\,,
\end{equation}
where $\reals^4$ corresponds to space-time translations and 
$\reals^3$ to boost. The action of $\Gal_+^\uparrow$ on 
$\reals^4$ is via the `defining representation', i.e. the obvious 
action of $4\times 4$ matrices of the form (\ref{eq:1plus3HomGal})
on $\reals^4$. Not that this 4-dimensional representation of 
$\Gal_+^\uparrow$ is reducible: it transforms the 3-dimensional 
subspace of `spatial' vectors $(0,\vec a)^\top$ into themselves. 
Hence the semi-direct product of $\Gal_+^\uparrow$ with 
the subgroup of pure spatial translations, isomorphic to 
$\reals^3$, is a proper subgroup of $\IGal_+^\uparrow$ that 
properly contains $\Gal_+^\uparrow$: $\Gal_+^\uparrow\subset%
\reals^3\rtimes\Gal_+^\uparrow\subset\IGal_+^\uparrow$. 
In other words: $\Gal_+^\uparrow$ is \emph{not} a maximal\footnote{%
A proper subgroup $G'\subsetneq G$ is \emph{maximal} if there is no 
subgroup $H$ of $G$ such that $G'\subsetneq H\subsetneq G$.} 
subgroup of $\IGal_+^\uparrow$. Hence another way to write 
$\IGal_+^\uparrow$ as semi-direct product is
\begin{equation}
\label{eq:InGalGroup2}
\IGal_+^\uparrow\cong(\reals^3\times\reals^3 )
          \rtimes (\reals\times\group{SO}(3))\,.
\end{equation}
where the first two $\reals^3$ on the right hand side correspond 
to spatial translations and boosts respectively, and the single 
$\reals$ to time translations. The action of 
$\reals\times\group{SO}(3)$ on $\reals^3\times\reals^3$
is the factor-wise standard action of $\group{SO}(3)$ on 
$\reals^3$ and the trivial action of $\reals$.  

At this point we can already anticipate some major group-theoretic 
differences between the Galilei and the Lorentz groups (denoted by 
$\Lor$). For example:
\begin{itemize}
\item[1.]
$\Lor_+^\uparrow$ is a simple group, that is, it does not contain 
any normal subgroup other than the trivial ones (itself and the 
unit element). The set of pure boost does not form a subgroup. 
In contrast, $\Gal_+^\uparrow$ is not even semi-simple, meaning that 
it contains a non-trivial Abelian normal subgroup, namely the boosts. 
This makes a big difference in the corresponding representation theories.  
\item[2.]
In $\ILor_+^\uparrow=\reals^4\rtimes\Lor_+^\uparrow$ the 
action of $\Lor_+^\uparrow$ on $\reals^4$ is irreducible 
and $\Lor_+^\uparrow$ is a maximal subgroup of $\ILor_+^\uparrow$,
in contrast to the Galilean case. This makes a difference for 
the existence of invariant equivalence relations on spacetime 
(cf. Sect.\,\ref{sec:GroupActions}), like, for example, 
absolute simultaneity structures. This will be further discussed
in Sect.\,\ref{sec:EinsteinSynch}. 
\end{itemize}

\subsection{Polar decomposition}
\label{sec:PolarDec}
In (\ref{eq:1plus3HomGal}) we have given an easy proof-by-inspection 
of the unique decomposability of any element in $\Gal_+^\uparrow$ into 
a product of a rotation and a boost. We now like to discuss the analog 
of this decomposition within $\Lor_+^\uparrow$, which is more difficult 
to obtain. We start by recalling the statement and proof of the 
`polar decomposition' of matrices:   
\begin{proposition}
\label{prop:PolarDecomp}
Let $X\in\group{GL}(n{,}\complex)$; then there exists a 
unique $R\in\group{U}(n)$ (i.e. $R^\dagger=R^{-1}$) and a unique 
positive-definite Hermitian matrix $B$ (i.e. $B =B^{\dagger}$
with strictly positive eigenvalues) such that 
\begin{equation}
\label{eq:PolarDec}
X=B\cdot R\,.
\end{equation}
If $X\in\group{GL}(n{,}\reals)$ then $B$ is real, symmetric, 
and positive definite. $R$ is real and orthogonal.  
\end{proposition}
\begin{proof} 
Let $A:=XX^{\dagger}$, which is positive-definite and Hermitean
(zero eigenvalues are excluded since $X$ is invertible). Recall 
that the square-root is a well defined bijective map (a homeomorphism 
in fact) of the space of positive-definite Hermitean matrices onto itself.     
Define $B:=\sqrt{A}$ and $R:=B^{-1}X$, then 
$R^{\dagger}=X^{\dagger}B^{-1}=X^{-1}B=R^{-1}$, where the first 
equality follows from Hermiticity of $B$ and the second from 
$B^2=XX^{\dagger}$. Hence $R$ is unitary and we have shown 
existence of a polar decomposition. To show uniqueness,
assume there exist two such decompositions: $X=B_1R_1=B_2R_2$. 
Then $B_1=B_2R_3$, where $R_3:=R_2R_1^{-1}$ is again unitary. 
Hermiticity of $B_{1,2}$ and unitarity of $R_3$ now imply 
$B_1^2=B_1B_1^{\dagger}=B_2R_3R_3^{\dagger}B_2^{\dagger}=B_2^2$
and hence $B_1=B_2$, since `squaring' is an injective map 
(a homeomorphism in fact) from the space of positive-definite 
Hermitean matrices onto itself. This, in turn, implies $R_1=R_2$ 
and hence uniqueness. Finally, if $X$ is real, then $B$ and 
consequently $R$ are also real.
\end{proof} 
We wish to apply this to 
$\Lor_+^\uparrow\subset\group{GL}(4{,}\reals)$. But note 
that polar decomposing an element in 
$\group{G}\subset\group{GL}(n{,}\complex)$ need not generally 
lead to factors in $\group{G}$. However, this is true in many 
cases. For example, we have 
\begin{proposition}
\label{prop:PolarDecGroups}
Let $E^{(p{,}q)}$ be the diagonal matrix whose first $p$ diagonal 
entries equal $+1$ and the remaining $q=n-p$ diagonal entries equal 
$-1$. We define the group 
\begin{equation}
\label{eq:Def-U(p,q)}
\group{U}(p{,}q):=
\{X\in\group{GL}(n,\complex)\ \mid\ 
X\cdot E^{(p{,}q)}\cdot X^\dagger=E^{(p{,}q)}\}\,.
\end{equation}
Restricting to matrices with real entries gives the group 
$\group{O}(p{,}q)$. Polar decomposing elements of 
$\group{U}(p{,}q)$ or $\group{O}(p{,}q)$ leads to factors 
within these groups respectively. The same is true if we 
restrict to the identity components of these groups.   
\end{proposition}
\begin{proof}
It is sufficient to prove that $B=\sqrt{XX^\dagger}$ is in 
$\group{U}(p{,}q)$ since this clearly implies that the product 
$R=B^{-1}X$ will also be in $\group{U}(p{,}q)$.  
Now, $E^{(p{,}q)}\in\group{U}(p{,}q)$ (it clearly satisfies the 
defining relation in (\ref{eq:Def-U(p,q)})) so that $X^\dagger$ 
and hence $XX^\dagger$ are elements in $\group{U}(p{,}q)$. But then 
$\sqrt{XX^\dagger}\in\group{U}(p{,}q)$, too. To see this, use e.g. 
the exponential map (cf. Sect.\,\ref{sec:ExpMap}), which defines 
a homeomorphism from the space of Hermitean to the space of 
positive-definite Hermitean matrices. Then 
$X=\exp(Y)\in\group{U}(p{,}q)\Leftrightarrow E^{(p{,}q)}\cdot Y\cdot
E^{(p{,}q)}=-Y^\dagger\Leftrightarrow \sqrt{X}=\exp(Y/2)\in\group{U}(p{,}q)$.
Finally it is clear from the explicit construction of the polar 
factors that if $X(s)$ is a continuous path connecting the 
identity to $X$, and if $X(s)=B(s) R(s)$ is the polar 
decomposition for each value of $s$, then $B(s)$ and $R(s)$ 
are continuous paths connecting $B$ and $R$ to the identity.   
\end{proof}

We will use this to decompose any proper orthochronous Lorentz 
transformation $L$ into a boost $B$ and a proper spatial 
rotation $R$.\footnote{Note that the analogous factorization 
(\ref{eq:1plus3HomGal}) of a homogeneous Galilei transformation 
into boost and rotation is \emph{not} given by polar decomposition, 
but rather by a decomposition into a lower triangular matrix with 
unit diagonal (the boost) and an orthogonal matrix. This is a 
special case of what is generally known as Iwasawa decomposition.}  
Let  
\begin{equation}
\label{eq:1plus3LT}
L=
\begin{pmatrix}
\gamma&\vec a^\top\\
\vec b&\mat{M} 
\end{pmatrix}
\end{equation}
be a Lorentz transformation. The defining relation 
(\ref{eq:DefRelLor}), as well as the relation 
$L^\alpha_\mu L^\beta_\nu g^{\mu\nu}=g^{\alpha\beta}$
which follows from it, are equivalent respectively to 
\begin{subequations}
\label{eq:1plus3Rel}
\begin{alignat}{6}
\label{eq:1plus3Rel1}
& \vec a^2&&=\gamma^2-1\,,\quad
&&\gamma\vec b&&=\mat{M}\cdot\vec a\,,\quad
&&\mat{M}\cdot\mat{M}^\top&&=\mat{1}_3+\vec b\otimes\vec b^\top\,,\\
\label{eq:1plus3Rel2}
& \vec b^2&&=\gamma^2-1\,,\quad
&&\gamma\vec a&&=\mat{M}^\top\cdot\vec b\,,\quad
&&\mat{M}^\top\cdot\mat{M}&&=\mat{1}_3+\vec a\otimes\vec a^\top\,.
\end{alignat}
\end{subequations}

The polar decomposition of the matrix $L\in\group{O(1,3)}_+^\uparrow$ 
in (\ref{eq:1plus3LT}) is given by    
\begin{equation}
\label{eq:PolDecLor1}
L=B\cdot R
\end{equation}
with 
\begin{equation}
\label{eq:PolDecLor2}
B=
\begin{pmatrix}
\gamma&\quad &\vec b^\top\\
\vec b&\quad &\mat{1}_3+\frac{\vec b\otimes\vec b^\top}{1+\gamma}
\end{pmatrix}\,,\qquad
R=
\begin{pmatrix}
1&\quad &\vec 0^\top\\
\vec 0&\quad &\mat{M}-\frac{\vec b\otimes\vec a^\top}{1+\gamma}
\end{pmatrix}\,.
\end{equation}
Tho check this, first verify that $L$ is indeed the product $B\cdot R$ 
using the relations (\ref{eq:1plus3Rel}). Next we note that $B$ is 
symmetric and that its eigenvalues (EV) are all positive:  
\begin{equation}
\label{eq:BisPosDef}
\text{EV}(B)=(\gamma+\sqrt{\gamma^2-1}\,,\,
\gamma-\sqrt{\gamma^2-1}\,,\,1\,,\,1)\ >\ 0\,.
\end{equation}
Finally one checks that $R$ is a spatial rotation, i.e. 
\begin{equation}
\label{eq:DisRot}
\mat{D}:= 
\mat{M}-\frac{\vec b\otimes\vec a^\top}{1+\gamma}\ \in\ \group{SO}(3)\,.
\end{equation}
Indeed, $\mat{D}\cdot\mat{D}^\top=\mat{1}_3$ is easily verified 
using the relations (\ref{eq:1plus3Rel}) and $\det(\mat{D})=1$ 
follows from $\det(L)=\det(B)=1$. Hence we have found the polar 
decomposition of $L\in\group{O}^\uparrow_+(1{,}3)$. 

We can now characterize the factors $B$ (boost) and $R$ (rotation) 
of $L$ in terms of the parameters $\gamma,\vec a,\vec b,\mat{M}$ 
in (\ref{eq:1plus3LT}). We start with $R$: Using the first and second 
equation in (\ref{eq:1plus3Rel1}) one readily 
shows that 
\begin{equation}
\label{eq:DRotPlane}
\mat{D}\cdot\vec a=\vec b\,.
\end{equation}
Hence the plane of rotation for $\mat{D}$ is 
$\Span\{\vec a,\vec b\}\subset\reals^3$. The rotation angle 
$\theta$ obeys 
\begin{subequations}
\label{eq:ThomRotAngle}
\begin{equation}
\label{eq:ThomRotAngle1}
\cos\theta=\frac{\vec a\cdot\vec b}{\gamma^2-1}\,,
\end{equation}
where we used $\vec a^2=\vec b^2=\gamma^2-1$ (first 
equations in (\ref{eq:1plus3Rel})). On the other hand, it 
evidently also obeys the general equation 
$1+2\cos\theta=\trace(\mat{D})$, i.e. 
\begin{equation}
\label{eq:ThomRotAngle2}
1+2\cos\theta=\trace(\mat{M})-\frac{\vec a\cdot\vec b}{1+\gamma}\,.
\end{equation}
Elimination of $\vec a\cdot\vec b$ via (\ref{eq:ThomRotAngle1}) 
gives
\begin{equation}
\label{eq:ThomRotAngle3}
\cos\theta=\frac{\trace(\mat{M})-1}{1+\gamma}\,.
\end{equation}
\end{subequations}

Next we set $\vec\beta:=\vec b/\gamma$, $\beta:=\Vert\vec\beta\Vert$,
and $\hat{\vec\beta}:=\vec\beta/\beta$; then    
\begin{equation}
\label{eq:LorTransPhysPar1}
\gamma=\gamma(\vec\beta):=1/\sqrt{1-\beta^2}\,,\qquad
\vec b=\gamma\vec\beta\,,\qquad
\vec a=\gamma \mat{D}^\top\cdot\vec\beta\,.
\end{equation}
Writing $B$ in terms of $\vec\beta$ explicitly shows that it is a 
boost with parameter $\vec\beta=\vec v/c$.  

The general Lorentz transformation (\ref{eq:1plus3LT}), instead of 
being considered as function of $\gamma,\vec a,\vec b,\mat{M}$ 
obeying (\ref{eq:1plus3Rel}), can now be considered as function of 
$\vec\beta$ and $\mat{D}$,
\begin{equation}
\label{eq:LorTransPhysPar2}
L(\vec\beta,\mat{D})=
\underbrace{
\begin{pmatrix}
\gamma &\quad & \gamma\vec\beta^\top\\
\gamma\vec\beta &\quad & \mat{1}_3+(\gamma-1)\,
\hat{\vec\beta}\otimes\hat{\vec\beta}^\top
\end{pmatrix}}_{\displaystyle =:B(\vec\beta)}
\underbrace{
\begin{pmatrix}
1&\vec 0^\top\\
\vec 0 &\mat{D}
\end{pmatrix}}_{\displaystyle =:R(\mat{D})}\,,
\end{equation}
where $\gamma$ is now understood as function of $\vec\beta$ as 
in (\ref{eq:LorTransPhysPar1}). The only restrictions on the 
parameters being that $\mat{D}\in\group{SO(3)}$ and 
$\vec\beta\in\mathcal{B}_1\subset\reals^3$,
where $\mathcal{B}_1$ denotes the ball of unit radius centered at 
the origin (cf. (\ref{eq:DefUnitBall})). The decomposition 
(\ref{eq:LorTransPhysPar2}) should be regarded as the analog of 
(\ref{eq:1plus3HomGal}).

It is easy to check directly that the boost are indeed equivariant 
with respect to rotations: 
\begin{equation}
\label{eq:EquivBoostRot}
R(\mat{D})\cdot B(\vec\beta)\cdot R(\mat{D^{-1}})=B(\mat{D}\cdot\vec\beta)\,.
\end{equation}
The polar decomposition is unique once the order of rotations and boosts
are fixed. In (\ref{eq:PolDecLor1}) we had put the rotations to the 
right, i.e. one first rotates and then boosts (we think actively). 
Had we chosen the opposite order the rotation parameter would still be  
$\mat{D}$ but the boost parameter would change to 
$\mat{D}^\top\cdot\vec\beta$. This follows immediately from 
(\ref{eq:EquivBoostRot}).

\subsection{The Lie algebras of the Lorentz and Galilei groups}
\label{sec:LorentzLiealgebra}
The commutation relations of the Lorentz Lie-algebra follow from the 
general formula (\ref{eq:LieAlgGenPoin}), where we have to set $\epsilon=1$. 
Here we shall rename the generators $M_{ab}$, where $a,b\in\{0,1,2,3\}$, 
in the way explained below (indices $i,j,k$ are in $\{1,2,3\}$). 
For direct comparison with  (\ref{eq:LieAlgGenPoin}) we also give their 
expression in terms of the defining representation (i.e. as elements 
of $\End(\reals^4)$). So let $\{e_a\}_{a=0\cdots 3}$ and 
$\{\eta^a\}_{a=0\cdots 3}$ be dual bases of $\reals^4$ and 
$\eta_a:=g_{ab}\eta^ b$, where $g_{ab}:=g(e_a,e_b)$ 
(cf. Sect.\,\ref{sec:IndexMoving}). Then:
\begin{subequations}
\label{eq:LorentzLieGenerators}
\begin{alignat}{3}
\label{eq:LorentzLieGenerators1}
& J_i 
 &&\,:=\,\tfrac{1}{2}\varepsilon_{ijk}M_{jk} 
 &&\,=\,\varepsilon_{ijk}e_j\otimes\eta_k\,,\\
\label{eq:LorentzLieGenerators2}
& K_i 
 &&\,:=\,\tfrac{1}{c}\,M_{i0}
 &&\,=\,\tfrac{1}{c}\,(e_i\otimes\eta_0-e_0\otimes\eta_i)\,,\\
\label{eq:LorentzLieGenerators3}
& P_i
 &&\,:=\,T_i
 &&\,=\, e_i\,,\\
\label{eq:LorentzLieGenerators4}
& E
 &&\,:=\,c\,T_0
 &&\,=\,c\,e_0\,.
\end{alignat}
\end{subequations}
These generate active rotations, boosts, translations in space,
and translations in time respectively. The reason for the factors 
of $1/c$ in (\ref{eq:LorentzLieGenerators2}) and $c$ in 
(\ref{eq:LorentzLieGenerators4}) is as follows: We wish the $K_i$ 
to be the generators of boosts with velocity parameters $v^i$ 
(rather than $\beta^i=v^i/c$), i.e. $\exp(M_{i0}\beta^i)=\exp(K_iv^i)$.
Similarly, we wish $E$ to be the generator of time translation 
with parameter $\Delta t$ (rather than $\Delta x^0=c\Delta t$), i.e. 
$\exp(T_0\Delta x^0)=\exp(E\Delta t)$. This puts the $K_i$ 
and $E$ in quantitative analogy to the corresponding generators 
in the Galilei group and hence facilitates a direct comparison.   

The relations (\ref{eq:LieAlgGenPoin}) now amount to 
\begin{subequations}
\label{eq:LorentzLieCommutators}
\begin{alignat}{2}
\label{eq:LorentzLieCommutatorsa}
& [J_i,J_j] &&\ =\ \varepsilon_{ijk}\,J_k        \\
\label{eq:LorentzLieCommutatorsb}
& [J_i,K_j] &&\ =\ \varepsilon_{ijk}\,K_k        \\
\label{eq:LorentzLieCommutatorsc}
& [K_i,K_j] &&\ =\ -\,\varepsilon_{ijk}\,J_k/c^2 \\
\label{eq:LorentzLieCommutatorsd}
& [J_i,P_j] &&\ =\ \varepsilon_{ijk}\,P_k        \\
\label{eq:LorentzLieCommutatorse}
& [J_i,E]   &&\ =\ 0                             \\
\label{eq:LorentzLieCommutatorsf}
& [K_i,P_j] &&\ =\ \delta_{ij}\,E/c^2            \\
\label{eq:LorentzLieCommutatorsg}
& [K_i,E]   &&\ =\ P_i                           \\
\label{eq:LorentzLieCommutatorsh}
& [P_i,P_j] &&\ =\ 0                             \\
\label{eq:LorentzLieCommutatorsi}
& [P_i,E]   &&\ =\ 0\,.
\end{alignat}
\end{subequations}
Those involving $J_i$ on the left hand side just tell us that the 
other quantity in the bracket is either a spatial vector
or scalar. According to (\ref{eq:LorentzLieCommutatorsa}) the $J_i$
form a Lie subalgebra but, as e.g. (\ref{eq:LorentzLieCommutatorsb}) 
shows, not an ideal (cf. Sect.\,\ref{sec:LieAlgebrasGenCons}).
In contrast, (\ref{eq:LorentzLieCommutatorsc})
shows that the $K_i$ do not form a Lie subalgebra. The $J_i,K_i$
span the Lie algebra of $\group{O}(1{,}3)$ and it is easy to prove from 
the first three relations above that it is simple (has no non-trivial 
ideals). Moreover, any of the ten generators appears on the right 
hand side of some relation (\ref{eq:LorentzLieCommutators}), i.e. 
can be written as a commutator. This means that the Lie algebra of 
the inhomogeneous Lorentz group is \emph{perfect} (i.e. generated 
by commutators). 

Another fact easily seen from 
(\ref{eq:LorentzLieCommutatorsa}-\ref{eq:LorentzLieCommutatorsc})
is that the Lie algebra of the homogeneous Lorentz group is the 
`complex double' (my terminology, see below) of the Lie algebra 
of $\group{SO}(3)$. Let us explain this. Given a real Lie algebra 
$L$ of dimension $n$, we consider the \emph{real} vector space 
$\complex\otimes L$ of dimension $2n$. Here 
$\complex$ is considered as two-dimensional real vector space 
and $\otimes$ is clearly also taken over $\reals$. $\complex\otimes L$
can be made into a real $2n$-dimensional Lie algebra by defining 
$[z_1\otimes X_1\,,\,z_2\otimes X_2]:=z_1z_2\otimes[X_1,X_2]$ and  
$\reals$-linear extension. This is easily checked to satisfy all 
axioms (\ref{eq:DefLieAlg}). The \emph{complex double} of $L$ is now 
defined to be the real Lie algebra $\complex\otimes L$. For sure, 
$\complex\otimes L$ has a natural complex structure, which allows 
to consider it as $n$-dimensional \emph{complex} Lie algebra. In 
this case we\footnote{The terminology used here is non-standard. 
Often the distinction between $\complex\otimes L$ and 
$L^{\complex}$ is not explicitly made, and even if it is, both 
are called `the complexification' of $L$.} would call it 
$L^{\complex}$, the \emph{complexification} of $L$. However, we 
are interested in Lie algebras of Lie groups, which a priori are 
always considered as real (cf. Sect.\,\ref{sec:LieAlgebras}), 
regardless of the possible existence of a complex structure. Now 
let $L$ be the Lie algebra of $\group{SO}(3)$, i.e. 
$L=\Span\{e_1,e_2,e_3\}$ where $[e_i,e_j]=\varepsilon_{ijk}e_k$. 
Consider $\complex\otimes L$ and set $R_j:=1\otimes e_j$ and 
$cK_j:=\text{i}\otimes e_j$, so that 
$\complex\otimes L=\Span\{R_1,R_2,R_3,K_1,K_2,K_3\}$. In this 
basis the Lie brackets are just given by    
(\ref{eq:LorentzLieCommutatorsa}-\ref{eq:LorentzLieCommutatorsc}),
showing that the homogeneous Lorentz Lie-algebra is indeed the 
complex double of the Lie algebra of $\group{SO}(3)$.  

The Lie algebra of the inhomogeneous Galilei group is formally 
obtained from (\ref{eq:LorentzLieCommutators}) by taking the 
limit $1/c^2\rightarrow 0$, to that the right hand sides of 
(\ref{eq:LorentzLieCommutatorsc}) and (\ref{eq:LorentzLieCommutatorsf}) 
are now replaced with zero. This causes big structural changes. 
For example, the generators of boosts now generate an Abelian ideal
in the homogeneous Galilei Lie-algebra (generated by $R_i,K_i$),
implying that it is not even semisimple, whereas we just said that 
the homogeneous Lorentz Lie-algebra is simple. In the inhomogeneous 
Galilei Lie-algebra the $K_i$ and $P_i$ together generate an Abelian 
ideal. It is not perfect since the $K_i$ and $E$ do not occur 
on the right hand sides anymore.   

One might argue that it is physically incorrect to take $E/c^2$ to 
zero in the limit $c\rightarrow\infty$. Rather, $E\rightarrow\infty$ 
as $c\rightarrow\infty$ since $E$ contains a contribution $mc^2$ 
from the rest-energy of the system ($m$ denotes the rest mass,
which we wish to keep at a finite value). 
Hence, for an isolated system, one should rather set 
$E=mc^2+E_0$ and therefore have $E/c^2\rightarrow m$ 
in the limit as $c\rightarrow\infty$. Then the right hand side of 
(\ref{eq:LorentzLieCommutatorsc}) is still zero in this limit but  
the right hand side of (\ref{eq:LorentzLieCommutatorsf}) becomes 
proportional $m\,\delta_{ij}$, where $m$ is now read as a new element 
of the Lie algebra that commutes with all other elements, i.e. lies 
in the center (in any irreducible representation it is therefore 
written as $m\mathbf{1}$ where $\mathbf{1}$ is the unit operator).
Also, due to $m$ being central, (\ref{eq:LorentzLieCommutatorsg}) is 
maintained with $E_0$ replacing $E$.
 
The 11-dimensional Lie algebra so obtained is well known. It is a 
central extension of the inhomogeneous Galilei Lie-algebra, out of 
a unique 1-parameter family of inequivalent central extensions, 
labeled by the value of $m$. As is well known, it is this 
extension (and the corresponding 11-dimensional Lie group, 
sometimes called the Schr\"odinger group), which implement the 
Galilean symmetries in quantum mechanics by proper representations,
whereas the inhomogeneous Galilei group only acts by 
ray-representations. Formally, the central element $m$ then gives 
rise to superselection rules. There are certain analogs of this 
on the classical level; see~\cite{Giulini:1996}. 

The formal process by which the (inhomogeneous and homogeneous)
Galilei Lie-algebra emerges from the Lorentz Lie-algebra is 
a special case of what is called a \emph{contraction}, which 
was introduced in \cite{InonuWigner:1953} just in order to understand 
precisely the way in which the Galilei Lie-algebra and group can be 
understood as limiting case of the Lorentz Lie-algebra and group 
respectively. The general idea can be briefly described as follows: 
Consider a Lie algebra $L$ with decomposition into two linear subspaces 
$L=H\oplus H'$, none of which we a priori assume to be a Lie 
subalgebra. Choose an adapted basis 
$\{X_1,\cdots X_n,X'_1,\cdots X'_{n'}\}$ such that the unprimed 
elements span $H$ and the primed elements $H'$. The Lie brackets 
have the general form
\begin{subequations}
\label{eq:LieBracketsOrig}
\begin{alignat}{3}
\label{eq:LieBracketsOrig1}
& [X_a,X_b]&&\,=\,C^c_{ab}X_c&&+\,C^{c'}_{ab}X'_{c'}\,,\\
\label{eq:LieBracketsOrig2}
& [X_a,X'_{b'}]&&\,=\,C^c_{ab'}X_c&&+\,C^{c'}_{ab'}X'_{c'}\,,\\
\label{eq:LieBracketsOrig3}
& [X'_{a'},X'_{b'}]&&\,=\,C^c_{a'b'}X_c&&+\,C^{c'}_{a'b'}X'_{c'}\,.
\end{alignat}
\end{subequations}
We now rescale the primed generators, leaving the unprimed ones 
untouched,  
\begin{subequations}
\label{eq:RescaleGen}
\begin{alignat}{4}
\label{eq:RescaleGen1}
& X_a && \mapsto && Y_a &&\,:=\,X_a\,,\\
\label{eq:RescaleGen2}
& X'_{a'} && \mapsto && Y'_{a'} &&\,:=\,\epsilon\,X'_{a'}
\end{alignat}
\end{subequations}
and write down (\ref{eq:LieBracketsOrig}) in terms of the new 
basis:
\begin{subequations}
\label{eq:LieBracketsResc}
\begin{alignat}{3}
\label{eq:LieBracketsResc1}
& [Y_a,Y_b]&&\,=\,C^c_{ab}Y_c
  &&+\ \epsilon^{-1}\,C^{c'}_{ab}Y'_{c'}\,,\\
\label{eq:LieBracketsResc2}
& [Y_a,Y'_{b'}]&&\,=\,\epsilon\,C^c_{ab'}Y_c
  &&+\ C^{c'}_{ab'}Y'_{c'}\,,\\
\label{eq:LieBracketsResc3}
& [Y'_{a'},Y'_{b'}]&&\,=\,\epsilon^2\,C^c_{a'b'}Y_c
&&+\ \epsilon\,C^{c'}_{a'b'}Y'_{c'}\,.
\end{alignat}
\end{subequations}
We wish to formally take the limit $\epsilon\rightarrow 0$. 
Clearly this cannot be done unless the terms $\propto\epsilon^{-1}$ 
in (\ref{eq:LieBracketsResc1}) all vanish, i.e. unless $C^{c'}_{ab}=0$, 
which is equivalent to saying that $H:=\Span\{X_1,\cdots X_n\}$ must be 
a Lie-subalgebra of $L$. Assuming that this is the case, the limit 
can be taken and the following Lie algebra emerges: 
\begin{subequations}
\label{eq:LieBracketsLimit}
\begin{alignat}{2}
\label{eq:LieBracketsLimit1}
& [Y_a,Y_b]&&\ =\ C^c_{ab}Y_c\,,\\
\label{eq:LieBracketsLimit2}
& [Y_a,Y'_{b'}]&&\ =\ C^{c'}_{ab'}Y'_{c'}\,,\\
\label{eq:LieBracketsLimit3}
& [Y'_{a'},Y'_{b'}]&&\ =\ 0\,.
\end{alignat}
\end{subequations}
Thus we see that in the limit the subalgebra $H$ survives whereas 
the linear space $H'$ turns into an Abelian ideal. Hence the 
limit Lie algebra is a semi-direct sum of the original 
Lie subalgebra $H$ with the Abelian ideal $H'$. It is called the 
\emph{contraction of $L$ over $H$}, since $H$ stays intact and the 
rest is contracted. On the level of Lie groups one might think of 
the contracted group (the group generated by $H'$) as an 
infinitesimal neighborhood of the group one contracts over 
(the group generated by $H$) within the full Lie group 
(the group generated by $L$).  

This applies to the transition Lorentz $\rightarrow$ Galilei 
as follows: In the homogeneous case, we decompose the Lorentz 
Lie-algebra into  the Lie subalgebra $H=\Span\{J_1,J_2,J_3\}$ 
and the linear subspace $H'=\Span\{K_1,K_2,K_3\}$, and then 
contract it over $H$ to obtain the homogeneous Galilei Lie-algebra. 
In the inhomogeneous case we set $H=\Span\{J_1,J_2,J_3,E\}$, which 
is indeed a Lie subalgebra as seen from (\ref{eq:LorentzLieCommutators}),
and $H'=\Span\{K_1,K_2,K_3,P_1,P_2,P_3\}$. Contracting over $H$ 
then just results in making $H'$ Abelian, i.e. annihilating the 
right hand sides of (\ref{eq:LorentzLieCommutatorsc}) and 
(\ref{eq:LorentzLieCommutatorsf}), which just results in the 
inhomogeneous Galilei Lie-algebra. Its structure as semi-direct 
sum with $H'$ as Abelian ideal is just the Lie-algebra analog 
of the semi-direct product structure (\ref{eq:InGalGroup2}).

\subsection{Composing boosts}
\label{sec:CompBoosts}
After this digression into Lie algebras we return to the level of 
groups. More specifically, we are now interested in the composition 
of two boosts, $B(\vec\beta_1)$ and $B(\vec\beta_2)$. The matrix 
product can be easily computed using the explicit form of 
$B(\vec\beta)$ as given in (\ref{eq:LorTransPhysPar2}). 
We set $\gamma_i:=\gamma(\vec\beta_i)$
for $i=1,2$ and denote by $\vec\beta_{i\Vert}$ and 
$\vec\beta_{i\perp}$ the components of $\vec\beta_i$ parallel and 
perpendicular to the other velocity respectively. The angle between 
$\vec\beta_1$ and $\vec\beta_2$ is denote by $\varphi$, i.e. 
$\hat{\vec\beta}_1\cdot\hat{\vec\beta}_2=\cos\varphi$. Then the 
matrix product has the general form (\ref{eq:1plus3LT}), where 
\begin{subequations}
\label{eq:CompBoostsParam}
\begin{alignat}{2}
\label{eq:CompBoostsParam1}
& \gamma &&\,=\,\gamma_1\gamma_1(1+\vec\beta_1\cdot\vec\beta_2)
           \,=\,\gamma_1\gamma_1(1+\beta_1\beta_2\cos\varphi)\,, \\
\label{eq:CompBoostsParam2}
& \vec a &&\,=\,\gamma_1\gamma_2\bigl(\vec\beta_2+
  \vec\beta_{1\Vert}+\gamma_2^{-1}\vec\beta_{1\perp}\bigr)\,,\\
\label{eq:CompBoostsParam3}
& \vec b &&\,=\,\gamma_1\gamma_2\bigl(\vec\beta_1+
  \vec\beta_{2\Vert}+\gamma_1^{-1}\vec\beta_{2\perp}\bigr)\,,\\
\label{eq:CompBoostsParam4}
& \mat{M}&&\,=\,\mat{1}_3
  +(\gamma_1-1)\,\hat{\vec\beta}_1\otimes\hat{\vec\beta}_1^\top
   +(\gamma_2-1)\,\hat{\vec\beta}_2\otimes\hat{\vec\beta}_2^\top\nonumber\\
& &&\hspace{0.8cm}+\bigl(\beta_1\gamma_1\beta_2\gamma_2+(\gamma_1-1)(\gamma_2-1)
   \hat{\vec\beta}_1\cdot\hat{\vec\beta}_2\bigr)\ \hat{\vec\beta}_1
   \otimes{\hat{\vec\beta}}^\top_2\,.
\end{alignat}
\end{subequations}
The resulting boost and rotation parameters will be called 
$\vec\beta=\vec\beta_1\star\vec\beta_2$ and 
$\mat{D}=\mat{T}[\vec\beta_1,\vec\beta_2]$ respectively. Hence we have: 
\begin{equation}
\label{eq:CompBoosts}
B(\vec\beta_1)\cdot B(\vec\beta_2)=
B(\vec\beta_1\star\vec\beta_2)\cdot
R(\mat{T}[\vec\beta_1,\vec\beta_2])\,.
\end{equation}
The operation $\star$ entails the law of how to compose velocities 
in SR. $T[\vec\beta_1,\vec\beta_2]$ is called the `Thomas rotation'. 
Its existence (i.e. it being non trivial) means that pure boosts 
do not form a subgroup in the Lorentz group, in contrast to the 
Galilei group.  

The functional form of the $\star$ operation follows from 
(\ref{eq:CompBoostsParam}), since 
$\vec\beta_1\star\vec\beta_2=\vec b/\gamma$:
\begin{equation}
\label{eq:StarOp}
\vec\beta_1\star\vec\beta_2=
\frac{\vec\beta_1+\vec\beta_{2\Vert}+
\gamma_1^{-1}\vec\beta_{2\perp}}{1+\vec\beta_1\cdot\vec\beta_2}\,.
\end{equation}
Comparing (\ref{eq:CompBoostsParam2}) with (\ref{eq:CompBoostsParam3})
shows $\vec a/\gamma=\vec\beta_2\star\vec\beta_1$. Equation 
(\ref{eq:DRotPlane}) then shows 
\begin{equation}
\label{eq:StarNonComm}
\vec\beta_1\star\vec\beta_2
=\mat{T}[\vec\beta_1,\vec\beta_2]\cdot (\vec\beta_2\star\vec\beta_1)\,,
\end{equation}
which in turn implies (we write $\mat{T}^{-1}[-,-]$ for the inverse matrix
$(\mat{T}[-,-])^{-1}$) 
\begin{equation}
\label{eq:ThomasInverse}
\mat{T}[\vec\beta_1,\vec\beta_2]
=\mat{T}^{-1}[\vec\beta_2,\vec\beta_1]\,.
\end{equation}
Let now $\mat{D}\in\group{SO(3)}$ be any rotation; then 
(\ref{eq:StarOp}) shows that $\star$ obeys 
\begin{equation}
\label{eq:SO3EqivStar}
(\mat{D}\cdot\vec\beta_1)\star(\mat{D}\cdot\vec\beta_2)=
\mat{D}\cdot(\vec\beta_1\star\vec\beta_2)\,,
\end{equation}
which combined with (\ref{eq:StarNonComm}) also shows that  
\begin{equation}
\label{eq:SO3EqivThom}
\mat{T}[\mat{D}\cdot\vec\beta_1,\mat{D}\cdot\vec\beta_2]=
\mat{D}\cdot\mat{T}[\vec\beta_1,\vec\beta_2]\cdot\mat{D}^{-1}\,.
\end{equation}

The Thomas rotation takes place in the plane 
$\Span\{\vec a,\vec b\}=\Span\{\vec\beta_1,\vec\beta_2\}$. The cosine 
of the angle of rotation, $\theta$, follows from (\ref{eq:ThomRotAngle3})
and (\ref{eq:CompBoostsParam4}): 
\begin{subequations}
\label{eq:CosThomas}
\begin{equation} 
\label{eq:CosThomas1}
\cos\theta=1-\frac{(\gamma_1-1)(\gamma_2-1)}{\gamma+1}\sin^2\varphi\,,
\end{equation}
where we used  (\ref{eq:CompBoostsParam1}) to eliminate a term 
$\propto\cos\varphi$. It shows that 
$\mat{T}[\vec\beta_1,\vec\beta_2]=\mat{1}_3$ iff $\vec\beta_1$
and $\vec\beta_2$ are either parallel ($\varphi=0$) or anti-parallel 
($\varphi=\pi$). We can now again  make use of (\ref{eq:CompBoostsParam1}) 
to eliminate $\gamma$ in favor of $\gamma_1,\gamma_2$, and 
$\cos\varphi$, so as to make $\cos\theta$ a function of the moduli 
$\beta_1,\beta_2$ of the velocities and the angle $\varphi$ between them: 
\begin{equation}
\label{eq:CosThomas2}
\cos\theta =
1-\frac{(\gamma_1-1)(\gamma_2-1)\sin^2\varphi}%
{1+\gamma_1\gamma_2+\sqrt{(\gamma_1^2-1)(\gamma_2^2-1)}\,\cos\varphi}\,.
\end{equation}
Alternatively we can use  (\ref{eq:CompBoostsParam1}) to express 
$\cos\theta$ as function of the tree moduli $\beta_1,\beta_2$, and 
$\beta=\Vert\vec\beta_1\star\vec\beta_2\Vert$, which assumes a 
nice symmetric form:\footnote{This derivation, albeit straightforward, 
is a little tedious. A more elegant derivation, using Clifford algebra, 
is given in~\cite{Urbantke:1990}.}
\begin{equation}
\label{eq:CosThomas3}
\cos\theta=\frac{(1+\gamma+\gamma_1+\gamma_2)^2}%
{(1+\gamma)(1+\gamma_1)(1+\gamma_2)}-1\,.
\end{equation}
\end{subequations}

\begin{figure}[htb]
\centering\epsfig{figure=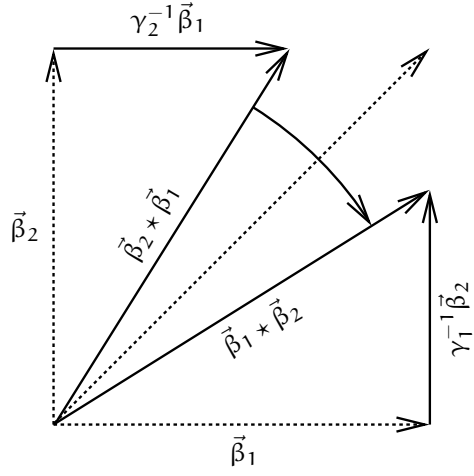,width=0.4\linewidth}
\put(-80,-10){\small $\vec\beta_1$}
\put(-163,75){\small $\vec\beta_2$}
\put(-80,30){\small\begin{rotate}{30}$\vec\beta_1\star\vec\beta_2$\end{rotate}}
\put(-117,154){\small $\gamma_2^{-1}\vec\beta_1$}
\put(9,30){\small\begin{rotate}{90}$\gamma_1^{-1}\vec\beta_2$\end{rotate}}
\put(-115,65){\small\begin{rotate}{60}$\vec\beta_2\star\vec\beta_1$\end{rotate}}
\caption{%
\label{fig:SenseThomas}
Addition of perpendicular velocities $\vec\beta_1$ and $\vec\beta_2$
of modulus $\beta_1=\beta_2=0.78$ so that $\gamma^{-1}_1=\gamma^{-1}_2=5/8$. 
In this case (\ref{eq:StarOp}) gives 
$\vec\beta_1\star\vec\beta_2=\vec\beta_1+\tfrac{5}{8}\vec\beta_2$
and likewise  
$\vec\beta_2\star\vec\beta_1=\vec\beta_2+\tfrac{5}{8}\vec\beta_1$.
For comparison, the dashed arrow corresponds to the `classically'
composed velocities (vector addition). According to 
(\ref{eq:StarNonComm}), the rotation $\mat{T}[\vec\beta_1,\vec\beta_2]$ 
turns $\vec\beta_2\star\vec\beta_1$ into $\vec\beta_1\star\vec\beta_2$, 
as indicated by the curved arrow.} 
\end{figure}
Figure\,\ref{fig:SenseThomas} illustrates the laws (\ref{eq:StarOp}) 
and (\ref{eq:StarNonComm}) of the Thomas rotation 
for a special case in which the two velocities are perpendicular. 
In such cases $\theta$ ranges between $0$ and $\pi/2$, as can be 
immediately deduced from (\ref{eq:CosThomas2}). The sense of the 
Thomas rotation in the $\vec\beta_1\vec\beta_2$-plane 
is negative (we orient this plane in the usual way, such that 
$\vec\beta_1\times\vec\beta_2$ defines the direction of the normal). 

Generally $\theta$ ranges between $0$ and $\pi$. More precisely, 
take fixed moduli $\beta_1$ and $\beta_2$ and consider $\cos\theta$ 
as function of $\varphi$ as given by (\ref{eq:CosThomas2}). 
For $\varphi=0$ and $\varphi=\pi$ this function has obvious maxima 
(where $\cos\theta=1$) and hence must have a minimum inbetween, which 
corresponds to a \emph{maximal} value of $\theta$. 
Using (\ref{eq:CosThomas2}) we compute that this maximum of $\theta$ 
occurs at a value $\varphi_m$ which obeys  
\begin{equation}
\label{eq:ThomasAngleMax1}
\cos\varphi_m
=\,-\,\sqrt{\frac{(\gamma_1-1)(\gamma_2-1)}{(\gamma_1+1)(\gamma_2+1)}}\,,
\end{equation}
(the negative sign shows that $\varphi_m>\pi/2$) and that the maximal 
value $\theta_m$ obeys 
\begin{equation}
\label{eq:ThomasAngleMax2}
\cos\theta_m
=1-2\ \frac{(\gamma_1-1)(\gamma_2-1)}{(\gamma_1+1)(\gamma_2+1)} 
=\,-\,cos(2\varphi_m)\,.
\end{equation}
Hence we see that $\theta$ becomes larger than $\pi/2$ for sufficiently 
large values of $\gamma_1$ and $\gamma_2$. 
For example, if $\beta_1=\beta_2=\beta$, i.e. 
$\gamma_1=\gamma_2=\gamma$, the value of $\beta$ above which 
$\theta_m$ exceeds $\pi/2$ is given by $2^{5/4}/(2^{1/2}+2)\approx 0.985$.
Equation (\ref{eq:ThomasAngleMax2}) also shows that $\theta_m$ 
approaches its maximal value, $\pi$, only if $\gamma_1$ \emph{and} 
$\gamma_2$ tend to infinity. In general, (\ref{eq:CosThomas2}) 
shows that in that limit $\cos\theta$ approaches $\cos\varphi$,
which means that $\theta$ approaches $2\pi-\varphi$, since the Thomas 
rotation is in the negative sense relative to the orientation of 
the $\vec\beta_1$-$\vec\beta_2$ plane. 

Finally, using (\ref{eq:EquivBoostRot}) and (\ref{eq:CompBoosts}), 
we can now write down the general composition law for Lorentz 
transformations:
\begin{equation}
\label{eq:GenCompLawLorentz}
L(\vec\beta_1,\mat{D}_1)\cdot 
L(\vec\beta_2,\mat{D}_2)=
L\bigl(\vec\beta_1\star\mat{D}_1\cdot\vec\beta_2\,,
\,\mat{T}[\vec\beta_1,\mat{D}_1\cdot\vec\beta_2]\cdot\mat{D}_1
\cdot\mat{D}_2\bigr)\,.
\end{equation}
Moreover, noting that $(B(\vec\beta))^{-1}=B(-\vec\beta)$,
equations~(\ref{eq:PolDecLor1},\ref{eq:EquivBoostRot}) also 
show that 
\begin{equation}
\label{eq:GenInvLawLorentz}
\bigl(L(\vec\beta,\mat{D})\bigr)^{-1}
=L(-\,\mat{D}^{-1}\cdot\vec\beta\,,\,\mat{D}^{-1})\,.
\end{equation}
Note that (\ref{eq:GenInvLawLorentz}) and (\ref{eq:GenInvLawGalilei}) 
are just the same analytic operations on the parameter spaces. The 
multiplication law (\ref{eq:GenCompLawLorentz}) now replaces the 
semi-direct product structure (\ref{eq:GenCompLawGalilei}) of the 
Galilei group, into which it turns in the limit $c\rightarrow\infty$.
Indeed, writing $\vec\beta=\vec v/c$, the operation $\star$ 
between the $\vec v$'s approaches $+$ and the Thomas rotation 
$\mat{T}[-,-]$ becomes the identity, as one e.g. sees from
(\ref{eq:CosThomas2}) for $\gamma_1,\gamma_2\rightarrow 1$.

\subsection{The algebraic structure of velocity composition}
\label{sec:AlgStrucVelAdd}
Let us say a little more about the algebraic structure behind 
(\ref{eq:StarOp}). First of all, $\star$ defines a map  
\begin{equation}
\label{eq:StarMap}
\star:\mathcal{B}_1\times\mathcal{B}_1\rightarrow\mathcal{B}_1\,,
\quad
(\vec\beta_1,\vec\beta_2)\mapsto\vec\beta_1\star\vec\beta_2\,,
\end{equation}
where 
\begin{equation}
\label{eq:DefUnitBall}
\mathcal{B}_1:=\{\vec\beta\in\reals^3\ \mid\ \Vert\vec\beta\Vert<1\}
\end{equation}
is the open ball in 3-dimensional Euclidean space (here space of 
velocities$/c$). That its image lies indeed in 
$\mathcal{B}_1\subset\reals^3$ follows from 
(\ref{eq:CompBoostsParam1}), which e.g. implies 
$\gamma<2\gamma_1\gamma_2$. Hence $\star$ makes $\mathcal{B_1}$
into a \emph{groupoid} (see below). Moreover, for each 
$\vec\beta\in\mathcal{B}_1$, we have 
\begin{equation}
\label{eq:StarUnit}
\vec 0\star\vec\beta=\vec\beta\star\vec 0=\vec\beta\,,
\end{equation}
so that $\vec 0$ is a unit with respect to $\star$. Each element 
also has an inverse (left and right): 
\begin{equation}
\label{eq:StarInverse}
\vec\beta\star(-\vec\beta)=(-\vec\beta)\star\vec\beta=\vec 0\,. 
\end{equation}

We already saw in (\ref{eq:StarNonComm}) that the Thomas rotation 
obstructs commutativity of $\star$. We now show that it also obstructs 
associativity. Consider the composition of three boosts 
$B(\vec\beta_1)\cdot B(\vec\beta_2)\cdot B(\vec\beta_3)$ and use 
associativity of matrix multiplication:      
\begin{subequations}
\label{eq:NonAssStar}
\begin{equation}
\label{eq:NonAssStar1}
 B(\vec\beta_1)\cdot\bigl(B(\vec\beta_2)\cdot B(\vec\beta_3)\bigr)
=\bigl(B(\vec\beta_1)\cdot B(\vec\beta_2)\bigr)\cdot B(\vec\beta_3)\,.
\end{equation}
Iterated application of (\ref{eq:CompBoosts}) shows that the left 
hand side is equal to 
\begin{equation}
\label{eq:NonAssStar2}
B\bigl(\vec\beta_1\star(\vec\beta_2\star\vec\beta_3)\bigr)\cdot
R(\mat{T}[\vec\beta_1,\vec\beta_2\star\vec\beta_3]\cdot
 \mat{T}[\vec\beta_2,\vec\beta_3])\,,
\end{equation}
whereas the right hand side equals (also making use of 
(\ref{eq:EquivBoostRot})), 
\begin{equation}
\label{eq:NonAssStar3}
B\bigl((\vec\beta_1\star\vec\beta_2)
\star(\mat{T}[\vec\beta_1,\vec\beta_2]\cdot\vec\beta_3)\bigr)
\cdot R\bigl(\mat{T}\bigl[\vec\beta_1\star\vec\beta_2\,,\,
 \mat{T}[\vec\beta_1,\vec\beta_2]\cdot\vec\beta_3\bigr]\cdot
 \mat{T}[\vec\beta_1,\vec\beta_2]\bigr)\,.
\end{equation}
\end{subequations}
Expressions (\ref{eq:NonAssStar2}) and (\ref{eq:NonAssStar3})
are in polar decomposed form. Uniqueness then implies equality 
of the boost and rotation factors separately. For the boosts this 
implies   
\begin{equation}
\label{eq:NonAssStar4}
\vec\beta_1\star(\vec\beta_2\star\vec\beta_3)
=(\vec\beta_1\star\vec\beta_2)\star
 \bigl(\mat{T}[\vec\beta_1,\vec\beta_2]\cdot\vec\beta_3\bigr)\,,
\end{equation}
which shows how the Thomas rotation obstructs associativity. 
The general identity obtained from equating the rotational parts 
of  (\ref{eq:NonAssStar2}) and (\ref{eq:NonAssStar3}) does not 
interest us here. Rather, we wish to consider the special case
where $\vec\beta_1=\vec\beta_3$. Then the product (\ref{eq:NonAssStar})
is a symmetric and positive definite matrix\footnote{For matrices it is 
generally true that if $B$ is positive definite and $A$ invertible, 
then $A\cdot B\cdot A^t$ is again positive definite. Note that 
here $A^t$ is the adjoint of $A$ with respect to the Euclidean inner 
product.}, that is, it is a pure boost and therefore (trivially) 
polar decomposed. Hence the rotational part in (\ref{eq:NonAssStar2}) 
must be the identity. This gives
\begin{equation}
\label{eq:LoopProperty}
\mat{T}[\vec\beta_1,\vec\beta_2]=
\mat{T}[\vec\beta_1,\vec\beta_2\star\vec\beta_1]=
\mat{T}[\vec\beta_1\star\vec\beta_2,\vec\beta_2]\,,
\end{equation}
where the second equality follows from the first by simultaneously 
taking the inverse and exchanging $\vec\beta_1$ and $\vec\beta_2$
(which leaves $\mat{T}[\vec\beta_1,\vec\beta_2]$ invariant 
according to (\ref{eq:ThomasInverse})).    

Now consider the following equation in $\vec\beta_1,\vec\beta_2$,
and $\vec\beta_3$:
\begin{subequations}
\label{eq:StarSolvability}
\begin{equation}
\label{eq:StarSolvability1}
\vec\beta_1\star\vec\beta_2=\vec\beta_3\,.
\end{equation}
Can we (uniquely) solve it for $\vec\beta_1$ given $\vec\beta_2$ and 
$\vec\beta_3$, or for $\vec\beta_2$ given $\vec\beta_1$ and 
$\vec\beta_3$? Since each $\vec\beta$ has an inverse, 
associativity would immediately answer this in the 
affirmative.\footnote{For then we could e.g. $\star$-multiply  
(\ref{eq:StarSolvability1}) with $-\vec\beta_2$ from the right and 
get on the left hand side $(\vec\beta_1\star\vec\beta_2)\star%
(-\vec\beta_2)=\vec\beta_1\star(\vec\beta_2\star(-\vec\beta_2))=%
\vec\beta_1$.} But associativity fails to hold. However, the 
answer is still affirmative: 

\begin{proposition}
\label{prop:StarSolvability}
The unique solutions of~(\ref{eq:StarSolvability1}) for $\vec\beta_1$
and $\vec\beta_2$ are given by 
\begin{alignat}{2}
\label{eq:StarSolvability2}   
& \vec\beta_1\,&&=\,
\vec\beta_3\star (-\mat{T}[\vec\beta_3,\vec\beta_2]\cdot\vec\beta_2)\,,\\
\label{eq:StarSolvability3}  
& \vec\beta_2\,&&=\, (-\vec\beta_1)\star\vec\beta_3\,.
\end{alignat}
\end{proposition}
\end{subequations}
\begin{proof}
(\ref{eq:StarSolvability3}) immediately follows from $\star$-multiplying 
(\ref{eq:StarSolvability1}) with $-\vec\beta_1$ from the left and using 
(\ref{eq:NonAssStar4}), taking into account that 
$\mat{T}[-\vec\beta_1,\vec\beta_1]=\mat{1}_3$. The proof of 
(\ref{eq:StarSolvability2}) is more difficult. One way that is not 
just `guessing and verifying', but rather arrives at the solution in 
a more systematic fashion, is to go back to the group level 
and consider the corresponding equation 
\begin{subequations}
\label{eq:StarSolvabilityProofA}
\begin{equation}
\label{eq:StarSolvabilityProofA1}
L(\vec\beta_1,\mat{D_1})\cdot L(\vec\beta_2,\mat{D}_2)=
L(\vec\beta_3,\mat{D}_3)\,,
\end{equation}
whose parameter form is   
\begin{alignat}{2}
\label{eq:StarSolvabilityProofA2}
& \vec\beta_3&&\,=\,\vec\beta_1\star\mat{D_1}\cdot\vec\beta_2\,,\\
\label{eq:StarSolvabilityProofA3}
& \mat{D_3} && \,=\,\mat{T}[\vec\beta_1,\mat{D}_1\cdot\vec\beta_2]%
\cdot\mat{D}_1\cdot\mat{D}_2\,.
\end{alignat}
\end{subequations}
The group structure now tells us that the unique solution for 
$L(\vec\beta_1,\mat{D}_1)$ is, using (\ref{eq:GenInvLawLorentz}),
\begin{subequations}
\label{eq:StarSolvabilityProofB}
\begin{equation} 
\label{eq:StarSolvabilityProofB1}
L(\vec\beta_1,\mat{D}_1)=L(\vec\beta_3,\mat{D}_3)\cdot 
L(-\mat{D}_2^{-1}\cdot\vec\beta_2,\mat{D}^{-1}_2)
\end{equation}
whose parameter form is 
\begin{alignat}{2}
\label{eq:StarSolvabilityProofB2}
& \vec\beta_1&&\,=\,\vec\beta_3\star(-\mat{D}_3\cdot%
  \mat{D}^{-1}_2\cdot\vec\beta_2)\,,\\
\label{eq:StarSolvabilityProofB3}
& \mat{D}_1 && \,=\,\mat{T}[\vec\beta_3\,,\,-\,\mat{D}_3\cdot%
  \mat{D}_2^{-1}\cdot\vec\beta_2]\cdot\mat{D}_3\cdot\mat{D}_2^{-1}\,.
\end{alignat}
\end{subequations}
Due to the group structure (\ref{eq:StarSolvabilityProofA}) and 
(\ref{eq:StarSolvabilityProofB}) are equivalent. In particular, 
(\ref{eq:StarSolvabilityProofB}) is a consequence of 
(\ref{eq:StarSolvabilityProofA}). We now specialize to the case
$\mat{D}_1=\mat{1}_3$, in which (\ref{eq:StarSolvabilityProofA2})
just becomes (\ref{eq:StarSolvability1}). Equation 
(\ref{eq:StarSolvabilityProofA3}) then becomes
\begin{alignat}{2}
\label{eq:StarSolvabilityC}
\mat{D}_3\cdot\mat{D}_2^{-1} 
&\,=\,\mat{T}[\vec\beta_1,\vec\beta_2] &&\nonumber\\
&\,=\,\mat{T}[\vec\beta_1\star\vec\beta_2\,,\,\vec\beta_2] 
  \qquad &&\text{using (\ref{eq:LoopProperty})}\nonumber\\
&\,=\,\mat{T}[\vec\beta_3,\vec\beta_2] 
  \qquad &&\text{using (\ref{eq:StarSolvabilityProofA2})\,.}
\end{alignat}
Inserting this into (\ref{eq:StarSolvabilityProofB2}) gives  
(\ref{eq:StarSolvability2}).   
\end{proof}

Let us relate these findings to some algebraic terminology. 
A \emph{groupoid} is a set $S$ with some map 
$\phi:S\times S\rightarrow S$. Hence $\star$ makes the open unit 
ball $\mathcal{B}_1\subset\reals^3$ into a groupoid. 
An associative groupoid is called a \emph{semigroup} (so we don't 
have a semigroup) . A groupoid $S$ is called a \emph{quasigroup}
if for any pair $(a,b)\in S\times S$ there is a unique pair 
$(x,y)\in S\times S$ such that $\phi(x,a)=b$ and $\phi(a,y)=b$. 
In our case we have just seen that the unique pair $(x,y)$ 
associated to $(a,b)=(\vec\beta_1,\vec\beta_2)$ is 
$x=\vec\beta_2\star(-\mat{T}[\vec\beta_2,\vec\beta_1]\cdot\vec\beta_1)$ 
and $y=(-\vec\beta_1)\star\vec\beta_2$. If a common unit element 
exists, as in (\ref{eq:StarUnit}), one calls it a 
\emph{quasigroup with unit} or simply a \emph{loop}. 
Note that in this case the existence of a unique inverse for each 
element follows. In some sense a loop is as close as you can get to 
the structure of a group if you drop associativity. This is the 
algebraic structure of velocity space in SR. Much original work 
on this has been done by A.\,Ungar, starting with \cite{Ungar:1988},
where e.g. the precise way in which strict associativity fails 
(i.e. (\ref{eq:NonAssStar4})) was first spelled out; 
see also his comprehensive treatise \cite{Ungar:2001} and 
references therein. In a more recent book \cite{Ungar:2004} the same 
author systematically develops the intimate relation to hyperbolic 
geometry. A brief history of the research on these generalized 
algebraic structures is given in~\cite{SexlUrbantke:RGP}. 
         
Let us briefly come back to the composition formulae 
(\ref{eq:StarSolvability}).  We interpret $\vec\beta_1$,
$\vec\beta_2$, and $\vec\beta_3$ as velocities of frames: 
$\vec\beta_1$ is the velocity of frame\,2 with respect to (i.e. 
measured in) frame\,1. $\vec\beta_2$ is the velocity of frame\,3 
with respect to frame\,2. Finally, $\vec\beta_3$ is the 
velocity of frame\,3 with respect to frame\,1. Then, using 
(\ref{eq:CompBoostsParam1}), it is easy to derive the following 
expressions for the moduli of $\vec\beta_3$ and $\vec\beta_2$:
\begin{subequations}
\label{eq:CompBoostModulus}
\begin{alignat}{2}
\label{eq:CompBoostModulus1}
&\beta_3^2&&\,=\,
\frac{(\vec\beta_1+\vec\beta_2)^2-(\vec\beta_1\times\vec\beta_2)^2}%
{(1+\vec\beta_1\cdot\vec\beta_2)^2}\,,\\
\label{eq:CompBoostModulus2}
&\beta_2^2&&\,=\,
\frac{(\vec\beta_3-\vec\beta_1)^2-(\vec\beta_3\times\vec\beta_1)^2}%
{(1-\vec\beta_3\cdot\vec\beta_1)^2}\,.
\end{alignat}
\end{subequations}
$\beta_2$ is the modulus of the relative velocity between frames 
2 and 3 as function of the velocities of these frames with respect 
to a third one (here frame\,1). It may either be interpreted as 
velocity of frame\,3 with respect to frame\,2, (as above) or as 
velocity of frame\,2 with respect to frame\,3 (reciprocity of frame 
velocities, see Sect.\,\ref{sec:ImpactRP}). Accordingly, the right 
hand side of (\ref{eq:CompBoostModulus2}) is symmetric under the 
exchange $\vec\beta_1\leftrightarrow\vec\beta_3$. 

\subsection{The geometric structure of velocity composition}
\label{sec:GeomStrucVelAdd}
Even though the discussion of the geometry behind velocity 
composition belongs, strictly speaking, to the next, the geometry 
section, it is so intimately related to the discussion just given 
that it seems more appropriate to place the two right next to each 
other.

More precisely, the composition law for velocities is intimately 
related with hyperbolic geometry (i.e. geometry on spaces with 
constant negative curvature), as was first pointed out by 
Sommerfeld~\cite{Sommerfeld:1909}, 
Vari{\v c}ak~\cite{Varicak:1910}\cite{Varicak:1912},
Robb~\cite{Robb:OptGeom}, and Borel~\cite{Borel:1913a}. More 
recently the subject was elaborated on by Ungar~\cite{Ungar:2004}. 
The general reason is that the space of four-velocities 
\begin{equation}
\label{eq:def:FourVelHyp}
\mathcal{H}_c:=\{u\in\reals^4\mid g(u,u)=c^2\}\subset\reals^4\,,
\end{equation}
is a 3-dimensional hyperbola in $(\reals^4,g)$, whose induced 
metric, $h_c$, is of constant negative curvature. Since $\mathcal{H}_c$
is spacelike,  we obtain the Riemannian metric $h_c$ by restricting 
$-g$ to the tangent bundle of $\mathcal{H}_c$). It is easy to write 
down $h_c$ in terms of the coordinates $\vec\beta\in\mathcal{B}_1$
(cf. (\ref{eq:DefUnitBall})). For this write $u=c\gamma(1,\vec\beta)$ 
and set $d\vec\beta=d\vec\beta_{\Vert}+d\vec\beta_{\perp}$, where  
\begin{subequations}
\label{eq:DefDbetaKomp}
\begin{alignat}{2}
\label{eq:DefDbetaKomp1}
&d\vec\beta_{\Vert}&&\,:=\,
\vec\beta(\vec\beta\cdot d\vec\beta)/\beta^2\,,\\ 
\label{eq:DefDbetaKomp2}
&d\vec\beta_{\perp}&&\,:=\,
d\vec\beta-d\vec\beta_{\Vert}\,.
\end{alignat}
\end{subequations}
Then we have 
\begin{equation}
\label{eq:FourVelDiff}
du=c\,\gamma^3\,(\vec\beta\cdot d\vec\beta_{\Vert}\,,\,
d\vec\beta_{\Vert}+\gamma^{-2}\,d\vec\beta_{\perp})\,,
\end{equation}
so that the Riemannian metric on the unit hyperbola is given by  
\begin{equation}
\label{eq:MetricUnitHyp}
h:=c^{-2}h_c:=-c^{-2}\,g_{\mu\nu}\,du^\mu\otimes du^\nu
=\gamma^4\,d\vec\beta^2_{\Vert}+\gamma^2\,d\vec\beta^2_{\perp}\,.
\end{equation}
Introducing spherical angular coordinates in the usual fashion this 
can be written in various standard forms, depending on the choice of 
the radial coordinate: 
\begin{subequations}
\label{eq:eq:MetricUnitHypPolar}
\begin{alignat}{1}
\label{eq:eq:MetricUnitHypPolar1}
h & \ =\
\frac{d\beta^2}{(1-\beta^2)^2}
+\frac{\beta^2}{1-\beta^2}\bigl(d\theta^2+\sin^2\theta\,d\varphi^2\bigr)\\
\label{eq:eq:MetricUnitHypPolar2}
& \ =\
\frac{dR^2}{1+R^2}
+R^2\,\bigl(d\theta^2+\sin^2\theta\,d\varphi^2\bigr)\\
\label{eq:eq:MetricUnitHypPolar3}
& \ =\
\frac{4r^2}{1-r^2}\ \Bigl\{
dr^2+r^2\,\bigl(d\theta^2+\sin^2\theta\,d\varphi^2\bigr)\Bigr\}\\
\label{eq:eq:MetricUnitHypPolar4}
& \ =\
d\rho^2+\sinh^2\rho\,\bigl(d\theta^2+\sin^2\theta\,d\varphi^2\bigr)\,,
\end{alignat}
\end{subequations}
where\footnote{The relation between (\ref{eq:eq:MetricUnitHypPolar2})
and the conformally flat form (\ref{eq:eq:MetricUnitHypPolar3}) is 
given by $R=2r/(1-r^2)$ and $r=R/\bigl(1+\sqrt{1+R^2}\bigr)$.}
\begin{subequations}
\begin{alignat}{4}
& R&&\,:=\,\frac{\beta}{\sqrt{1-\beta^2}}\qquad 
   &&\text{with range}\quad &&[0,\infty)\,,\\
& r&&\,:=\,\frac{\beta}{1+\sqrt{1-\beta^2}}\qquad 
   &&\text{with range}\quad &&[0,1)\,,\\
&\rho&&\,:=\,\tanh^{-1}\beta\qquad 
   &&\text{with range}\quad &&[0,\infty)\,.
\end{alignat}
\end{subequations}
The forms 
(\ref{eq:eq:MetricUnitHypPolar2}-\ref{eq:eq:MetricUnitHypPolar4})
correspond to three standard ways of writing a metric of constant 
sectional curvature $-1$, familiar e.g. from relativistic cosmology,
where they appear as spatial part of the $k=-1$ Friedmann-Robertson-Walker
metric. From (\ref{eq:eq:MetricUnitHypPolar4}) we learn that the 
rapidity $\rho$ (cf. (\ref{eq:def:Rapidity})) is just the geodesic 
distance\footnote{In (\ref{eq:eq:MetricUnitHypPolar3}) $\rho$ appears 
as geodesic distance to the center of the chosen system of spherical 
polar coordinates. However, since $\Lor_+^{\uparrow}$ acts as a 
transitive group of isometries on $\mathcal{H}_c$, 
(\ref{eq:eq:MetricUnitHypPolar3}) is valid no matter what point 
on $\mathcal{H}_c$ is chosen for the center of coordinates.} on the 
hyperbola $\mathcal{H}_c$ of four-velocities with respect to the 
rescaled metric $h=c^{-2}h_c$. This explains why the composition of 
velocities in the same direction is just ordinary addition if written 
in terms of rapidities; compare the remark following equation 
(\ref{eq:def:Rapidity}). 

The metric $h$ is also inherent in formula (\ref{eq:CompBoostModulus2}), 
which may also be read as endowing $\mathcal{B}_1$ with a Riemannian
metric when applied to two infinitesimally nearby velocities 
$\vec\beta=\vec\beta_1$ and $\vec\beta+d\vec\beta=\vec\beta_3$. 
Then $\beta_2^2$ gives us the square of their distance, $ds^2$, 
for which we obtain 
\begin{equation}
ds^2=\frac{d\vec\beta^2-(\vec\beta\times d\vec\beta)^2}%
{\bigl(1-\beta^2\bigr)^2}
=\gamma^4\,d\vec\beta^2_{\Vert}+\gamma^2\,d\vec\beta^2_{\perp}
\end{equation}
which coincides with (\ref{eq:MetricUnitHyp}). Equation 
(\ref{eq:CompBoostModulus2}) is the starting point in Fock's 
discussion of the hyperbolic geometry of velocity space 
(\cite{Fock:STG} \S\,17). We also draw attention to a recent 
pedagogical discussion in \cite{RhodesSemon:2005} and the 
systematic treatment in Ch.\,7 of \cite{Ungar:2004}. 

In this geometric setting the law (\ref{eq:CompBoostModulus1}) for 
the modulus of the composed velocities just turns into the law for 
the length of the third side of a geodesic triangle as function of 
the length of the two other sides and the angle between them. This 
is most easily read off from (\ref{eq:CompBoostsParam1}) if rewritten 
in terms of rapidities, i.e. $\gamma_i=\cosh\rho_i$ and 
$\beta_i\gamma_i=\sinh\rho_i$:
\begin{equation}
\label{eq:HypCosLaw}
\cosh\rho_3=
\cosh\rho_1\,\cosh\rho_1+
\sinh\rho_1\,\sinh\rho_2\,\cos\varphi\,.
\end{equation}
This is just the well known `cosine-law' for hyperbolic triangles, 
the connection of which with the law of composing velocities in SR 
was first pointed out by Sommerfeld~\cite{Sommerfeld:1909} and later, 
independently, by Borel~\cite{Borel:1913a}. Again we refer to 
\cite{Ungar:2004} for a modern and comprehensive treatment. 

A beautiful application of the hyperbolic geometry of velocity 
space (\ref{eq:def:FourVelHyp}) concerns Thomas rotation
\cite{Urbantke:1990}. Suppose a torque-free gyro is carried along 
the worldline $z(\tau)$ of an observer. The \emph{hodograph} is 
the curve $\dot z(\tau)$ on $\mathcal{H}_c$ and $\dot z(\tau)^\perp$ 
can be identified with the tangent plane to $\mathcal{H}_c$ at 
$\dot z(\tau)$. At each instant the gyro's angular-momentum vector 
lies in this tangent plane and along the worldline it is 
Fermi-Walker transported. We recall that given a vector field 
$X$ along the worldline $z$, the Fermi-Walker derivative of $X$ 
along $z$ is defined by
\begin{equation}
\label {eq:DefFermiWalker} 
F_{\dot z}X:=
(\nabla_{\dot z}X_\Vert)_{\Vert}+
(\nabla_{\dot z}X_\perp)_{\perp}\,,
\end{equation}
where $\Vert$ and $\perp$ denote the $g$-orthogonal projections 
parallel and perpendicular to the worldline's tangent direction 
$\dot z$. Applied to the gyro's angular momentum vector one sees 
that the law of Fermi-Walker transportation along $z$ turns into 
the law of parallel propagation along the hodograph on $\mathcal{H}_c$ 
with respect to the Levi-Civita connection for the hyperbolic metric 
that $\mathcal{H}_c$ inherits from its embedding into Minkowski 
space.\footnote{Generally, the Levi-Civita covariant derivative of a 
submanifold is obtained from the (covariant) derivative of the 
ambient manifold by restricting it to tangent vectors and subsequently 
projecting the result tangentially to the submanifold.}
Applied to spatially periodic orbits the holonomy of their closed 
hodographs in the tangent bundle of $\mathcal{H}_c$ is then just 
Thomas' rotation. This neat geometric idea goes back to 
Borel~\cite{Borel:1913a}, who sketched it almost 15 years 
before Thomas' paper~\cite{Thomas:1927} appeared.   

\section{Geometric structures in Minkowski space}
\label{sec:GeomStructMink}
\subsection{Preliminaries}
Let us generally consider $n$ dimensional Minkowski space $\mathbb{M}^n$, 
that is, the affine space over an $n$-dimensional, real vector space 
$V$ with a non-degenerate bilinear form $g$ of signature $(1,n-1)$
(compare Sec.\,\ref{sec:AffineSpaces} and Sec.\,\ref{sec:VectorSpaces}
respectively). We introduce the following notations:
\begin{equation}
\label{eq:def:MinkProd}
v\cdot w\,:=\,g(v,w)\,\qquad\text{and}\qquad
\Vert v\Vert_g:=\sqrt{\vert g(v,v)\vert}\,.
\end{equation}
We shall also simply write $v^2$ for $v\cdot v$. A vector $v\in V$ 
is called \emph{timelike, lightlike}, or \emph{spacelike} according 
to $v^2$ being $>0$, $=0$, or $<0$ respectively. Non-spacelike vectors 
are also called \emph{causal} and their set, $\mathcal{\bar C}\subset V$, 
is called the \emph{causal-doublecone}. Its interior, $\mathcal{C}$, is 
called the \emph{chronological-doublecone} and its boundary, $\mathcal{L}$,
the \emph{light-doublecone}:    
\begin{subequations}
\label{eq:VecDC}
\begin{alignat}{2}
\label{eq:def:VecCausalDC}
& \mathcal{\bar C}&&:\,=\,\{v\in V\mid v^2\geq 0\}\,,\\
\label{eq:def:VecChronDC}
& \mathcal{C}&&:\,=\,\{v\in V\mid v^2> 0\}\,,\\
\label{eq:def:VecLightDC}
& \mathcal{L}&&:\,=\,\{v\in V\mid v^2=0\}\,.
\end{alignat}
\end{subequations}

A linear subspace $V'\subset V$ is called timelike, lightlike, 
or spacelike according to $g\big\vert_{V'}$ being 
indefinite, negative semi-definite but not negative definite, 
or negative definite respectively. 
Instead of the usual Cauchy-Schwarz-inequality we have 
\begin{subequations}
\label{eq:CauchySchwarz}
\begin{alignat}{3}
\label{eq:CauchySchwarz1}
& v^2w^2&&\,\leq\, (v\cdot w)^2\quad &&\text{for $\Span\{v,w\}$ timelike}\,,\\
\label{eq:CauchySchwarz2}
& v^2w^2&&\,=\,(v\cdot w)^2\quad &&\text{for $\Span\{v,w\}$ lightlike}\,,\\ 
\label{eq:CauchySchwarz3}
& v^2w^2&&\,\geq\, (v\cdot w)^2\quad &&\text{for $\Span\{v,w\}$ spacelike}\,.
\end{alignat}
\end{subequations}

Given a set $W\subset V$ (not necessarily a subspace\footnote{By a 
`subspace' of a vector space we always understand a sub vector-space.}), 
its $g$-orthogonal complement is the subspace
\begin{equation}
\label{eq:OrthCompl}
W^\perp:=\{v\in V\mid v\cdot w=0,\,\forall w\in W\}\,.
\end{equation}
If $v\in V$ is lightlike then $v\in v^\perp$. In fact, $v^\perp$ is 
the unique lightlike hyperplane containing $v$. On the other hand, 
if $v$ is timelike/spacelike $v^\perp$ is spacelike/timelike 
and $v\not\in v^\perp$.   

Given any subset $W\subset V$, we can attach it to a point $p$ 
in $\mathbb{M}^n$:
\begin{equation}
\label{eq:AffineSubset}
W_p:=p+W:=\{p+w\mid w\in W\}\,.
\end{equation}
In particular, the causal-, chronological-, and light-doublecones at 
$p\in\mathbb{M}^n$ are given by: 
\begin{subequations}
\label{eq:AffDC}
\begin{alignat}{2}
\label{eq:def:AffCausalDC}
& \mathcal{\bar C}_p&&:\,=\,p+\mathcal{\bar C}\,,\\
\label{eq:def:AffPointChronDC}
& \mathcal{C}_p&&:\,=\,p+\mathcal{C}\,,\\
\label{eq:def:AffLightDC}
& \mathcal{L}_p&&:\,=\,p+\mathcal{L}\,.
\end{alignat}
\end{subequations}

If $W$ is a subspace of $V$ then $W_p$ is an affine subspace of 
$\mathbb{M}^n$ over $W$. If $W$ is time-, light-, or spacelike 
then $W_p$ is also called time-, light-, or spacelike. 
Of particular interest are the hyperplanes $v_p^\perp$
which are timelike, lightlike, or spacelike according to $v$ 
being spacelike, lightlike, or timelike respectively. 

Two points $p,q\in\mathbb{M}^n$ are said to be timelike-, lightlike-, 
or spacelike separated if the line joining them (equivalently: the 
vector $p-q$) is timelike, lightlike, or spacelike respectively. 
Non-spacelike separated points are also called causally separated and 
the line though them is called a causal line.  

It is easy to show that the relation $v\sim w\Leftrightarrow v\cdot w>0$ 
defines an equivalence relation on the set of timelike vectors.
(Only transitivity is non-trivial, i.e. if $u\cdot v>0$ and $v\cdot w>0$
then $u\cdot w>0$. To show this, decompose $u$ and $w$ into their 
components parallel and perpendicular to $v$.). Each of the two 
equivalence classes is a \emph{cone} in $V$, that is, closed under 
addition and multiplication with positive numbers. Vectors in the same 
class are said to have the same time orientation. In the same fashion 
the relation $v\sim w\Leftrightarrow v\cdot w\geq 0$ defines an 
equivalence relation on the set of causal vectors, with both 
equivalence classes being again cones. The existence of these 
equivalence relations is expressed by saying that $\mathbb{M}^n$ is 
\emph{time orientable}. Picking one of the two possible time 
orientations is then equivalent to specifying a single timelike 
reference vector, $v_*$, whose equivalence class of directions may 
be called the \emph{future}. This being done we can speak of the future 
(or forward) $(+)$ and past (or backward) $(-)$ cones:  
\begin{subequations}
\label{eq:VecC}
\begin{alignat}{2}
\label{eq:def:VecCausalC}
& \mathcal{\bar C}^\pm &&:\,=\,
  \{v\in\mathcal{\bar C}\mid v\cdot v_*\gtrless 0\}\,,\\
\label{eq:def:VecChronC}
& \mathcal{C}^\pm &&:\,=\,
  \{v\in\mathcal{\bar C}\mid v\cdot v_*\gtrless 0\}\,,\\
\label{eq:def:VecLightC}
& \mathcal{L}^\pm &&:\,=\,
  \{v\in \mathcal{\bar L}\mid v\cdot v_*\gtrless 0\}\,.
\end{alignat}
\end{subequations}
Note that $\mathcal{\bar C}^\pm=\mathcal{C}^\pm\cup\mathcal{L}^\pm$
and $\mathcal{C}^\pm\cap\mathcal{L}^\pm=\emptyset$. Usually 
$\mathcal{L}^+$ is called the future and $\mathcal{L}^-$ the past 
lightcone. Mathematically speaking this is an abuse of language 
since, in contrast to $\mathcal{\bar C}^\pm$ and $\mathcal{C}^\pm$, 
they are not cones: They are each invariant (as sets) under 
multiplication with positive real numbers, but adding to vectors in 
$\mathcal{L}^\pm$ will result in a vector in $\mathcal{C}^\pm$ unless 
the vectors were parallel.

As before, these cones can be attached to the points in $\mathbb{M}^n$.
We write in a straightforward manner: 
\begin{subequations}
\label{eq:AffC}
\begin{alignat}{2}
\label{eq:def:AffCausalC}
& \mathcal{\bar C}_p^\pm &&:\,=\,p+\mathcal{\bar C}^\pm\,,\\
\label{eq:def:AffChronC}
& \mathcal{C}_p^\pm &&:\,=\,p+\mathcal{C}^\pm\,,\\
\label{eq:def:AffLightC}
& \mathcal{L}_p^\pm &&:\,=\,p+\mathcal{L}^\pm\,.
\end{alignat}
\end{subequations}

The Cauchy-Schwarz inequalities (\ref{eq:CauchySchwarz}) result in various 
cases for generalized triangle inequalities. Clearly, for spacelike 
vectors, one just has the ordinary triangle inequality. But for causal or 
timelike vectors one has to distinguish the cases according to the relative 
time orientations. For example, for timelike vectors of equal time 
orientation, one obtains the reversed triangle inequality:
\begin{equation}
\label{sec:InvTriangleIneq}
\Vert v+w\Vert_g\geq\Vert v\Vert_g +\Vert w\Vert_g\,,
\end{equation}
with equality iff $v$ and $w$ are parallel. It expresses the geometry 
behind the `twin paradox'. 

Before we turn to the next section, we remark that any bijective 
map $\phi:\mathbb{M}^n\rightarrow\mathbb{M}^n$ that satisfies 
$d(p,q)=d(\phi(p),\phi(q))$, where $d(p,q):=\Vert p-q\Vert_g$, is 
necessarily affine linear. This follows immediately from the 
corresponding statement for vector spaces, as given in 
Proposition\,\ref{prop:IsometriesV}. The results in the following
section should be considered as strengthenings of this statement.

\subsection{Causality relations and the Lorentz group}
\label{sec:CausalityRelations}
The family of cones $\{\mathcal{\bar C}_q^+\mid q\in\mathbb{M}^n\}$ 
defines a partial order relation, denoted by $ \geq$ 
(cf. Sec.\,\ref{sec:GroupActions}), on spacetime as follows:
$p\geq q$ iff $p\in\mathcal{\bar C}^+_q$, i.e. iff $p-q$ is causal 
and future pointing. Similarly, the family 
$\{\mathcal{C}^+_q\mid q\in\mathbb{M}^n\}$ defines a strict partial 
order, denoted by $>$ (cf. Sec.\,\ref{sec:GroupActions}): $p>q$ iff 
$p\in\mathcal{C}^+_q$, i.e. if $p-q$ is 
timelike and future pointing. There is a third relation, called $\gtrdot$,
defined as follows: $p\gtrdot q$ iff $p\in\mathcal{L}_q^+$, i.e. $p$ is 
on the future lightcone at $q$. It is not a partial order due to the lack 
of transitivity, which, in turn, is due to the lack of the lightcone 
being a cone (in the proper mathematical sense explained above). 
Replacing the future ($+$) with the past ($-$) cones gives the 
relations $\leq$, $<$, and $\lessdot$.

It is obvious that the action of $\ILor^\uparrow$ (spatial reflections 
are permitted) on $\mathbb{M}^n$ maps each of the six families of 
cones (\ref{eq:AffC}) into itself and therefore leave each of the 
six relations invariant. For example: Let $p>q$ and $f\in\ILor^\uparrow$, 
then $(p-q)^2>0$ and $p-q$ future pointing, but also $(f(p)-f(q))^2>0$
and $f(p)-f(q)$ future pointing, hence $f(p)>f(q)$. Another set of `obvious' 
transformations of $\mathbb{M}^n$ leaving these relations invariant is 
given by all dilations: 
\begin{equation}
\label{eq:def:Dilations}
d_{(\lambda,m)}:\mathbb{M}^n\rightarrow\mathbb{M}^n\,,\quad
p\mapsto d_{(\lambda,m)}(p):=\lambda(p-m)+m\,,
\end{equation}
where $\lambda\in\mathbb{R }_+$ is the constant dilation-factor and 
$m\in\mathbb{M}^n$ the center. This follows from  
$\bigl(d_{\lambda,m}(p)-d_{\lambda,m}(q)\bigr)^2=\lambda^2(p-q)^2$,
$\bigl(d_{\lambda,m}(p)-d_{\lambda,m}(q)\bigr)\cdot v_*=\lambda(p-q)\cdot v_*$,
and the positivity of $\lambda$. Since translations are already 
contained in $\ILor^\uparrow$, the group generated by $\ILor^\uparrow$
and all $d_{\lambda,m}$ is the same as the group generated by 
$\ILor^\uparrow$ and all $d_{\lambda,m}$ for fixed $m$. 

A seemingly difficult question is this: What are the most general 
transformations of $\mathbb{M}^n$ that preserve those relations? 
Here we understand `transformation' synonymously with `bijective map', 
so that each transformation $f$ has in inverse $f^{-1}$. `Preserving 
the relation' is taken to mean that $f$ \emph{and} $f^{-1}$ 
preserve the relation. Then the somewhat surprising answer to the 
question just posed is that, in three or more spacetime dimensions, 
there are no other such transformations besides those already listed: 

\begin{theorem}
Let $\succ$ stand for any of the relations $\geq,>,\gtrdot$ and 
let $f$ be a bijection of $\mathbb{M}^n$ with $n\geq 3$, such that 
$p\succ q$  implies  $f(p)\succ f(q)$ and 
$f^{-1}(p)\succ f^{-1}(q)$. Then $f$ is the composition of an 
Lorentz transformation in $\ILor^\uparrow$ with a dilation.
\end{theorem}
\begin{proof}
These results were proven by A.D.\,Alexandrov and independently 
by E.C. Zeeman. A good review of Alexandrov's results is 
\cite{Alexandrov:1975}; Zeeman's paper is \cite{Zeeman:1964}.
The restriction to $n\geq 3$ is indeed necessary, as for $n=2$ 
the following possibility exists: Identify $\mathbb{M}^2$
with $\reals^2$ and the bilinear form $g(z,z)=x^2-y^2$,
where $z=(x,y)$. Set $u:=x-y$ and $v:=x+y$ and define 
$f:\reals^2\rightarrow \reals^2$ by $f(u,v):=(h(u),h(v))$,
where $h:\reals\rightarrow\reals$ is any smooth function 
with $h'>0$. This defines an orientation preserving diffeomorphism
of $\reals^2$ which transforms the set of lines $u=$ const. and 
$v=$ const. respectively into each other. Hence it preserves the 
families of cones (\ref{eq:def:AffCausalC}). Since these 
transformations need not be affine linear they are not generated by 
dilations and Lorentz transformations.
\end{proof}

These results may appear surprising since without a continuity 
requirement one might expect all sorts of wild behavior to allow 
for more possibilities. However, a little closer inspection reveals 
a fairly obvious reason for why continuity is implied here.  
Consider the case in which a transformation $f$ preserves the 
families $\{\mathcal{C}^+_q\mid q\in\mathbb{M}^n\}$ and 
$\{\mathcal{C}^-_q\mid q\in\mathbb{M}^n\}$. The open 
diamond-shaped sets (usually just called `open diamonds'),
\begin{equation}
\label{eq:OpenDiamonds}
U(p,q):=(\mathcal{C}^+_p\cap\mathcal{C}^-_q)\cup
        (\mathcal{C}^+_q\cap\mathcal{C}^-_p)\,,
\end{equation}
are obviously open in the standard topology of $\mathbb{M}^n$ 
(which is that of $\reals^n$). Note that at least one of the 
intersections in (\ref{eq:OpenDiamonds}) is always empty. 
Conversely, is is also easy to see that each open set of 
$\mathbb{M}^n$ contains an open diamond. Hence the topology 
that is defined by taking the $U(p,q)$ as subbase (the basis 
being given by their finite intersections) is equivalent to the 
standard topology of $\mathbb{M}^n$. But, by hypothesis, $f$ and 
$f^{-1}$ preserves the cones $\mathcal{C}^\pm_q$ and therefore 
open sets, so that $f$ must, in fact, be a homeomorphism. 

There is no such \emph{obvious} continuity input if one 
makes the strictly weaker requirement that instead of the cones 
(\ref{eq:AffC}) one only preserves the doublecones 
(\ref{eq:AffDC}). Does that allow for more transformations, except 
for the obvious time reflection? The answer is again in the negative.
The following result was shown by  Alexandrov (see his review 
\cite{Alexandrov:1975}) and later, in a different fashion, by 
Borchers and Hegerfeld~\cite{BorchersHegerfeld:1972}: 
\begin{theorem}
\label{thm:BorchersHegerfeld}
Let $\sim$ denote any of the relations: $p\sim q$ iff $(p-q)^2\geq 0$,
 $p\sim q$ iff $(p-q)^2> 0$, or $p\sim q$ iff $(p-q)^2=0$. Let $f$ be a 
bijection of $\mathbb{M}^n$ with $n\geq 3$, such that $p\sim q$  
implies  $f(p)\sim f(q)$ and $f^{-1}(p)\sim f^{-1}(q)$. Then $f$ 
is the composition of an Lorentz transformation in $\ILor$ with a 
dilation.
\end{theorem}

All this shows that, up to dilations, Lorentz transformations can be 
characterized by the causal structure of Minkowski space. 
Let us focus on a particular subcase of 
Theorem\,\ref{thm:BorchersHegerfeld}, which says that any bijection 
$f$ of $\mathbb{M}^n$ with $n\geq 3$, which satisfies 
$\Vert p-q\Vert_g=0\Leftrightarrow\Vert f(p)-f(q)\Vert_g=0$ must be 
the composition of a dilation and a transformation in $\ILor$. 
This is sometimes referred to as \emph{Alexandrov's theorem}. 
It is, to my knowledge, the closest analog in Minkowskian geometry to 
the famous theorem of Beckman and Quarles \cite{BeckmanQuarles:1953},
which refers to Euclidean geometry and reads as 
follows\footnote{In fact, Beckman and Quarles proved 
the conclusion of Theorem\,\ref{thm:BeckmanQuarles} under slightly 
weaker hypotheses: They allowed the map $f$ to be `many-valued', 
that is, to be a map $f:\reals^n\rightarrow\mathcal{S}^n$, 
where $\mathcal{S}^n$ is the set of non-empty subsets of $\reals^n$,
such that $\Vert x-y\Vert=\delta\Rightarrow \Vert x'-y'\Vert=\delta$
for any $x'\in f(x)$ and any $y'\in f(y)$. However, given the 
statement of Theorem\,\ref{thm:BeckmanQuarles}, it is immediate 
that such `many-valued maps' must necessarily be single-valued. To see 
this, assume that $x_*\in\reals^n$ has the two image points 
$y_1,y_2$ and define $h_i:\reals^n\rightarrow\reals^n$ for 
$i=1,2$ such that $h_1(x)=h_2(x)\in f(x)$ for all $x\ne x_*$ and 
$h_i(x_*)=y_i$. Then, according to Theorem\,\ref{thm:BeckmanQuarles},
$h_i$ must both be Euclidean motions. Since they are continuous and 
coincide for all $x\ne x_*$, they must also coincide at $x_*$.}:

\begin{theorem}[Beckman and Quarles 1953]
\label{thm:BeckmanQuarles}
Let $\reals^n$ for $n\geq 2$ be endowed with the standard Euclidean 
inner product $\langle\cdot\mid\cdot\rangle$. The associated norm is 
given by $\Vert x\Vert:=\sqrt{\langle x\mid x\rangle}$. 
Let $\delta$ be any fixed positive real number and 
$f:\reals^n\rightarrow\reals^n$ any map such that 
$\Vert x-y\Vert=\delta\Rightarrow \Vert f(x)-f(y)\Vert=\delta$; then 
$f$ is a Euclidean motion, i.e. $f\in\reals^n\rtimes\group{O}(n)$.
\end{theorem}
Note that there are three obvious points which let the result of 
Beckman and Quarles in Euclidean space appear somewhat stronger than 
the theorem of Alexandrov in Minkowski space: 
\begin{itemize}
\item[1.]
The conclusion of Theorem\,\ref{thm:BeckmanQuarles} holds for 
\emph{any} $\delta\in\reals_+$, whereas Alexandrov's theorem 
singles out lightlike distances.
\item[2.]
In Theorem\,\ref{thm:BeckmanQuarles}, $n=2$ is not excluded. 
\item[3.]
In Theorem\,\ref{thm:BeckmanQuarles}, $f$ is not required to be a 
bijection, so that we did not assume the existence of an inverse map 
$f^{-1}$. Correspondingly, there is no assumption 
that $f^{-1}$ also preserves the distance $\delta$. 
\end{itemize}

\subsection{Einstein synchronization}
\label{sec:EinsteinSynch}
We start by characterizing those cases in which a strict inverted 
Cauchy-Schwarz inequality holds:  
\begin{lemma}
\label{lemma:StrictICSI}
Let $V$ be of dimension $n>2$ and $v\in V$ be some 
non-zero vector. The strict inverted Cauchy-Schwarz inequality,
\begin{equation}
\label{eq:StrictInvCSI}
v^2w^2<(v\cdot w)^2\,,
\end{equation}
holds for all $w\in V$ linearly independent of $v$ iff $v$ is timelike.  
\end{lemma}
\begin{proof}
Obviously $v$ cannot be spacelike, for then we would violate 
(\ref{eq:StrictInvCSI}) with any spacelike $w$. If $v$ is 
lightlike then $w$ violates (\ref{eq:StrictInvCSI}) iff it is 
in the set $v^\perp-\Span\{v\}$, which is non-empty iff 
$n>2$. Hence $v$ cannot be lightlike if $n>2$. If $v$ is timelike
we decompose $w=av+w'$ with $w'\in v^\perp$ so that $w'^2\leq 0$,
with equality iff $v$ and $w$ are linearly dependent. Hence 
\begin{equation}  
\label{eq:LemmaStrictICSI1}
(v\cdot w)^2-v^2w^2=-v^2\,w'^2\geq 0\,,
\end{equation}
with equality iff $v$ and $w$ are linearly dependent. 
\end{proof}
 
\begin{figure}[h!]
\centering\epsfig{figure=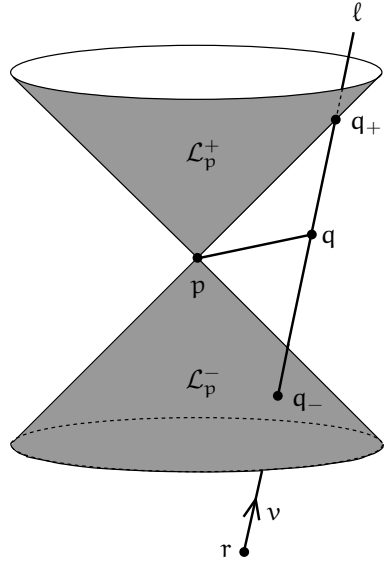,width=0.38\linewidth}
\put(-14,162){\small $q_+$}
\put(-25,119){\small $q$}
\put(-36,57){\small $q_-$}
\put(-74,100){\small $p$}
\put(-63,0){\small $r$}
\put(-45,14){\small $v$}
\put(-76,150){\small $\mathcal{L}^+_p$}
\put(-76,64){\small $\mathcal{L}^-_p$}
\put(-13,203){\small $\ell$}
\caption{\label{fig:Robb}
A timelike line $\ell=\{r+\lambda v\mid\lambda\in\reals\}$ intersects 
the light-cone with vertex $p\not\in\ell$ in two points: $q_+$, its  
intersection with the future light-cone and $q_-$, its intersection 
with past the light cone. $q$ is a point inbetween $q_+$ and $q_-$.}
\end{figure}

\noindent
The next Lemma deals with the intersection of a causal line
with a light cone, a situation depicted in Fig.\,\ref{fig:Robb}.

\begin{lemma}
\label{lemma:IntLineCone}
Let $\mathcal{L}_p$ be the light-doublecone with vertex $p$ and 
$\ell:=\{r+\lambda v\mid r\in\reals\}$ be a non-spacelike line, 
i.e. $v^2\geq 0$, through $r\not\in\mathcal{L}_p$. If $v$ is timelike 
$\ell\cap\mathcal{L}_p$ consists of two points. If $v$ is lightlike 
this intersection consists of one point if $p-r\not\in v^\perp$
and is empty if $p-r\in v^\perp$. Note that the latter two statements 
are independent of the choice of $r\in\ell$---as they must be---, i.e. 
are invariant under $r\mapsto r':=r+\sigma v$, where $\sigma\in\reals$.
\end{lemma}
\begin{proof}
We have $r+\lambda v\in\mathcal{L}_p$ iff
\begin{equation}
\label{eq:LemmaIntLineCone1}
(r+\lambda v-p)^2=0\ \Longleftrightarrow\
\lambda^2 v^2+2\lambda v\cdot(r-p)+(r-p)^2=0\,.
\end{equation}
For $v$ timelike we have $v^2>0$ and (\ref{eq:LemmaIntLineCone1})
has two solutions 
\begin{equation}
\label{eq:LemmaIntLineCone2}
\lambda_{1{,}2}=\frac{1}{v^2}\left\{
-v\cdot (r-p)\pm\sqrt{\bigl(v\cdot(r-p)\bigr)^2-v^2(r-p)^2}\,\right\}\,.
\end{equation}
Indeed, since $r\not\in\mathcal{L}_p$, the vectors $v$ and $r-p$ 
cannot be linearly dependent so that Lemma\,\ref{lemma:StrictICSI} 
implies the positivity of the expression under the square root. 
If $v$ is lightlike (\ref{eq:LemmaIntLineCone1}) becomes a linear 
equation which is has one solution if $v\cdot(r-p)\ne 0$ and no solution if 
$v\cdot (r-p)=0$ [note that $(r-p)^2\ne 0$ since $q\not\in \mathcal{L}_p$
by hypothesis].
\end{proof}

\begin{proposition}
\label{prop:EinsteinSynch}
Let $\ell$ and $\mathcal{L}_p$ as in Lemma\,\ref{lemma:IntLineCone} with $v$ 
timelike. Let $q_+$ and $q_-$ be the two intersection points of 
$\ell$ with $\mathcal{L}_p$ and $q\in\ell$ a point between them. Then 
\begin{equation}
\label{eq:prop:EinsteinSynch1}
\Vert q-p\Vert_g^2=\Vert q_+-q\Vert_g\,\Vert q-q_-\Vert_g\,.
\end{equation}
Moreover, $\Vert q_+-q\Vert_g=\Vert q-q_-\Vert_g$ iff 
$p-q$ is perpendicular to $v$. 
\end{proposition}
\begin{proof}
The vectors $(q_+-p)=(q-p)+(q_+-q)$ and $(q_--p)=(q-p)+(q_--q)$ 
are lightlike, which gives (note that $q-p$ is spacelike):
\begin{subequations}
\label{eq:prop:EinsteinSynch23}
\begin{alignat}{3}
\label{eq:prop:EinsteinSynch2}
& \Vert q-p\Vert_g^2 
&&\,=\,-(q-p)^2 
&&\,=\,(q_+-q)^2+2(q-p)\cdot(q_+-q)\,,\\
\label{eq:prop:EinsteinSynch3}
& \Vert q-p\Vert_g^2 
&&\,=\,-(q-p)^2
&&\, =\,(q_--q)^2+2(q-p)\cdot(q_--q)\,.
\end{alignat}
\end{subequations}
Since $q_+-q$ and $q-q_-$ are parallel we have 
$q_+-q=\lambda (q-q_-)$ with $\lambda\in\reals_+$ so that 
$(q_+-q)^2=\lambda\Vert q_+-q\Vert_g\Vert q-q_-\Vert_g$ and
$\lambda (q_--q)^2=\Vert q_+-q\Vert_g\Vert q-q_-\Vert_g$.
Now, multiplying (\ref{eq:prop:EinsteinSynch3}) with $\lambda$ 
and adding this to (\ref{eq:prop:EinsteinSynch2}) immediately 
yields 
\begin{equation}
\label{eq:prop:EinsteinSynch4}
(1+\lambda)\,\Vert q-p\Vert_g^2
=(1+\lambda)\,\,\Vert q_+-q\Vert_g\Vert q-q_-\Vert_g\,.\\
\end{equation}
Since $1+\lambda\ne 0$ this implies (\ref{eq:prop:EinsteinSynch1}).
Finally, since $q_+-q$ and $q_--q$ are antiparallel, 
$\Vert q_+-q\Vert_g=\Vert q_--q\Vert_g$ iff $(q_+-q)=-(q_--q)$. 
Equations (\ref{eq:prop:EinsteinSynch23}) now show that this is the 
case iff $(q-p)\cdot(q_\pm-q)=0$, i.e. iff $(q-p)\cdot v=0$. 
Hence we have shown  

\begin{equation}
\label{eq:prop:EinsteinSynch5}
\Vert q_+-q\Vert_g=\Vert q-q_-\Vert_g\ \Longleftrightarrow\ 
(q-p)\cdot v=0\,.
\end{equation}
In other words, $q$ is the midpoint of the segment $\overline{q_+q_-}$ 
iff the line through $p$ and $q$ is perpendicular (wrt. $g$) to $\ell$.
\end{proof}
The somewhat surprising feature of the first statement of this 
proposition is that (\ref{eq:prop:EinsteinSynch1}) holds for 
\emph{any} point of the segment $\overline{q_+q_-}$, not just 
the midpoint, as it would have to be the case for the corresponding 
statement in Euclidean geometry.

The second statement of Proposition\,\ref{prop:EinsteinSynch} gives 
a convenient geometric characterization of Einstein-simultaneity.
Recall that an event $q$ on a timelike line $\ell$ (representing an 
inertial observer) is defined to be Einstein-simultaneous with an 
event $p$ in spacetime iff $q$ bisects the segment $\overline{q_+q_-}$ 
between the intersection points $q_+,q_-$ of $\ell$ with the 
double-lightcone at $p$. Hence Proposition\,\ref{prop:EinsteinSynch} 
implies
\begin{corollary}
\label{corol:EinsteinSim}
Einstein simultaneity with respect to a timelike line $\ell$ is 
an equivalence relation on spacetime, the equivalence classes of 
which are the spacelike hyperplanes orthogonal (wrt. $g$) to $\ell$. 
\end{corollary}
The first statement simply follows from the fact that the family 
of parallel hyperplanes orthogonal to $\ell$ form a partition 
(cf. Sect.\,\ref{sec:GroupActions}) of spacetime. 

From now on we shall use the terms `timelike line' and `inertial 
observer' synonymously. Note that Einstein simultaneity is only defined 
relative to an inertial observer. Given two inertial observers, 
\begin{subequations}
\label{eq:TwoObservers}
\begin{alignat}{3}
\label{eq:TwoObservers1}
&\ell&&\,=\,\{r+\lambda v\mid\lambda\in\reals\}\qquad
&&\text{first observer}\,,\\
\label{eq:TwoObservers2}
&\ell'&&\,=\,\{r'+\lambda' v'\mid\lambda'\in\reals\}\qquad
&&\text{second observer}\,,
\end{alignat}
\end{subequations}
we call the corresponding Einstein-simultaneity relations 
\emph{$\ell$-simultaneity} and \emph{$\ell'$-simultaneity}. Obviously 
they coincide iff $\ell$ and $\ell'$ are parallel ($v$ and $v'$ 
are linearly dependent). 
In this case $q'\in\ell'$ is $\ell$-simultaneous to $q\in\ell$ iff
$q\in\ell$ is $\ell'$-simultaneous to $q'\in\ell'$. If $\ell$ and 
$\ell'$are not parallel (skew or intersecting in one point) it is 
generally not true that if $q'\in\ell'$ is $\ell$-simultaneous to 
$q\in\ell$ then $q\in\ell$ is also $\ell'$-simultaneous to $q'\in\ell'$. 
In fact, we have
\begin{proposition}
\label{prop:SkewSim}
Let $\ell$ and $\ell'$ two non-parallel timelike likes. 
There exists a unique pair $(q,q')\in\ell\times\ell'$ so that $q'$ is 
$\ell$-simultaneous to $q$ and $q$ is $\ell'$ simultaneous to $q'$.
\end{proposition}
\begin{proof}
We parameterize $\ell$ and $\ell'$ as in  (\ref{eq:TwoObservers}).  
The two conditions for $q'$ being $\ell$-simultaneous to $q$ and $q$ 
being $\ell'$-simultaneous to $q'$ are $(q-q')\cdot v=0=(q-q')\cdot v'$.
Writing $q=r+\lambda v$ and $q'=r'+\lambda'v'$ this takes the form 
of the following matrix equation for the two unknowns $\lambda$ and 
$\lambda'$: 
\begin{equation}
\label{eq:prop:SkewSim1}
\begin{pmatrix}%
v^2&-v\cdot v'\\
v\cdot v'&-v'^2
\end{pmatrix}
\begin{pmatrix}
\lambda\\
\lambda'
\end{pmatrix}=
\begin{pmatrix}
(r'-r)\cdot v\\
(r'-r)\cdot v'
\end{pmatrix}\,.
\end{equation}
This has a unique solution pair $(\lambda,\lambda')$, since for 
linearly independent timelike vectors $v$ and $v'$  
Lemma\,\ref{lemma:StrictICSI} implies $(v\cdot v')^2-v^2v'^2>0$.
Note that if $\ell$ and $\ell'$ intersect
$q=q'=\text{intersection point}$.
\end{proof} 

Clearly, Einstein-simultaneity is conventional and physics proper 
should not depend on it. For example, the fringe-shift in the 
Michelson-Morley experiment is independent of how we choose to 
synchronize clocks. In fact, it does not even make use of any clock. 
So what is the general definition of a `simultaneity structure'?
It seems obvious that it should be a relation on spacetime that is 
at least symmetric (each event should be simultaneous to itself). 
Going from one-way simultaneity to the mutual synchronization of 
two clocks, one might like to also require reflexivity (if $p$ is 
simultaneous to $q$ then $q$ is simultaneous to $p$), though this 
is not strictly required in order to one-way synchronize each clock 
in a set of clocks with one preferred `master clock', which is 
sufficient for many applications. 

Moreover, if we like to speak of the mutual simultaneity of sets of 
more than two events we need an equivalence relation on spacetime. 
The equivalence relation should be such that each inertial observer 
intersect each equivalence class precisely once. Let us call such 
a simultaneity structure `admissible'. Clearly there are zillions of 
such structures: just partition spacetime into any set of 
appropriate\footnote{For example, the hypersurfaces should not be 
asymptotically hyperboloidal, for then a constantly accelerated observer 
would not intersect all of them.} spacelike 
hypersurfaces (there are more possibilities at this point, like 
families of forward or backward lightcones). An \emph{absolute} 
admissible simultaneity structure would be one which is invariant 
(cf. Sect.\,\ref{sec:GroupActions}) under the automorphism group of 
spacetime. We have 
\begin{proposition}
\label{prop:AbsSim}
There exits precisely one admissible simultaneity 
structure which is invariant under the inhomogeneous proper 
orthochronous Galilei group and none that is invariant under the 
inhomogeneous proper orthochronous Lorentz group.
\end{proposition}
\begin{proof}
See \cite{Giulini:2002a}.
\end{proof}
\noindent
There is a group-theoretic reason that highlights this 
existential difference: 
\begin{proposition}
\label{prop:AbsSimMath} 
Let $G$ be a group with transitive action on a set $S$. Let 
$\Stab(p)\subset G$ be the stabilizer subgroup for $p\in S$
(due to transitivity all stabilizer subgroups are conjugate). 
Then $S$ admits a $G$-invariant equivalence relation 
$R\subset S\times S$ iff $\Stab(p)$ is \emph{not} maximal, 
that is, iff $\Stab(p)$ is properly contained in a proper 
subgroup $H$ of $G$: $\Stab(p)\subsetneq H\subsetneq G$. 
\end{proposition}
\begin{proof}
See Theorem\,1.12 in \cite{Jacobson:BasicAlgebraI}.
\end{proof}
Regarding the action of the inhomogeneous Galilei and Lorentz groups 
on spacetime their stabilizers are the corresponding homogeneous 
groups. As already discussed at the end of 
Sect.\,\ref{sec:LorentzGalileiGroup}, the homogeneous Lorentz group 
is maximal in the inhomogeneous one, whereas the homogeneous Galilei 
group is not maximal in the inhomogeneous one. This, according to 
Proposition\,\ref{prop:AbsSimMath}, is the group theoretic origin of 
the absence of any invariant simultaneity structure in the Lorentzian 
case.

\subsection{The lattice structure of causally and chronologically complete sets}
\label{sec:CausCompSets}
Here we wish to briefly discuss another important structure associated 
with causality relations in Minkowski space, which plays a fundamental 
r\^ole in modern Quantum Field Theory (see e.g. \cite{Haag:LocQuantPhys}). 
Let $S_1$ and $S_2$ be subsets of $\mathbb{M}^n$. We say that $S_1$ and 
$S_2$ are \emph{causally disjoint} or \emph{spacelike separated} iff $p_1-p_2$ 
is spacelike, i.e. $(p_1-p_2)^2<0$, for any $p_1\in S_1$ and $p_2\in S_2$. 
Note that because a point is not spacelike separated from itself, 
causally disjoint sets are necessarily disjoint in the 
ordinary set-theoretic sense---the converse being of course not true. 

For any subset $S\subseteq\mathbb{M}^n$ we denote by $S'$ the largest 
subset of $\mathbb{M}^n$ which is causally disjoint to $S$. The set 
$S'$ is called the \emph{causal complement} of $S$. The procedure of 
taking the causal complement can be iterated and we set $S'':=(S')'$
etc. 
$S''$ is called the \emph{causal completion} of $S$. It also follows 
straight from the definition that $S_1\subseteq S_2$ implies 
$S'_1\supseteq S'_2$ and 
also $S''\supseteq S$. If $S''=S$ we call $S$ \emph{causally complete}. 
We note that the causal complement $S'$ of any given $S$ is 
automatically causally complete. Indeed, from $S''\supseteq S$ we 
obtain  $(S')''\subseteq S'$, but the first inclusion applied to 
$S'$ instead of $S$ leads to $(S')''\supseteq S'$, showing $(S')''=S'$. 
Note also that for any subset $S$ its causal completion, $S''$, is the 
smallest causally complete subset containing $S$, for if 
$S\subseteq K\subseteq S''$ with $K''=K$, we derive from the first 
inclusion by taking ${}''$ that $S''\subseteq K$, so that the second 
inclusion yields $K=S''$. Trivial examples of causally complete 
subsets of $\mathbb{M}^n$ are the empty set, single points, and the 
total set $\mathbb{M}^n$. Others are the open diamond-shaped 
regions (\ref{eq:OpenDiamonds}) as well as their closed counterparts:      
\begin{equation}
\label{eq:ClosedDiamonds}
\bar{U}(p,q):=(\mathcal{\bar C}^+_p\cap\mathcal{\bar C}^-_q)\cup
        (\mathcal{\bar C}^+_q\cap\mathcal{\bar C}^-_p)\,.
\end{equation}

We now focus attention to the set $\mathrm{Caus}(\mathbb{M}^n)$ 
of causally complete subsets of $\mathbb{M}^n$, including the 
empty set, $\emptyset$, and the total set, $\mathbb{M}^n$, which are 
mutually causally complementary. It is partially ordered by ordinary  
set-theoretic inclusion $(\subseteq)$ (cf. Sect.\,\ref{sec:GroupActions})
and carries the `dashing operation' $(')$ of taking the causal complement.
Moreover, on $\mathrm{Caus}(\mathbb{M}^n)$ we can define 
the operations of `meet' and `join', denoted by $\wedge$ and $\vee$ 
respectively, as follows: Let $S_i\in\mathrm{Caus}(\mathbb{M}^n)$ where 
$i=1,2$, then $S_1\wedge S_2$ is the largest causally complete subset in the 
intersection $S_1\cap S_2$ and $S_1\vee S_2$ is the smallest 
causally complete set containing the union $S_1\cup S_2$.
 
The operations of $\wedge$ and $\vee$ can be characterized in terms 
of the ordinary set-theoretic intersection $\cap$ together with the 
dashing-operation. To see this, 
consider two causally complete sets, $S_i$ where $i=1,2$, and note 
that the set of points that are spacelike separated from $S_1$ and 
$S_2$ are obviously given by $S'_1\cap S'_2$, but also by 
$(S_1\cup S_2)'$, 
so that 
\begin{subequations}
\label{eg:CapCup}
\begin{alignat}{2}
\label{eq:CapCup1}
& S'_1\cap S'_2&&\,=\,(S_1\cup S_2)'\,,\\
\label{eq:CapCup2}
& S_1\cap S_2&&\,=\,(S'_1\cup S'_2)'\,.
\end{alignat}
\end{subequations}
Here (\ref{eq:CapCup1}) and (\ref{eq:CapCup2}) are equivalent 
since any $S_i\in\mathrm{Caus}(\mathbb{M}^n)$ can be written as 
$S_i=P'_i$, namely $P_i=S'_i$. If $S_i$ runs through all 
sets in $\mathrm{Caus}(\mathbb{M}^n)$ so does $P_i$. Hence any 
equation that holds generally for all $S_i\in\mathrm{Caus}(\mathbb{M}^n)$
remains valid if the $S_i$ are replaced by $S'_i$. 

Equation (\ref{eq:CapCup2}) immediately shows that $S_1\cap S_2$
is causally complete (since it is the $'$ of something). Taking the 
causal complement of (\ref{eq:CapCup1}) we obtain the desired relation 
for $S_1\vee S_2:=(S_1\cup S_2)''$. Together we have  
\begin{subequations}
\label{eg:MeetJoin}
\begin{alignat}{2}
\label{eg:MeetJoin1}
& S_1\wedge S_2&&\,=\,S_1\cap S_2\,,\\
\label{eg:MeetJoin2}
& S_1\vee S_2&&\,=\,(S'_1\cap S'_2)'\,.
\end{alignat}
\end{subequations}
From these we immediately derive  
\begin{subequations}
\label{eg:Negating}
\begin{alignat}{2}
\label{eg:Negating1}
& (S_1\wedge S_2)'&&\,=\,S'_1\vee S'_2\,,\\
\label{eg:Negating2}
& (S_1\vee S_2)'&&\,=\,S'_1\wedge S'_2\,.
\end{alignat}
\end{subequations}

All what we have said so far for the set $\mathrm{Caus}(\mathbb{M}^n)$
could be repeated verbatim for the set $\mathrm{Chron}(\mathbb{M}^n)$
of \emph{chronologically complete} subsets. We say that $S_1$ and 
$S_2$ are \emph{chronologically disjoint} or \emph{non-timelike separated}, 
iff $S_1\cap S_2=\emptyset$ and $(p_1-p_2)^2\leq 0$ for any $p_1\in S_1$ 
and $p_2\in S_2$. $S'$, the \emph{chronological complement} of $S$, is 
now the largest subset of $\mathbb{M}^n$ which is chronologically 
disjoint to $S$. The only difference between the causal and the 
chronological complement of $S$ is that the latter now contains 
lightlike separated points outside $S$. 
A set $S$ is \emph{chronologically complete} iff $S=S''$, where the 
dashing now denotes the operation of taking the chronological complement. 
Again, for any set $S$ the set $S'$ is automatically chronologically 
complete and $S''$ is the smallest  chronologically complete subset 
containing $S$. Single points are chronologically complete subsets.
All the formal properties regarding ${}'$, $\wedge$, and $\vee$ stated 
hitherto for $\mathrm{Caus}(\mathbb{M}^n)$ are the same for 
$\mathrm{Chron}(\mathbb{M}^n)$. 

One major difference between $\mathrm{Caus}(\mathbb{M}^n)$ and 
$\mathrm{Chron}(\mathbb{M}^n)$ is that the types of diamond-shaped 
sets they contain are different. For example, the closed ones, 
(\ref{eq:ClosedDiamonds}), are members of both. The open ones, 
(\ref{eq:OpenDiamonds}), are contained in $\mathrm{Caus}(\mathbb{M}^n)$ 
but \emph{not} in $\mathrm{Chron}(\mathbb{M}^n)$. Instead, 
$\mathrm{Chron}(\mathbb{M}^n)$ contains the closed diamonds whose 
`equator'\footnote{By `equator' we mean the $(n-2)$--sphere in which the 
forward and backward light-cones in (\ref{eq:ClosedDiamonds}) intersect. 
In the two-dimensional drawings the `equator' is represented by just 
two points marking the right and left corners of the diamond-shaped set.}
have been removed. An essential structural difference between 
$\mathrm{Caus}(\mathbb{M}^n)$ and $\mathrm{Chron}(\mathbb{M}^n)$
will be stated below, after we have introduced the notion of a 
lattice to which we now turn. 

To put all these formal properties into the right frame we recall 
the definition of a lattice. Let $(L,\leq)$ be a partially ordered set 
and $a,b$ any two elements in $L$. Synonymously with $a\leq b$ we 
also write $b\geq a$ and say that $a$ is smaller than $b$, $b$ is 
bigger than $a$, or $b$ majorizes $a$. We also write $a<b$ if 
$a\leq b$ and $a\ne b$. If, with respect to $\leq$, their greatest 
lower and least upper bound exist, they are denoted by 
$a\wedge b$---called the `meet of $a$ and $b$'---and $a\vee b$---called 
the `join of $a$ and $b$'---respectively. A partially ordered set for 
which the greatest lower and least upper bound exist for any pair 
$a,b$ of elements from $L$ is called a \emph{lattice}.

We now list some of the most relevant additional structural elements 
lattices can have: A lattice is called \emph{complete} if greatest 
lower and least upper bound exist for any subset $K\subseteq L$. 
If $K=L$ they are called $0$ (the smallest element in the lattice)
and $1$ (the biggest element in the lattice) respectively. An 
\emph{atom} in a lattice is an element $a$ which majorizes only $0$, 
i.e. $0\leq a$ and if $0\leq b\leq a$ then $b=0$ or $b=a$. The lattice 
is called \emph{atomic} if each of its elements different from $0$ 
majorizes an atom. An atomic lattice is called \emph{atomistic} if 
every element is the join of the atoms it majorizes. An element 
$c$ is said to \emph{cover} $a$ if $a<c$ and if $a\leq b\leq c$ 
either $a=b$ or $b=c$. An atomic lattice is said to have the 
\emph{covering property} if, for every element $b$ and every atom $a$
for which $a\wedge b=0$, the join $a\vee b$ covers $b$.  

The subset $\{a,b,c\}\subseteq L$ is called a 
\emph{distributive triple} if 
\begin{subequations}
\label{eq:DistTriple}
\begin{alignat}{3}
\label{eq:DistTriple1}
& a\wedge (b\vee c) &&\,=\, (a\wedge b)\vee(a\wedge c)
\quad&&\text{and $(a,b,c)$ cyclically permuted}\,,\\
\label{eq:DistTriple2}
& a\vee (b\wedge c) &&\,=\, (a\vee b)\wedge(a\vee c)
\quad&&\text{and $(a,b,c)$ cyclically permuted}\,.   
\end{alignat}
\end{subequations}
\begin{definition}
A lattice is called \emph{distributive} or \emph{Boolean} if every triple 
$\{a,b,c\}$ is distributive. It is called \emph{modular} if every triple
$\{a,b,c\}$ with $a\leq b$ is distributive.
\end{definition} 
It is straightforward to check from (\ref{eq:DistTriple})
that modularity is equivalent to the following single condition: 
\begin{equation}
\label{eq:Modularity}
\text{modularity}\Leftrightarrow 
a\vee (b\wedge c)=b\wedge(a\vee c)\quad
\text{for all $a,b,c\in L$ s.t. $a\leq b$.}
\end{equation}

If in a lattice with smallest element $0$ and greatest element $1$ 
a map $L\rightarrow L$, $a\mapsto a'$, exist such that 
\begin{subequations}
\label{eq:Orthocomplement}
\begin{alignat}{1}
\label{eq:Orthocomplement1}
& a'':=(a')'=a\,,\\
\label{eq:Orthocomplement2}
& a\leq b\Rightarrow b'\leq a'\,,\\
\label{eq:Orthocomplement3}
& a\wedge a'=0\,,\quad a\vee a'=1\,,
\end{alignat}
\end{subequations}
the lattice is called \emph{orthocomplemented}. It follows that whenever 
the meet and join of a subset $\{a_i\mid i\in I\}$ ($I$ is some index set) 
exist one has De\,Morgan's laws\footnote{From these laws it also appears that 
the definition (\ref{eq:Orthocomplement3}) is redundant, as each of its 
two statements follows from the other, due to $0'=1$.}:   
\begin{subequations}
\label{eq:GenDeMorgan}
\begin{alignat}{2}
\label{eq:GenDeMorgan1}
&\bigl(\textstyle{\bigwedge_{i\in I}}\ a_i\bigr)'&&\,=\,
\textstyle{\bigvee_{i\in I}}\ a'_i\,,\\
\label{eq:GenDeMorgan2}
&\bigl(\textstyle{\bigvee_{i\in I}}\ a_i\bigr)'&&\,=\,
\textstyle{\bigwedge_{i\in I}}\ a'_i\,.
\end{alignat}
\end{subequations}

For orthocomplemented lattices there is a still weaker version 
of distributivity than modularity, which turns out to be physically 
relevant in various contexts: 
\begin{definition}
\label{def:Orthomodular}
An orthocomplemented lattice is called \emph{orthomodular} if every 
triple $\{a,b,c\}$ with $a\leq b$ and $c\leq b'$ is distributive. 
\end{definition}
From (\ref{eq:Modularity}) and using that $b\wedge c=0$ for $b\leq c'$ 
one sees that this is equivalent to the single condition (renaming 
$c$ to $c'$): 
\begin{subequations}
\label{eq:Orthomodularity}
\begin{alignat}{3}
\label{eq:Orthomodularity1}
\text{orthomod.}\
&\Leftrightarrow \quad
&&a\,=\,b\wedge(a\vee c')
\quad&&\text{for all $a,b,c\in L$ s.t. $a\leq b\leq c$\,,}\\
\label{eq:Orthomodularity2}
&\Leftrightarrow 
&&a\,=\,b\vee(a\wedge c')
\quad&&\text{for all $a,b,c\in L$ s.t. $a\geq b\geq c$\,,}
\end{alignat}
\end{subequations}
where the second line follows from the first by taking its 
orthocomplement and renaming $a',b',c$ to $a,b,c'$. It turns out 
that these conditions can still be simplified by making them 
independent of $c$. In fact, (\ref{eq:Orthomodularity}) are 
equivalent to 
\begin{subequations}
\label{eq:AltOrthomodularity}
\begin{alignat}{3}
\label{eq:AltOrthomodularity1}
\text{orthomod.}\
&\Leftrightarrow \quad
&&a=b\wedge(a\vee b')
\quad&&\text{for all $a,b\in L$ s.t. $a\leq b$\,,}\\
\label{eq:AltOrthomodularity2}
&\Leftrightarrow 
&&a=b\vee(a\wedge b')
\quad&&\text{for all $a,b\in L$ s.t. $a\geq b$\,.}
\end{alignat}
\end{subequations}
It is obvious that (\ref{eq:Orthomodularity}) implies 
(\ref{eq:AltOrthomodularity}) (set $c=b$). But the converse is 
also true. To see this, take e.g. (\ref{eq:AltOrthomodularity2})
and choose any $c\leq b$. Then $c'\geq b'$, $a\geq b$ (by hypothesis), 
and $a\geq a\wedge c'$ (trivially), so that $a\geq b\vee(a\wedge c')$. 
Hence $a\geq b\vee(a\wedge c')\geq b\vee(a\wedge b')=a$, which proves 
(\ref{eq:Orthomodularity2}).

Complete orthomodular atomic lattices are automatically atomistic. 
Indeed, let $b$ be the join of all atoms majorized by $a\ne 0$. Assume
$a\ne b$ so that necessarily $b<a$, then (\ref{eq:AltOrthomodularity2}) 
implies $a\wedge b'\ne 0$. Then there exists an atom $c$ majorized by 
$a\wedge b'$. This implies $c\leq a$ and $c\leq b'$, hence also 
$c\not\leq b$. But this is a contradiction, since $b$ is by definition 
the join of all atoms majorized by $a$.    
 
Finally we mention the notion of \emph{compatibility} or \emph{commutativity}, 
which is a symmetric, reflexive, but generally not transitive relation 
$R$ on an orthomodular lattice (cf. Sec.\,\ref{sec:GroupActions}). 
We write $a\natural b$ for $(a,b)\in R$ and define: 
\begin{subequations}
\label{eq:DefCompatibility}
\begin{alignat}{3}
\label{eq:DefCompatibility1}
a\natural b\quad
&\Leftrightarrow\quad
&&a&&\,=\,(a\wedge b)\vee (a\wedge b')\,,\\
\label{eq:DefCompatibility2}
&\Leftrightarrow\quad
&&b&&\,=\,(b\wedge a)\vee (b\wedge a')\,.
\end{alignat}
\end{subequations}
The equivalence of these two lines, which shows that the relation of 
being compatible is indeed symmetric, can be demonstrated using 
orthomodularity as follows: Suppose (\ref{eq:DefCompatibility1}) holds; 
then $b\wedge a'=b\wedge(b'\vee a')\wedge(b\vee a')=b\wedge(b'\vee a')$,
where we used  the orthocomplement of (\ref{eq:DefCompatibility1}) to 
replace $a'$ in the first expression and the trivial identity 
$b\wedge(b\vee a')=b$ in the second step. Now, applying 
(\ref{eq:AltOrthomodularity2}) to $b\geq a\wedge b$ we get 
$b=(b\wedge a)\vee[b\wedge(b'\vee a')]=(b\wedge a)\vee (b\wedge a')$, 
i.e. (\ref{eq:DefCompatibility2}). The converse,  
$(\ref{eq:DefCompatibility2})\Rightarrow(\ref{eq:DefCompatibility1})$, 
is of course entirely analogous. 
 
From (\ref{eq:DefCompatibility}) a few things are immediate: 
$a\natural b$ is equivalent to $a\natural b'$, $a\natural b$ is 
implied by $a\leq b$ or $a\leq b'$, and the elements $0$ and $1$ 
are compatible with all elements in the lattice. The \emph{center} 
of a lattice is the set of elements which are compatible with all 
elements in the lattice. In fact, the center is a Boolean sublattice. 
If the center contains no other elements than $0$ and $1$ the 
lattice is said to be \emph{irreducible}. The other extreme is a 
Boolean lattice, which is identical to its own center. Indeed, 
if $(a,b,b')$ is a distributive triple, one has 
$a=a\wedge 1=a\wedge (b\vee b')=(a\wedge b)\vee (a\wedge b')
\Rightarrow(\ref{eq:DefCompatibility1})$.

\begin{figure}[htb]
\begin{minipage}[b]{0.38\linewidth}
\centering\epsfig{figure=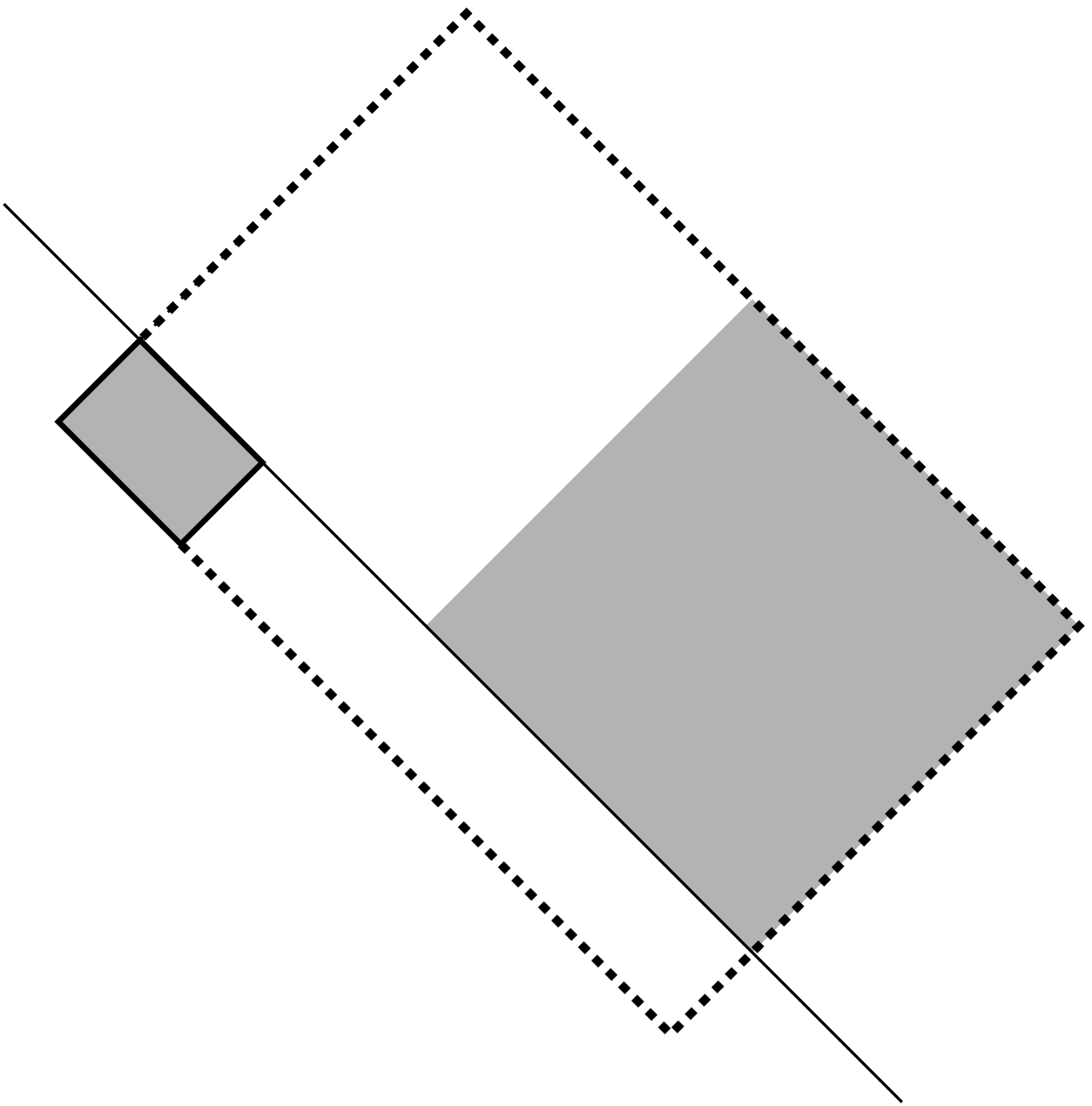,width=0.9\linewidth} 
\end{minipage}  
\hspace{0.8cm}
\begin{minipage}[b]{0.56\linewidth}
\centering\epsfig{figure=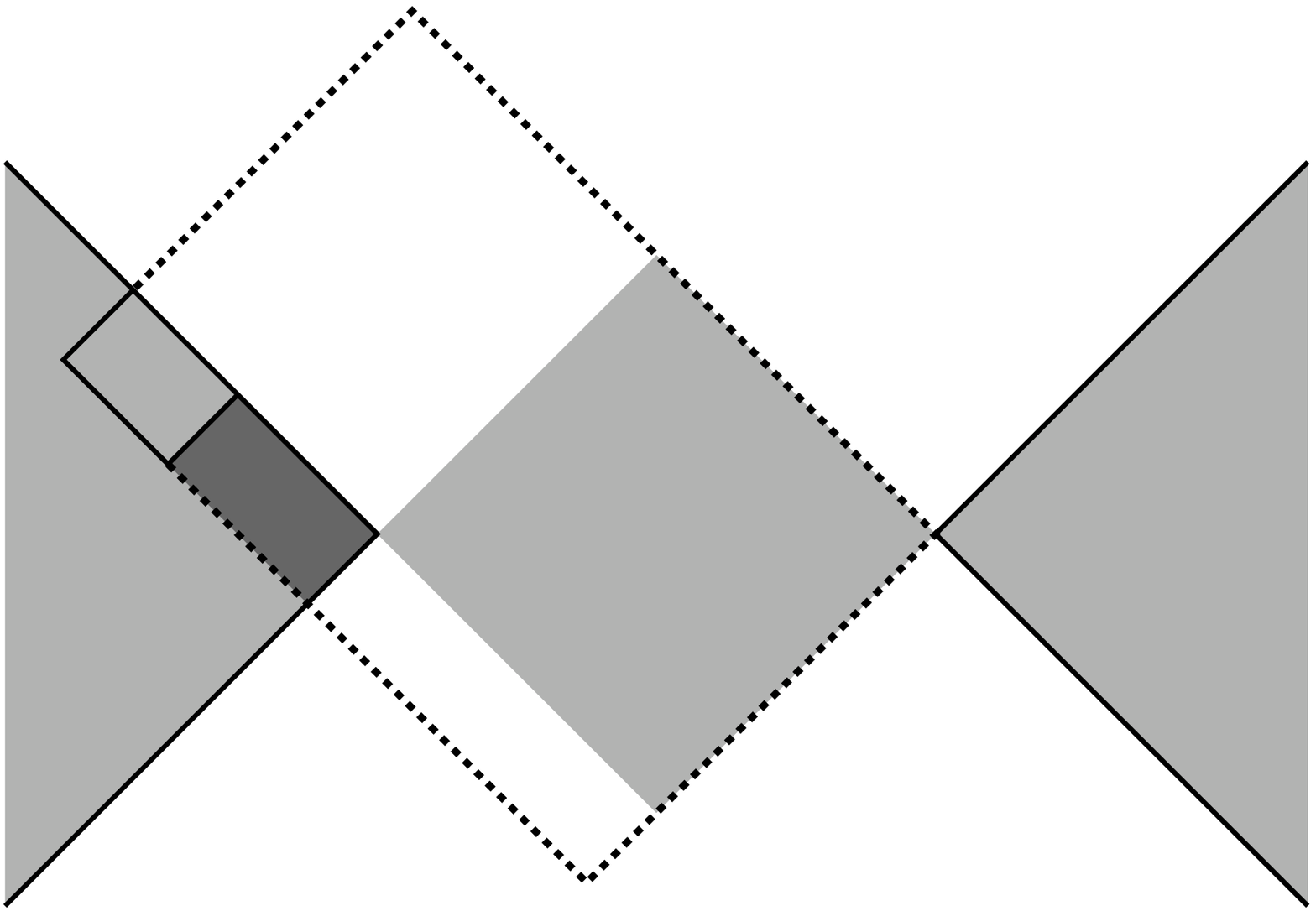,width=0.9\linewidth}
\end{minipage}
\put(-380,108){\large $\ell$}
\put(-358,76){\large $a$}
\put(-289,54){\large $b'$}
\put(-335,90){\large $a\vee b'$}
\put(-158,90){\large $a\vee b'$}
\put(-190,51){\large $b$}
\put(-30,51){\large $b$}
\put(-110,51){\large $b'$}
\put(-182,75){\large $a$}
\caption{\label{fig:CountEx}%
The two figures show that $\text{Caus}(\mathbb{M}^n)$ is not orthomodular. 
The first thing to note is that $\text{Caus}(\mathbb{M}^n)$ contains 
open (\ref{eq:OpenDiamonds}) as well as closed (\ref{eq:ClosedDiamonds})
diamond sets. In the left picture we consider the join of a small closed 
diamond $a$ with a large open diamond $b'$. (Closed sets are indicated by 
a solid boundary line.) 
Their edges are aligned along the lightlike line $\ell$. Even though 
these regions are causally disjoint, their causal completion is much 
larger than their union and given by the open (for $n>2$) enveloping 
diamond $a\vee b'$ framed by the dashed line.  (This also shows that 
the join of two regions can be larger than the domain of dependence 
of their union; compare footnote\,\ref{foot:Haag}.) . Next we consider 
the situation depicted on the right side. The closed double-wedge region 
$b$ contains the small closed diamond $a$. The causal 
complement $b'$ of $b$ is the open diamond in the middle. $a\vee b'$ is, 
according to the first picture, given by the large open diamond enclosed 
by the dashed line. The intersection of $a\vee b'$ with $b$ is strictly 
larger than $a$, the difference being the dark-shaded region in the left 
wedge of $b$ below $a$. Hence $a\ne b\wedge(a\vee b')$, in contradiction 
to (\ref{eq:AltOrthomodularity1}).}
\end{figure}

After these digression into elementary notions of lattice theory we 
come back to our examples of the sets $\mathrm{Caus}(\mathbb{M}^n)$
$\mathrm{Chron}(\mathbb{M}^n)$. Our statements above amount 
to saying that they are complete, atomic, and orthocomplemented 
lattices. The partial order relation $\leq$ is given by $\subseteq$ 
and the extreme elements $0$ and $1$ correspond to the empty set 
$\emptyset$ and the total set $\mathbb{M}^n$, the points of which 
are the atoms. Neither the covering property nor modularity 
is shared by any of the two lattices, as can be checked by way of 
elementary counterexamples.\footnote{An immediate counterexample for 
the covering property is this: Take two timelike separated points 
(i.e. atoms) $p$ and $q$. Then $\{p\}\wedge\{q\}=\emptyset$ whereas 
$\{p\}\vee\{q\}$ is given by the closed diamond (\ref{eq:ClosedDiamonds}). 
Note that this is true in $\mathrm{Caus}(\mathbb{M}^n)$ \emph{and} 
$\mathrm{Chron}(\mathbb{M}^n)$. But, clearly, $\{p\}\vee\{q\}$ does 
not cover either $\{p\}$ or $\{q\}$.} In particular, neither of them 
is Boolean. However, in \cite{CeglaJadczyk:1977} it was shown that 
$\mathrm{Chron}(\mathbb{M}^n)$ is orthomodular; see also \cite{Casini:2002} 
which deals with more general spacetimes. Note that by the argument 
given above this implies that $\mathrm{Chron}(\mathbb{M}^n)$ is 
atomistic. In contrast, $\mathrm{Caus}(\mathbb{M}^n)$ is definitely 
\emph{not} orthomodular, as is e.g. seen by the counterexample given 
in Fig.\,\ref{fig:CountEx}.\footnote{\label{foot:Haag}
Regarding this point, there are some conflicting statements in the 
literature. The first edition of \cite{Haag:LocQuantPhys} states 
orthomodularity of $\mathrm{Chron}(\mathbb{M}^n)$ in 
Proposition\,4.1.3, which is removed in the second edition without 
further comment. The proof offered in the first edition uses 
(\ref{eq:AltOrthomodularity1}) as definition of orthomodularity, 
writing $K_1$ for $a$ and $K_2$ for b. The crucial step is the claim 
that any spacetime event in the set $K_2\wedge(K_1\vee K_2')$ lies in 
$K_2$ and that any causal line through 
it must intersect either $K_1$ or $K_2'$.  The last statement is, 
however, not correct since the join of two sets (here $K_1$ and $K_2'$) 
is generally larger than the domain of dependence of their ordinary 
set-theoretic union; compare Fig.\,\ref{fig:CountEx}. :
(Generally, the domain of dependence of a subset $S$ of spacetime $M$ 
is the largest subset $D(S)\subseteq M$ such that any inextensible 
causal curve that intersects $D(S)$ also intersects $S$.)} 
It is also not difficult to prove that $\mathrm{Chron}(\mathbb{M}^n)$ 
is irreducible.\footnote{In general spacetimes $M$, the failure of 
irreducibility of $\mathrm{Chron}(M)$ is directly related to the 
existence of closed timelike curves; see~\cite{Casini:2002}.}
\noindent

It is well known that the lattices of propositions for classical systems 
are Boolean, whereas those for quantum systems are merely orthomodular.
In classical physics the elements of the lattice are measurable subsets 
of phase space, with $\leq$ being ordinary set-theoretic inclusion 
$\subseteq$, and $\wedge$ and $\vee$ being ordinary set-theoretic 
intersection $\cap$ and union $\cup$ respectively. The orthocomplement 
is the ordinary set-theoretic complement.  In Quantum Mechanics 
the elements of the lattice are the closed subspaces of Hilbert space, 
with $\leq$ being again ordinary inclusion, $\wedge$ ordinary 
intersection, and $\vee$ is given by 
$a\vee b:=\overline{\Span\{a,b\}}$. The orthocomplement of a 
closed subset is the orthogonal complement in Hilbert space. For 
comprehensive discussions see \cite{Jauch:FoundQM} and 
\cite{Beltrametti:LogicQM}.

One of the main questions in the foundations of Quantum Mechanics is 
whether one could understand (derive) the usage of Hilbert spaces
and complex numbers from somehow more fundamental principles. Even 
though it is not a priori clear what ones measure of fundamentality 
should be at this point, an interesting line of attack consists in 
deriving the mentioned structures from the properties of the lattice 
of propositions (Quantum Logic). It can be shown that a lattice that 
is complete, atomic, irreducible, orthomodular, and that satisfies the 
covering property is isomorphic to the lattice of closed subspaces 
of a linear space with Hermitean inner product. The complex numbers 
are selected if additional technical assumptions are added. 
For the precise statements of these reconstruction theorems see
\cite{Beltrametti:LogicQM}.

It is now interesting to note that, on a formal level, there is a similar 
transition in going from Galilei invariant to Lorentz invariant causality 
relations. In fact, in Galilean spacetime one can also define a chronological 
complement: Two points are chronologically related if they are connected 
by a worldline of finite speed and, accordingly, two subsets in spacetime 
are chronologically disjoint if no point in one set is chronologically 
related to a point of the other. For example, the chronological 
complement of a point $p$ are all points simultaneous to, but different 
from, $p$. More general, it is not hard to see that the chronologically 
complete sets are just the subsets of some $t=\text{const.}$ hypersurface. 
The lattice of chronologically complete sets is then the continuous 
disjoint union of sublattices, each of which is isomorphic to the 
Boolean lattice of subsets in $\reals^3$. For details see 
\cite{CeglaJadczyk:1976}.    

As we have seen above, $\text{Chron}(\mathbb{M}^n)$ is complete, 
atomic, irreducible, and orthomodular (hence atomistic). The main 
difference to the lattice of propositions in Quantum Mechanics, as 
regards the formal aspects discussed here, is that 
$\text{Chron}(\mathbb{M}^n)$ does \emph{not} satisfy the covering 
property. Otherwise the formal similarities are intriguing and it 
is tempting to ask whether there is a deeper meaning to this. 
In this respect it would be interesting to know whether one could 
give a lattice-theoretic characterization for $\text{Chron}(M)$ 
($M$ some fixed spacetime), comparable to the 
characterization of the lattices of closed subspaces in Hilbert space 
alluded to above. Even for $M=\mathbb{M}^n$ such a characterization 
seems, as far as I am aware, not to be known.

\subsection{Rigid motion}
\label{sec:RigidMotion}
As is well known, the notion of a rigid body, which proves so useful 
in Newtonian mechanics, is incompatible with the existence of a 
universal finite upper bound for all signal velocities~\cite{Laue:1911}. 
As a result, the notion of a perfectly rigid body does not exist within 
the framework of SR. However, the notion of a \emph{rigid motion} does 
exist. Intuitively speaking, a body moves rigidly if, locally, the 
relative spatial distances of its material constituents are unchanging. 

The motion of an extended body is described by a normalized timelike 
vector field $u:\Omega\rightarrow\reals^n$, where $\Omega$ is an
open subset of Minkowski space, consisting of the events where the 
material body in question `exists'. We write $g(u,u)=u\cdot u=u^2$ 
for the Minkowskian scalar product. Being normalized now means that
$u^2=c^2$ (we do \emph{not} choose units such that $c=1$). The Lie derivative 
with respect to $u$ is denoted by $L_u$. 

For each material part of the body in motion its local rest space 
at the event $p\in\Omega$ can be identified with the hyperplane 
through $p$ orthogonal to $u_p$:
\begin{equation}
\label{eq:LocalRestSpace}
H_p:=p+u_p^\perp\,.
\end{equation}
$u_p^\perp$ carries a Euclidean inner product, $h_p$, given by the 
restriction of $-g$ to $u_p^\perp$. Generally we can write 
\begin{equation}
\label{eq:HorMetric}
h=c^{-2}\,u^\flat\otimes u^\flat-g\,,
\end{equation}
where $u^\flat=g^\downarrow(u):=g(u,\cdot)$ is the one-form associated 
(`index-lowered', cf. Sec.\,\ref{sec:IndexMoving}) to $u$. 
Following~\cite{Born:1909} the precise definition of `rigid  motion' 
can now be given as follows:  
\begin{definition}[Born 1909]
\label {def:RigidMotion}
Let $u$ be a normalized timelike vector field $u$. The motion described 
by its flow  is \emph{rigid} if
\begin{equation}
\label{eq:def:RigMo}
L_u h=0\,.
\end{equation}
\end{definition}
\noindent
Note that, in contrast to the Killing equations $L_ug=0$,
these equations are non linear due to the dependence of $h$ 
upon $u$.   

We write $\Proj_h:=\text{id}-c^{-2}\,u\otimes u^\flat\in\End(\reals^n)$ 
for the tensor field over spacetime that pointwise projects vectors 
perpendicular to $u$. It acts on one forms 
$\alpha$ via $\Proj_h(\alpha):=\alpha\circ\Proj_h$ and accordingly on all 
tensors. The so extended projection map will still be denoted by 
$\Proj_h$. Then we e.g. have 
\begin{equation}
\label{eq:ProjMetric}
h=-\Proj_h g:=-g(\Proj_h\cdot,\Proj_h\cdot)\,. 
\end{equation}

It is not difficult to derive the following two 
equations:\footnote{\label{foot:LieVel} Equation (\ref{eq:RigMo2})
simply follows from $L_u\Proj_h=-c^{-2}u\otimes L_u u^\flat$, so that 
$g((L_u\Proj_h)X,\Proj_h Y)=0$ for all $X,Y$. In fact, $L_uu^\flat=a^\flat$,
where $a:=\nabla_uu$ is the spacetime-acceleration. This follows from 
$L_uu^\flat(X)=L_u(g(u,X))-g(u,L_uX)=g(\nabla_uu,X)+g(u,\nabla_uX-[u,X])=
g(a,X)-g(u,\nabla_Xu)=g(a,X)$, where $g(u,u)=\text{const.}$ was used 
in the last step.}  
\begin{alignat}{2}
\label{eq:RigMo1}
& L_{fu}h&&\,=\,fL_uh\,,\\
\label{eq:RigMo2}
& L_uh&&\,=\,-L_u(\Proj_h g)=-\Proj_h(L_ug)\,,
\end{alignat}
where $f$ is any differentiable real-valued function on $\Omega$.

Equation (\ref{eq:RigMo1}) shows that the normalized vector field 
$u$ satisfies (\ref{eq:def:RigMo}) iff any rescaling $fu$ with a 
nowhere vanishing function $f$ does. Hence the normalization 
condition for $u$ in (\ref{eq:def:RigMo}) is really irrelevant.  
It is the geometry in spacetime of the flow lines and not their 
parameterization which decide on whether motions (all, i.e. for any 
parameterization, or none) along them are rigid. This has be the 
case because, generally speaking, there is no distinguished family 
of sections (hypersurfaces) across the bundle of flow lines that 
would represent `the body in space', i.e. mutually simultaneous 
locations of the body's points. Distinguished cases are those 
exceptional ones in which $u$ is hypersurface orthogonal. 
Then the intersection of $u$'s flow lines with the orthogonal 
hypersurfaces consist of mutually \emph{Einstein synchronous} 
locations of the points of the body. An example is discussed below. 

Equation (\ref{eq:RigMo2}) shows that the rigidity condition is 
equivalent to the `spatially' projected Killing equation. 
We call the flow of the timelike normalized vector field $u$ a 
\emph{Killing motion} (i.e. a spacetime isometry) if there is a 
Killing field $K$ such that $u=cK/\sqrt{K^2}$. Equation 
(\ref{eq:RigMo2}) immediately implies that Killing motions are 
rigid. What about the converse? Are there rigid motions that are 
not Killing? This turns out to be a difficult question. Its answer 
in Minkowski space is: `yes, many, but not as many as naively 
expected.' 

Before we explain this, let us give an illustrative example for a 
Killing motion, namely that generated by the boost Killing-field
in Minkowski space. We suppress all but one spatial directions and 
consider boosts in $x$ direction in two-dimensional Minkowski space 
(coordinates $ct$ and $x$; metric $ds^2=c^2dt^2-dx^2$). The Killing 
field is\footnote{Here we adopt the standard notation from differential 
geometry, where $\partial_{\mu}:=\partial/\partial x^\mu$ denote the 
vector fields naturally defined by the coordinates 
$\{x^\mu\}_{\mu=0\cdots n-1}$. Pointwise the dual basis to 
$\{\partial_\mu\}_{\mu=0\cdots n-1}$ is $\{dx^\mu\}_{\mu=0\cdots n-1}$.}
\begin{equation}
\label{eq:BoostKill1}
K=x\,\partial_{ct}+ct\,\partial_x\,,
\end{equation}       
which is timelike in the region $\vert x\vert >\vert ct\vert$. 
We focus on the `right wedge' $x >\vert ct\vert$, which is 
now our region $\Omega$. Consider a rod of length $\ell$ which at $t=0$ 
is represented by the interval $x\in(r,r+\ell)$, where $r>0$. 
The flow of the normalized field $u=cK/\sqrt{K^2}$ is 
\begin{subequations}
\label{eq:BoostKill2}
\begin{alignat}{2}
\label{eq:BoostKill2a}
& ct(\tau)&&\,=\,x_0\,\sinh\bigl(c\tau/x_0)\,,\\
\label{eq:BoostKill2b}
&  x(\tau)&&\,=\,x_0\,\cosh\bigl(c\tau/x_0)\,,
\end{alignat}
\end{subequations}
where $x_0=x(\tau=0)\in(r,r+\ell)$ labels the elements of the 
rod at $\tau=0$. We have $x^2-c^2t^2=x_0^2$, showing that the 
individual elements of the rod move on hyperbolae (`hyperbolic 
motion'). $\tau$ is the proper time along each orbit, 
normalized so that the rod lies on the $x$ axis at $\tau=0$.

The combination 
\begin{equation}
\label{eq:KillingTime}
\lambda:=c\tau/x_0
\end{equation}
is just the flow parameter for $K$ (\ref{eq:BoostKill1}), sometimes 
referred to as `Killing time' (though it is dimensionless).  
From (\ref{eq:BoostKill2}) we can solve for $\lambda$ and $\tau$ 
as functions of $ct$ and $x$:
\begin{subequations}
\label{eq:LambdaTau}
\begin{alignat}{3}
\label{eq:LambdaTau1}
& \lambda&&\,=\,
f(ct,x)&&\,:=\,\tanh^{-1}\bigl(ct/x\bigr)\,,\\
\label{eq:LambdaTau2}
& \tau&&\,=\,
\hat f(ct,x)&&\,:=\,\underbrace{\sqrt{(x/c)^2-t^2}}_{x_0/c}\,\tanh^{-1}\bigl(ct/x\bigr)\,,
\end{alignat}
\end{subequations}
from which we infer that the hypersurfaces of constant $\lambda$ 
are hyperplanes which all intersect at the origin. Moreover, 
we also have $df=K^\flat/K^2$ ($d$ is just the ordinary exterior 
differential) so that the hyperplanes of constant 
$\lambda$ intersect all orbits of $u$ (and $K$) orthogonally.
Hence the hyperplanes of constant $\lambda$ qualify as the 
equivalence classes of mutually Einstein-simultaneous events in 
the region $x>\vert ct\vert$ for a family of observers moving along 
the Killing orbits. This does not hold for the hypersurfaces of 
constant $\tau$, which are curved.  

The modulus of the spacetime-acceleration (which is the same as the 
modulus of the spatial acceleration measured in the local rest frame) 
of the material part of the rod labeled by $x_0$ is 
\begin{equation}
\label{eq:AccRod}
\Vert a\Vert_g=c^2/x_0\,.
\end{equation} 
As an aside we generally infer from this that, given a timelike 
curve of local acceleration (modulus) $\alpha$, infinitesimally 
nearby orthogonal hyperplanes intersect at a spatial 
distance $c^2/\alpha$. This remark will become relevant in the 
discussion of part\,2 of the Noether-Herglotz theorem given below.

In order to accelerate the rod to the uniform velocity $v$ without
deforming it, its material point labeled by $x_0$ has to accelerate 
for the eigentime (this follows from (\ref{eq:BoostKill2}))
\begin{equation}
\label{eq:EigTimeRod}
\tau=\frac{x_0}{c}\tanh^{-1}(v/c)\,,
\end{equation} 
which depends on $x_0$. In contrast, the Killing time is the same 
for all material points and just given by the final rapidity. 
In particular, judged from the local observers moving with the rod, 
a rigid acceleration requires accelerating the rod's trailing end 
harder but shorter than pulling its leading end. 

In terms of the coordinates $(\lambda,x_0)$, which are comoving with 
the flow of $K$, and $(\tau,x_0)$, which are comoving with the flow 
of $u$, we just have $K=\partial/\partial\lambda$ and 
$u=\partial/\partial\tau$ respectively. The spacetime metric $g$ 
and the projected metric $h$ in terms of these coordinates are:
\begin{subequations}
\label{eq:MetricsComovCoord}
\begin{alignat}{2}
\label{eq:MetricsComovCoord1}
& h&&\,=\,dx_0^2\,,\\
\label{eq:MetricsComovCoord2}
& g&&\,=\,x_0^2\,d\lambda^2-dx_0^2
=c^2\bigl(d\tau-(\tau/x_0)\,dx_0\bigr)^2-dx_0^2\,.
\end{alignat}
\end{subequations}
Note the simple form $g$ takes in terms of $x_0$ and $\lambda$,
which are also called the `Rindler coordinates' for the region 
$\vert x\vert>\vert ct\vert$ of Minkowski space. They are the 
analogs in Lorentzian geometry to polar coordinates (radius $x_0$, 
angle $\lambda$) in Euclidean geometry.  

Let us now return to the general case. We decompose the derivative 
of the velocity one-form $u^\flat:=g^\downarrow(u)$ as follows:
\begin{equation}
\label{eq:DecVelOneform}
\nabla u^\flat=\theta+\omega+ c^{-2}\,u^\flat\otimes a^\flat\,,
\end{equation}
where $\theta$ and $\omega$ are the projected symmetrized and
antisymmetrized derivatives respectively\footnote{We denote the 
symmetrized and antisymmetrized tensor-product (not including 
the factor $1/n!$) by $\vee$ and $\wedge$ respectively and the 
symmetrized and antisymmetrized (covariant-) derivative by 
$\nabla\vee$ and $\nabla\wedge$. For example, 
$(u^\flat\wedge v^\flat)_{ab}=u_av_b-u_bv_a$ and 
$(\nabla\vee u^\flat)_{ab}=\nabla_a u_b+\nabla_bu_a$.
Note that $(\nabla\wedge u^\flat)$ is the same as the ordinary 
exterior differential $du^\flat$. Everything we say in the sequel 
applies to curved spacetimes if $\nabla$ is read as covariant 
derivative with respect to the Levi-Civita connection.}  
\begin{subequations}
\label{eq:ThetaOmega}
\begin{alignat}{3}
\label{eq:ThetaOmega1}
& 2\theta 
&&\,=\, \Proj_h(\nabla\vee u^\flat)
&&\,=\,\nabla\vee u^\flat-c^{-2}\,u^\flat\vee a^\flat\,,\\
\label{eq:ThetaOmega2}
& 2\omega 
&&\,=\,\Proj_h(\nabla\wedge u^\flat)
&&\,=\,\nabla\wedge u^\flat-c^{-2}\, u^\flat\wedge a^\flat\,.
\end{alignat}
\end{subequations}
The symmetric part, $\theta$, is usually further decomposed into 
its traceless and pure trace part, called the \emph{shear} and 
\emph{expansion} of $u$ respectively. The antisymmetric part 
$\omega$ is called the \emph{vorticity} of $u$. 

Now recall that the Lie derivative of $g$ is just twice the 
symmetrized derivative, which in our notation reads: 
\begin{equation}
\label{eq:LieDerNabla}
L_ug=\nabla\vee u^\flat\,.
\end{equation}
This implies in view of (\ref{eq:def:RigMo}), (\ref{eq:RigMo2}), and 
(\ref{eq:ThetaOmega1}) 
\begin{proposition}
Let $u$ be a normalized timelike vector field $u$. The motion described 
by its flow  is rigid iff $u$ is of vanishing shear and expansion, i.e. 
iff $\theta=0$. 
\end{proposition}  

Vector fields generating rigid motions are now classified according to 
whether or not they have a vanishing vorticity $\omega$: if $\omega=0$ 
the flow is called \emph{irrotational}, otherwise \emph{rotational}. 
The following theorem is due to Herglotz \cite{Herglotz:1910} 
and Noether~\cite{Noether:1910}:
\begin{theorem}[Noether \& Herglotz, part\,1]
\label{thm:NoetherHerglotz1}
A rotational rigid motion in Minkowski space must be a Killing motion.
\end{theorem}
An example of such a rotational motion is given by the Killing 
field\footnote{\label{foot:MetricCyl} We now use standard cylindrical 
coordinates $(z,\rho,\varphi)$, in terms of which  
$ds^2=c^2dt^2-dz^2-d\rho^2-\rho^2\,d\varphi^2$.}  
\begin{equation}
\label{eq:RotKill1}
K=\partial_t+\kappa\,\partial_{\varphi}
\end{equation}
inside the region 
\begin{equation}
\label{eq:RegionOmega}
\Omega=\{(t,z,\rho,\varphi)\mid\kappa\rho<c\}\,,
\end{equation}
where $K$ is timelike. This motion corresponds to a rigid rotation 
with constant angular velocity $\kappa$ which, without loss of generality, 
we take to be positive. Using the comoving angular coordinate 
$\psi:=\varphi-\kappa t$, the split (\ref{eq:HorMetric}) is now 
furnished by 
\begin{subequations}
\label{eq:RotSplitMetric}
\begin{alignat}{2}
\label{eq:RotSplitMetric1}
& u^\flat&&\,=\,
c\,\sqrt{1-(\kappa\rho/c)^2}
\left\{
c\,dt-\frac{\kappa\rho/c}{1-(\kappa\rho/c)^2}\ \rho\,d\psi
\right\}\,,\\
\label{eq:RotSplitMetric2}
& h&&\,=\,
dz^2+d\rho^2+\frac{\rho^2\,d\psi^2}{1-(\kappa\rho/c)^2}\,.
\end{alignat}
\end{subequations}
The metric $h$ is curved (cf. Lemma\,\ref{lemma:NoetherHerglotz1-1}). 
But the rigidity condition (\ref{eq:def:RigMo}) means that $h$, and 
hence its curvature, cannot change along the motion. Therefore, even 
though we can keep a body in uniform rigid rotational motion, we cannot 
put it into this state from rest by purely rigid motions, since this would 
imply a transition from a flat to a curved geometry of the body. 
This was first pointed out by Ehrenfest~\cite{Ehrenfest:1909}. 
Below we will give a concise analytical expression of this fact 
(cf. equation~(\ref{eq:OmegaLieZero})). All this is in contrast to the 
translational motion, as we will also see below.

The proof of Theorem\,\ref{thm:NoetherHerglotz1} relies on 
arguments from differential geometry proper and is somewhat 
tricky. Here we present the essential steps, basically following 
\cite{PiraniWilliams:1962} and \cite{Trautman:1965} in a 
slightly modernized notation. Some straightforward calculational 
details will be skipped. The argument itself is best broken 
down into several lemmas. 

At the heart of the proof lies the following general 
construction: Let $M$ be the spacetime manifold 
with metric $g$ and $\Omega\subset M$ the open region in which the 
normalized vector field $u$ is defined. We take $\Omega$ to be simply 
connected. The orbits of $u$ foliate $\Omega$ and hence define an 
equivalence relation on $\Omega$ given by $p\sim q$ iff $p$ and $q$ lie 
on the same orbit. The quotient space $\hat\Omega:=\Omega/\!\!\sim$ is 
itself a manifold. Tensor fields on $\hat\Omega$ can be represented by 
(i.e. are in bijective correspondence to) tensor fields $T$ on 
$\Omega$ which obey the two conditions: 
\begin{subequations}
\label{eq:Projectibility}
\begin{alignat}{2}
\label{eq:Projectibility1}
&\Proj_h\, T&&\,=\,T\,,\\
\label{eq:Projectibility2}     
& L_uT&&\,=\,0\,.
\end{alignat}
\end{subequations}
Tensor fields satisfying (\ref{eq:Projectibility1}) are called 
\emph{horizontal}, those satisfying both conditions 
(\ref{eq:Projectibility}) are called \emph{projectable}. 
The $(n-1)$-dimensional metric tensor $h$, defined in 
(\ref{eq:HorMetric}), is an example of a projectable tensor if 
$u$ generates a rigid motion, as assumed here. It turns $(\hat\Omega,h)$ 
into a $(n-1)$-dimensional Riemannian manifold. The covariant derivative 
$\hat\nabla$ with respect to the Levi-Civita connection of $h$ is 
given by the following operation on projectable tensor fields: 
\begin{equation}
\label{eq:ProjCovDer}
\hat\nabla:=\Proj_h\circ\nabla
\end{equation}
i.e. by first taking the covariant derivative $\nabla$ (Levi-Civita
connection in $(M,g)$) in spacetime and then projecting the result 
horizontally. This results again in a projectable tensor, as a 
straightforward calculation shows. 

The horizontal projection of the spacetime curvature tensor can 
now be related to the curvature tensor of $\hat\Omega$ (which is a 
projectable tensor field). Without proof we state 
\begin{lemma}
\label{lemma:NoetherHerglotz1-1}
Let $u$ generate a rigid motion in spacetime. Then the horizontal 
projection of the totally covariant (i.e. all indices down) curvature 
tensor $R$ of $(\Omega,g)$ is related to the totally covariant 
curvature tensor $\hat R$ of $(\hat\Omega,h)$ by the following 
equation\footnote{$\hat R$ 
appears with a minus sign on the right hand side of 
(\ref{eq:ProjectedCurvatureRelation}) because the first index on 
the hatted curvature tensor is lowered with $h$ rather than $g$. 
This induces a minus sign due to (\ref{eq:HorMetric}), i.e. as a 
result of our `mostly-minus'-convention for the signature of the 
spacetime metric.}:
\begin{equation}
\label{eq:ProjectedCurvatureRelation} 
\Proj_h R=-\hat R-3\,(\text{\rm id}-\Proj_\wedge)\omega\otimes\omega\,,
\end{equation}
where $\Proj_\wedge$ is the total antisymmetrizer, which here 
projects tensors of rank four onto their totally antisymmetric 
part. 
\end{lemma}
\noindent
Formula (\ref{eq:ProjectedCurvatureRelation}) is true in any spacetime 
dimension $n$. Note that the projector $(\text{id}-\Proj_\wedge)$
guarantees consistency with the first Bianchi identities for $R$ and 
$\hat R$, which state that the total antisymmetrization in their last 
three slots vanish identically. This is consistent with 
(\ref{eq:ProjectedCurvatureRelation}) since for tensors of rank 
four with the symmetries of $\omega\otimes\omega$ the total 
antisymmetrization on tree slots is identical to $\Proj_\wedge$,
the symmetrization on all four slots. The claim now simply follows from 
$\Proj_\wedge\circ(\text{id}-\Proj_\wedge)=\Proj_\wedge-\Proj_\wedge=0$.

We now restrict to spacetime dimensions of four or less, i.e. $n\leq 4$.
In this case $\Proj_\wedge\circ\Proj_h=0$ since $\Proj_h$ makes the 
tensor effectively live over $n-1$ dimensions, and any totally 
antisymmetric four-tensor in three or less dimensions must vanish. 
Applied to (\ref{eq:ProjectedCurvatureRelation}) this 
means that $\Proj_\wedge(\omega\otimes\omega)=0$, for horizontality 
of $\omega$ implies $\omega\otimes\omega=\Proj_h(\omega\otimes\omega)$.
Hence the right hand side of  (\ref{eq:ProjectedCurvatureRelation}) 
just contains the pure tensor product $-3\,\omega\otimes\omega$. 

Now, in our case $R=0$ since $(M,g)$ is flat Minkowski space. 
This has two interesting consequences: First, $(\hat\Omega,h)$
is curved iff the motion is rotational, as exemplified above. 
Second, since $\hat R$ is projectable, its Lie derivative with 
respect to $u$ vanishes. Hence (\ref{eq:ProjectedCurvatureRelation}) 
implies $L_u\omega\otimes\omega+\omega\otimes L_u\omega=0$, which is 
equivalent to\footnote{In more than four spacetime dimensions one only 
gets $(\text{id}-\Proj_\wedge)(L_u\omega\otimes\omega+\omega\otimes L_u\omega)=0$.}
\begin{equation}
\label{eq:OmegaLieZero}  
L_u\omega=0\,.
\end{equation}
This says that the vorticity cannot change along a rigid motion 
in flat space. It is the precise expression for the remark above 
that you cannot rigidly \emph{set} a disk into rotation. Note 
that it also provides the justification for the global 
classification of rigid motions into rotational and irrotational 
ones. 

A sharp and useful criterion for whether a rigid motion is Killing
or not is given by the following 
\begin{lemma}
\label{lemma:NoetherHerglotz1-2}
Let $u$ be a normalized timelike vector field on a region 
$\Omega\subseteq M$. The motion generated by $u$ is Killing iff 
it is rigid and $a^\flat$ is exact on $\Omega$.  
\end{lemma}
\begin{proof}
That the motion generated by $u$ be Killing is equivalent to the 
existence of a positive function $f:\Omega\rightarrow\reals$ such 
that $L_{fu}g=0$, i.e. $\nabla\vee(fu^\flat)=0$. 
In view of (\ref{eq:ThetaOmega1}) this is equivalent to
\begin{equation}
\label{eq:Matteo1}
2\theta+(d\ln f+c^{-2} a^\flat)\vee u^\flat=0\,,
\end{equation}
which, in turn, is equivalent to $\theta=0$ and $a^\flat=-c^2\,d\ln f$. 
This is true since $\theta$ is horizontal, $\Proj_h\theta=\theta$, whereas 
the first term in (\ref{eq:Matteo1}) vanishes upon applying $\Proj_h$.
The result now follows from reading this equivalence both ways: 
1)~The Killing condition for $K:=fu$ implies rigidity for $u$
and exactness of $a^\flat$. 2)~Rigidity of $u$ and $a^\flat=-d\Phi$ 
imply that $K:=fu$ is Killing, where $f:=\exp(\Phi/c^2)$.  
\end{proof}

We now return to the condition (\ref{eq:OmegaLieZero})
and express $L_u\omega$ in terms of $du^\flat$. For this 
we recall that $L_uu^\flat=a^\flat$ 
(cf. footnote\,\ref{foot:LieVel}) and that Lie derivatives 
on forms commute with exterior derivatives\footnote{This is most 
easily seen by recalling that on forms the Lie derivative can be 
written as $L_u=d\circ i_u+i_u\circ d$, where $i_u$ is the map of 
inserting $u$ in the first slot.}. Hence we have 
\begin{equation}
\label{eq:LieOmega1}
2\,L_u\omega
=L_u(\Proj_h du^\flat)
=\Proj_h da^\flat
=da^\flat-c^{-2}u^\flat\wedge L_ua^\flat\,.
\end{equation}
Here we used the fact that the additional terms that result 
from the Lie derivative of the projection tensor $\Proj_h$ vanish, 
as a short calculation shows, and also that on forms the 
projection tensor $\Proj_h$ can be written as 
$\Proj_h=\text{id}-c^{-2}u^\flat\wedge i_u$, where $i_u$ denotes 
the map of insertion of $u$ in the first slot. 

Now we prove 
\begin{lemma}
\label{lemma:NoetherHerglotz1-3} 
Let $u$ generate a rigid motion in flat space such that 
$\omega\ne 0$, then  
\begin{equation}
\label{eq:LieAccZero}
L_ua^\flat=0\,.
\end{equation}
\end{lemma}
\begin{proof}
Equation (\ref{eq:OmegaLieZero}) says that $\omega$ is 
projectable (it is horizontal by definition). Hence 
$\hat\nabla\omega$ is projectable, which implies 
\begin{equation}
\label{eq:HatNablaOmega1}
L_u\hat\nabla\omega=0\,.
\end{equation} 
Using (\ref{eq:DecVelOneform}) with $\theta=0$ one has
\begin{equation}
\label{eq:HatNablaOmega2}
\hat\nabla\omega
=\Proj_h\nabla\omega
=\Proj_h\nabla\nabla u^\flat-c^{-2}\Proj_h(\nabla u^\flat\otimes a^\flat)\,.
\end{equation}
Antisymmetrization in the first two tensor slots makes the first 
term on the right vanish due to the flatness on $\nabla$. The 
antisymmetrized right hand side is hence equal to 
$-c^{-2}\omega\otimes a^\flat$. Taking the Lie derivative of 
both sides makes the left hand side vanish due to 
(\ref{eq:HatNablaOmega1}), so that 
\begin{equation}
\label{eq:HatNablaOmega3}
L_u(\omega\otimes a^\flat)=\omega\otimes L_u a^\flat=0
\end{equation}
where we also used (\ref{eq:OmegaLieZero}). 
So we see that $L_ua^\flat=0$ if $\omega\ne 0$.\footnote{We 
will see below that (\ref{eq:LieAccZero}) is generally 
not true if $\omega=0$; see equation 
(\ref{eq:proof:NoeterHerglotz2i}).}
\end{proof}

The last three lemmas now constitute a proof for 
Theorem\,\ref{thm:NoetherHerglotz1}. Indeed, using 
(\ref{eq:LieAccZero}) in (\ref{eq:LieOmega1}) together with 
(\ref{eq:OmegaLieZero}) shows $da^\flat=0$, which,
according to Lemma\,\ref{lemma:NoetherHerglotz1-2}, implies 
that the motion is Killing. 

Next we turn to the second part of the theorem of Noether 
and Herglotz, which reads as follows:   
\begin{theorem}[Noether \& Herglotz, part\,2]
\label{thm:NoetherHerglotzP2}
All irrotational rigid motions in Minkowski space are given by 
the following construction: take a twice continuously differentiable 
curve $\tau\mapsto z(\tau)$ in Minkowski space, where w.l.o.g 
$\tau$ is the eigentime, so that ${\dot z}^2=c^2$. Let 
$H_\tau:=z(\tau)+({\dot z(\tau)})^\perp$ be the hyperplane through 
$z(\tau)$ intersecting the curve $z$  perpendicularly. Let $\Omega$ 
be a the tubular neighborhood of $z$ in which no two hyperplanes 
$H_\tau,H_{\tau'}$ intersect for any pair $z(\tau),z(\tau')$ of 
points on the curve. In $\Omega$ define $u$ as the unique (once 
differentiable) normalized timelike vector field perpendicular to 
all $H_\tau\cap\Omega$. 
The flow of $u$ is the sought-for rigid motion.    
\end{theorem}
\begin{proof}
We first show that the flow so defined is indeed rigid, even though 
this is more or less obvious from its very definition, since we just 
defined it by `rigidly' moving a hyperplane through spacetime. 
In any case, analytically we have, 
\begin{equation}
\label{eq:proof:NoeterHerglotz2a}
H_\tau=\{x\in\mathbb{M}^n\mid f(\tau,x):=
{\dot z}(\tau)\cdot\bigl(x-z(\tau)\bigr)=0\}\,.
\end{equation}
In $\Omega$ any $x$ lies on exactly one such hyperplane, $H_\tau$,
which means that there is a function $\sigma:\Omega\rightarrow\reals$ 
so that $\tau=\sigma(x)$ and hence $F(x):=f(\sigma(x),x)\equiv 0$. 
This implies $dF=0$. Using the expression for $f$ from 
(\ref{eq:proof:NoeterHerglotz2a}) this is equivalent to 
\begin{equation}
\label{eq:proof:NoeterHerglotz2b}
d\sigma=\dot z^\flat\circ \sigma/[c^2-({\ddot z}\circ\sigma)
\cdot(\text{id}-z\circ\sigma)]\,,
\end{equation}
where `$\text{id}$' denotes the `identity vector-field', 
$x\mapsto x^\mu\partial_\mu$, in Minkowski space. Note that in 
$\Omega$ we certainly have $\partial_\tau f(\tau,x)\ne 0$ and hence 
$\ddot z\cdot(x-z)\ne c^2$. In $\Omega$ we now define the normalized 
timelike vector field\footnote{Note that, by definition of 
$\sigma$, $(\dot z\circ\sigma)\cdot(\text{id}-z\circ\sigma)\equiv 0$.} 
\begin{equation}
\label{eq:proof:NoeterHerglotz2c}
u:=\dot z\circ\sigma\,.
\end{equation}
Using (\ref{eq:proof:NoeterHerglotz2b}), its derivative is given by  
\begin{equation}
\label{eq:proof:NoeterHerglotz2d}
\nabla u^\flat=
d\sigma\otimes({\ddot z}^\flat\circ\sigma)=
\bigl[({\dot z}^\flat\circ\sigma)\otimes({\ddot z}^\flat\circ\sigma)
\bigr]/(N^2c^2)\,,
\end{equation}
where
\begin{equation}
\label{eq:proof:NoeterHerglotz2e}
N:=1-(\ddot z\circ\sigma)\cdot(\text{id}-z\circ\sigma)/c^2\,.
\end{equation}
This immediately shows that $\Proj_h\nabla u^\flat=0$ (since 
$\Proj_h{\dot z}^\flat=0$) and therefore that $\theta=\omega=0$. 
Hence $u$, as defined in (\ref{eq:proof:NoeterHerglotz2c}), 
generates an irrotational rigid motion.

For the converse we need to prove that any irrotational rigid 
motion is obtained by such a construction. So suppose $u$ is a
normalized timelike vector field such that $\theta=\omega=0$. 
Vanishing $\omega$ means $\Proj_h(\nabla\wedge u^\flat)=\Proj_h(d u^\flat)=0$. 
This is equivalent to $u^\flat\wedge du^\flat=0$, which according 
to the Frobenius theorem in differential geometry is equivalent 
to the integrability of the distribution\footnote{`Distribution'
is here used in the differential-geometric sense, where for a 
manifold $M$ it denotes an assignment of a linear subspace 
$V_p$ in the tangent space $T_pM$ to each point $p$ of $M$. 
The distribution $u^\flat=0$ is defined by  
$V_p=\{v\in T_pM\mid u^\flat_p(v)=u_p\cdot v=0\}$. A distribution 
is called (locally) integrable if (in the neighborhood of each point)
there is a submanifold $M'$ of $M$ whose tangent space at any 
$p\in M'$ is just $V_p$.} 
$u^\flat=0$, i.e. the hypersurface orthogonality of $u$. We wish 
to show that the hypersurfaces orthogonal to $u$ are hyperplanes. 
To this end consider a spacelike curve $z(s)$, where $s$ is the 
proper length, running within one hypersurface perpendicular to $u$. 
The component of its second $s$-derivative parallel to the 
hypersurface is given by (to save notation we now simply write $u$ and 
$u^\flat$ instead of $u\circ z$ and $u^\flat\circ z$)  
\begin{equation}
\label{eq:proof:NoeterHerglotz2f}
\Proj_h\ddot z
=\ddot z-c^{-2}u\, u^\flat(\ddot z)  
=\ddot z+c^{-2}u\,\theta(\dot z,\dot z) =\ddot z\,,
\end{equation}
where we made a partial differentiation in the second step
and then used $\theta=0$. Geodesics in the hypersurface are 
curves whose second derivative with respect to proper length 
have vanishing components parallel to the hypersurface. Now, 
(\ref{eq:proof:NoeterHerglotz2f}) implies that geodesics in 
the hypersurface are geodesics in Minkowski space (the 
hypersurface is `totally geodesic'), i.e. given by straight 
lines. Hence the hypersurfaces are hyperplanes.
\end{proof}

Theorem\,\ref{thm:NoetherHerglotzP2} precisely corresponds to 
the Newtonian counterpart: The irrotational motion of a rigid 
body is determined by the worldline of any of its points, and 
any timelike worldline determines such a motion. We can rigidly 
put an extended body into any state of translational  motion, as 
long as the size of the body is limited by $c^2/\alpha$, where 
$\alpha$ is the modulus of its acceleration. This also shows 
that (\ref{eq:LieAccZero}) is generally not valid for 
irrotational rigid motions. In fact, the acceleration one-form
field for (\ref{eq:proof:NoeterHerglotz2c}) is 
\begin{equation}
\label{eq:proof:NoeterHerglotz2g}
a^\flat=(\ddot z^\flat\circ\sigma)/N
\end{equation}
from which one easily computes
\begin{equation}
\label{eq:proof:NoeterHerglotz2h}
da^\flat=(\dot z^\flat\circ\sigma)\wedge\left\{
(\Proj_h\dddot z^\flat\circ\sigma)+(\ddot z^\flat\circ\sigma)\frac{(\Proj_h\dddot z\circ\sigma)\cdot(\text{id}-z\circ\sigma)}{Nc^2}\right\}N^{-2}c^{-2}\,.
\end{equation}
From this one sees, for example, that for 
\emph{constant acceleration}, defined by $\Proj_h\dddot z=0$ (constant 
acceleration in time as measured in the instantaneous rest frame), 
we have $da^\flat=0$ and hence a Killing motion. Clearly, this is just 
the motion (\ref{eq:BoostKill2}) for the boost Killing field 
(\ref{eq:BoostKill1}). The Lie derivative of $a^\flat$ is now 
easily obtained: 
\begin{equation}
\label{eq:proof:NoeterHerglotz2i}
L_ua^\flat=i_uda^\flat
=(\Proj_h\dddot z^\flat\circ\sigma)N^{-2}\,,
\end{equation}
showing explicitly that it is not zero except for motions
of constant acceleration, which were just seen to be Killing 
motions. 

In contrast to the irrotational case just discussed, we have seen 
that we cannot put a body rigidly into rotational motion. 
In the old days this was sometimes expressed by saying that the rigid 
body in SR has only three instead of six degrees of freedom. 
This was clearly thought to be paradoxical as long as one assumed 
that the notion of a perfectly rigid body should also make sense in 
the framework of SR. However, this hope was soon realized to be 
physically untenable~\cite{Laue:1911}.

\subsection{Geometry of space and time in rotating reference frames}
\label{sec:NonInertialSync}
We have seen above that there is a generalization of
Einstein simultaneity for the case of rigid linear accelerations.
The hypersurfaces of simultaneity were given by the hyperplanes 
of constant Killing time $\lambda$, which are different from the 
(curved) hypersurfaces of constant proper time $\tau$. 
This worked because the Killing field was (locally) hypersurface 
orthogonal.

Note that in terms of the co-rotating coordinates $(ct,z,\rho,\psi)$
(recall that $\psi=\varphi-\kappa t$) the Killing field 
(\ref{eq:RotKill1}) is just $K=\partial_t$. It is convenient to 
rewrite the spacetime metric $g=c^{-2}u^\flat\otimes u^\flat-h$ 
in the following form  
\begin{equation}
\label{eq:MetricKKform}
g=c^2\,\exp(2\Phi/c^2)\,\,A\otimes A-h\,,
\end{equation}
where $h$ is given by (\ref{eq:RotSplitMetric2}) and, using 
(\ref{eq:RotSplitMetric1}), we have the following expressions
for $\Phi$ and $A$:
\begin{subequations}
\label{eq:KKfields}
\begin{alignat}{3}
\label{eq:KKfields1}
& \Phi&&\,:=\,
c^2\,\ln\left\{\sqrt{K^2/c^2}\right\}&&\,=\,
\tfrac{c^2}{2}\ln\left\{1-(\kappa\rho/c)^2\right\}\,,\\
\label{eq:KKfields2}
& A&&\,:=\qquad
K^\flat/K^2 &&\,=\,
dt-\frac{\kappa\rho^2/c^2}{1-(\kappa\rho/c)^2}\,d\psi\,.
\end{alignat}
\end{subequations}
The physical interpretation of $\Phi$ appears from calculating the 
acceleration $a$ that an observer experiences who moves along the 
Killing orbit: 
\begin{equation}
\label{eq:PhiInt}
a^\flat:=\nabla_uu^\flat= -d\Phi\,.
\end{equation}
Hence $\Phi$ is the Newtonian potential that is accelerating the 
Killing observer.  

The rotational Killing field is clearly not hypersurface orthogonal. 
The obstruction is just given by the vorticity $\omega$. A simple 
calculation gives 
\begin{equation}
\label{eq:LineCurvature}
F:=dA=2c^{-2}\exp(-\Phi/c^2)\,\omega\,. 
\end{equation}
Hence the obstruction for hypersurface orthogonality is likewise 
faithfully measured by $A$. Moreover, as we shall see below, the 
1-form $A$ has an interesting physical and geometric interpretation, 
which is the actual reason why we introduced it here. 

Inside the region $\Omega$ (defined in (\ref{eq:RegionOmega})) $K$ is 
a complete and nowhere vanishing timelike vector field. This means that 
the flow $f:\reals\times\Omega\rightarrow\Omega$ of $K$ 
defines a free action of the additive group $\reals$ on $\Omega$
that makes $\Omega$ the total space of a principle bundle with 
fiber $\reals$ and base $\hat\Omega=\Omega/\!\!\sim$. Here $\sim$ is
again the equivalence relation which declares two points in 
$\Omega$ to be equivalent iff they lie on the same $K$ orbit. 
Hence $\hat\Omega$, which is obviously diffeomorphic to the solid 
cylinder $\{(z,\rho,\varphi)\mid \rho<c/\kappa\}$, is the space of 
$K$ orbits. Since $\Omega$ is endowed with the metric $g$ and since 
$K$ acts by isometries, the distribution of hyperplanes 
(\ref{eq:LocalRestSpace}) orthogonal to the Killing orbits define 
a connection on the principal bundle whose corresponding 1-form 
is just $A$.\footnote{The connection 1-form $A$ associated to the 
distribution of `horizontal' subspaces has to  fulfill two 
conditions: 1)~vectors tangential to the horizontal subspaces are 
annihilated by $A$ and 2) $A(K)=1$, where $K$ is a `fundamental 
vector field' which generates the action of the structure group 
$\reals$. Both conditions are satisfied in our case.
See e.g. \cite{Bleecker:GaugeTheory} for a lucid discussion of 
these notions.} Accordingly, the bundle curvature is given by $F=dA$. 
Note that $F$ can be considered as 2-form on $\hat\Omega$ since 
$i_KF=0$ and $L_KF=i_KdF=0$.\footnote{More precisely, there is a 
unique 2-form $\hat F$ on $\hat\Omega$ such that $\pi^*\hat F=F$, 
where $\pi:\Omega\rightarrow\hat\Omega$ is the bundle projection.}

Now, parallel transport defined by the connection $A$ has a direct 
physical interpretation: it is just transportation of time according 
to Einstein synchronization. Since $F\ne 0$ this transportation is not 
path independent. In particular this implies that synchronization along
fixed paths is not a transitive operation anymore.
Given two points in $\hat\Omega$ connected by a spatial path
$\hat\gamma$ in $\hat\Omega$, parallel transportation along 
$\hat\gamma$ requires that we lift $\hat\gamma$ to a path $\gamma$ 
in $\Omega$ whose tangent vectors are annihilated by $A$, that 
is, which runs orthogonally to the orbits of $K$. But this is 
just what we mean by saying that the points on the curve $\gamma$ 
are locally Einstein synchronized, in the sense that any two 
infinitesimally nearby points on $\gamma$ are Einstein synchronized. 
Hence the integral of $A$ along $\gamma$ vanishes. Using 
(\ref{eq:KKfields2}) this is equivalent to 
\begin{equation}
\label{eq:IntegralA1}
\Delta t:=\int_\gamma dt
=\frac{\kappa}{c^2}\int_{\hat\gamma}\frac{\rho^2}{1-(\kappa\rho/c)^2}\ 
d\psi\,,
\end{equation}
where we interpreted $\rho$ and $\psi$ as coordinates on $\hat\Omega$ 
so that the right hand side could be written as integral along the 
curve $\hat\gamma$ in $\hat\Omega$. This means that if we Einstein 
synchronize clocks along $\hat\gamma$ in space, the clock at the 
final point of $\hat\gamma$ shows a lapse $\Delta t$ of 
coordinate-time as compared to the clock at the initial point of 
$\hat\gamma$. A striking consequence of the non-transitivity of 
Einstein synchronization is the non-zero lapse of coordinate time 
that one obtains for spatially closed curves. These lapses are just 
the holonomies of the connection $A$. If, for simplicity, we choose 
$\hat\gamma$ to be a closed planar loop of constant $\rho$ and $z$, 
(\ref{eq:IntegralA1}) immediately leads to 
\begin{equation}
\label{eq:IntegralA2}
\Delta t
=\frac{2\pi}{\kappa}\frac{(\kappa\rho/c)^2}{1-(\kappa\rho/c)^2}
\approx (2\kappa/c^2)\mathcal{S}\,,
\end{equation}
where $\mathcal{S}$ is explained below. The lapse in 
proper time, $\Delta\tau$, is obtained by multiplying this result 
with $\exp(\Phi/c^2)$, which merely amounts to replacing the 
denominator $1-(\kappa\rho/c)^2$ in (\ref{eq:IntegralA2}) with 
its square root. This time lapse is directly related to the 
\emph{Sagnac effect}. In fact, the observed phase shift in the Sagnac 
effect is obtained by multiplying the expression for the time
lapse with twice\footnote{The factor 2 results 
simply from the fact that the Sagnac effect measures the \emph{sum} of 
the moduli of time lapses for a closed curve traversed in both 
directions.} the light's frequency~$\nu$. 

In (\ref{eq:IntegralA2}) $\mathcal{S}$ denotes the area of the 
2-disk spanned by the planar loop. Note that this area is only 
approximately given by $\pi\rho^2$ since the geometry in 
$\hat\Omega$, determined with co-rotating rods and clocks, is 
given by the metric $h$; see (\ref{eq:RotSplitMetric2}). The 
precise expressions for the circumference, $\mathcal{C}$ and 
area, $\mathcal{S}$, of the planar loop of constant $\rho$ 
follow from (\ref{eq:RotSplitMetric2}):
\begin{subequations}
\label{eq:AreaCircum}
\begin{alignat}{4}
\label{eq:AreaCircum1}
& \mathcal{C} 
&&\,=\,\int_0^{2\pi}\frac{d\psi\,\rho}{\sqrt{1-(\kappa\rho/c)^2}}
&&\,=\,\frac{2\pi\rho}{\sqrt{1-(\kappa\rho/c)^2}}
&&\ >\ 2\pi\rho\,,\\
\label{eq:AreaCircum2}
& \mathcal{S}
&&\,=\,\int_0^{2\pi}\hspace{-0.5ex}\int_0^\rho\frac{d\psi\,d\rho'\rho'}%
  {\sqrt{1-(\kappa\rho'/c)^2}}
&&\,=\,\frac{2\pi c^2}{\kappa^2}\left\{1-\sqrt{1-(\kappa\rho/c)^2}\right\}
&&\ >\ \pi\rho^2\,.
\end{alignat}
\end{subequations}
The circumference grows faster than $\propto\rho$ and the area 
faster than $\propto\rho^2$. Note that according to 
(\ref{eq:RotSplitMetric2}) $\rho$ is the geodesic radial distance. 
Hence the two-dimensional hypersurfaces of constant $z$ in 
$\hat\Omega$ are negatively curved. In fact, the Gaussian curvature, 
$\mathcal{K}$, of these hypersurfaces turns out to be 
\begin{equation}
\label{eq:GaussCurvature}
\mathcal{K}=
\frac{-\,3\,(\kappa/c)^2}{\bigl\{1-(\kappa\rho/c)^2\bigr\}^2}\,,
\end{equation}
which is strictly negative, approximately constant for 
$\rho\ll c/\kappa$, and unbounded as $\rho$ approaches the 
critical radius $c/\kappa$. In contrast, according 
to (\ref{eq:RotSplitMetric2}), the metrics induced by $h$ on the 
hypersurfaces of constant $\psi$ are flat. 
 
The bundle curvature of $F$ for the connection $A$ and the Riemannian 
curvature for $(\hat\Omega,h)$ are indeed intimately linked through 
identities which arise by calculating the Riemannian curvature of 
$(\Omega,g)$, where $g$ is parameterized as in 
(\ref{eq:MetricKKform}), 
and noting that $g$ is flat (Minkowski metric). 
One such identity is the so called \emph{Kaluza-Klein identity}, 
which expresses the scalar curvature (Ricci scalar) of $g$ in terms 
of the scalar curvature $R_h$ of $h$, $\Phi$, and 
$\Vert F\Vert^2_h=h^{ik}h^{jl}F_{ij}F_{kl}$. Since the scalar 
curvature of $g$ is zero, one obtains: 
\begin{equation}
\label{eq:KKidentity}
R_h=2\exp(-\Phi/c^2)\Delta_h\exp(\Phi/c^2)
-\tfrac{1}{4}\,c^2\exp(2\Phi/c^2)\,\Vert F\Vert^2_h\,,
\end{equation}
where $\Delta_h$ is the Laplace operator on $(\hat\Omega,h)$. 

It is an interesting historical fact that it was Kaluza who 
pointed out that `space' in rotating reference frames cannot 
be identified with a submanifold perpendicular to the Killing orbits
(because such a submanifold does not exist) but rather has to be
constructed as the quotient manifold $\hat\Omega$ which carries the 
curved metric $h$~\cite{Kaluza:1911}. He also discussed the 
non-integrability of Einstein synchronization. This he did in 1910, 
ten years before he applied the very same mathematical ideas 
to the five-dimensional setting known as Kaluza-Klein theories. 

\newpage
\appendix
\section{Appendices}

For the interested reader this appendix collects some mathematical 
background material related to various points discussed in the 
main text.  

\subsection{Sets and group actions}
\label{sec:GroupActions}
Given a set $S$, recall that an \emph{equivalence relation} is a 
subset $R\subset S\times S$ such that for all $p,q,r\in S$ the 
following conditions hold: 
1)~$(p,p)\in R$ (called `reflexivity'), 
2)~if $(p,q)\in R$ then $(q,p)\in R$ (called `symmetry'), and 
3)~if $(p,q)\in R$ and  $(q,r)\in R$ then $(p,r)\in R$ 
(called `transitivity'). Once $R$ is given, one often conveniently 
writes $p\sim q$ instead of $(p,q)\in R$. Given $p\in S$, its 
\emph{equivalence class}, $[p]\subseteq S$, is given by all 
points $R$-related to $p$, i.e. $[p]:=\{q\in S\mid (p,q)\in R\}$. 
One easily shows that equivalence classes are either identical or 
disjoint. Hence they form a \emph{partition} of $S$, that is, 
a covering by mutually disjoint subsets. Conversely, given a partition 
of a set $S$, it defines an equivalence relation by declaring two 
points as related iff they are members of the same cover set.  
Hence there is a bijective correspondence  between partitions of 
and equivalence relations on a set $S$. The set of equivalence 
classes is denoted by $S/R$ or $S/\!\!\sim$. There is a natural surjection 
$S\rightarrow S/R$, $p\mapsto [p]$.   

If in the definition of equivalence relation we exchange symmetry
for antisymmetry, i.e. $(p,q)\in R$ and $(q,p)\in R$ implies $p=q$, 
the relation is called a \emph{partial order}, usually written as 
$p\geq q$ for $(p,q)\in R$. If, instead, reflexivity is dropped 
and symmetry is replaced by asymmetry, i.e.  $(p,q)\in R$ 
implies $(q,p)\not\in R$, one obtains a relation called a 
\emph{strict partial order}, usually denoted by $p> q$ for 
$(p,q)\in R$.  
   
An \emph{left action} of a group $G$ on a set $S$ is a map
$\phi:G\times S\rightarrow S$, such that $\phi(e,s)=s$ 
($e=$ group identity) and $\phi(gh,s)=\phi(g,\phi(h,s))$. 
If instead of the latter equation we have $\phi(gh,s)=\phi(h,\phi(g,s))$
one speaks of a \emph{right action}. For left actions one sometimes 
conveniently writes $\phi(g,s)=:g\cdot s$, for right actions 
$\phi(g,s)=:s\cdot g$. An action is called \emph{transitive} if 
for every pair $(s,s')\in S\times S$ there is a $g\in G$ such that 
$\phi(g,s)=s'$, and \emph{simply transitive} if, in addition,  
$(s,s')$ determine $g$ uniquely, that is, $\phi(g,s)=\phi(g',s)$
for some $s$ implies $g=g'$. The action is called \emph{effective} 
if $\phi(g,s)=s$ for all $s$ implies $g=e$ (`every $g\ne e$ moves 
something') and \emph{free} if $\phi(g,s)=s$ for some $s$ implies 
$g=e$ (`no $g\ne e$ has a fixed point'). It is obvious that simple 
transitivity implies freeness and that, conversely, freeness and 
transitivity implies simple transitivity. Moreover, for Abelian groups, 
effectivity and transitivity suffice to imply simple transitivity. 
Indeed, suppose $g\cdot s=g'\cdot s$ holds for some $s\in S$, then 
we also have  $k\cdot(g\cdot s)=k\cdot(g'\cdot s)$ for all $k\in G$
and hence $g\cdot(k\cdot s)=g'\cdot(k\cdot s)$ by commutativity.
This implies that $g\cdot s=g'\cdot s$ holds, in fact, for all $s$. 

For any $s\in S$ we can consider the \emph{stabilizer subgroup}
\begin{equation}
\label{eq:def:Stab}
\Stab(s):=\{g\in G\mid\phi(g,s)=s\}\subseteq G\,.
\end{equation} 
If $\phi$ is transitive, any two stabilizer subgroups are conjugate: 
$\Stab(g\cdot s)=g\Stab(s)g^{-1}$. By definition, if $\phi$ is free 
all stabilizer subgroups are trivial (consist of the identity element 
only). In general, the intersection 
$G':=\bigcap_{s\in S}\Stab(s)\subseteq G$ is the normal subgroup of 
elements acting trivially on $S$. If $\phi$ is an action of $G$ on 
$S$, then there is an effective action $\hat\phi$ of $\hat G:= G/G'$ 
on $S$, defined by $\hat\phi([g],s):=\phi(g,s)$, where $[g]$ denotes 
the $G'$-coset of $G'$ in $G$.  

The \emph{orbit} of $s$ in $S$ under the action $\phi$ of $G$ 
is the subset
\begin{equation}
\label{eq:def:Orb}
\Orb(s):=\{\phi(g,s)\mid g\in G\}\subseteq S\,.
\end{equation} 
It is easy to see that group orbits are either disjoint or 
identical. Hence they define a partition of $S$, that is, 
an equivalence relation.   

A relation $R$ on $S$ is said to be invariant under the self map 
$f:S\rightarrow S$ if $(p,q)\in R\Leftrightarrow (f(p),f(q))\in R$.
It is said to be invariant under the action $\phi$ of $G$ on $S$ if 
$(p,q)\in R\Leftrightarrow (\phi(g,p),\phi(g,q))\in R$ for all 
$g\in G$. If $R$ is such a $G$-invariant equivalence relation, there is 
an action $\phi'$ of $G$ on the set $S/R$ of equivalence classes, 
defined by $\phi'(g,[p]):=[\phi(g,p)]$. A general theorem states that 
invariant equivalence relations exist for transitive group actions, 
iff the stabilizer subgroups (which in the transitive case are all 
conjugate) are maximal (e.g. Theorem\,1.12 in \cite{Jacobson:BasicAlgebraI}).

\subsection{Structures on vector and affine spaces }
\label{sec:VectorSpaces}
\subsection{Non degenerate bilinear forms}
\label{sec:NonDegForms}
Consider a vector space $V$ of dimension $n$ over $\field$ (here 
denoting $\reals$ or $\complex$). Let it be endowed with a 
non-degenerate bilinear 
form $\omega:V\times V\rightarrow\field$. No assumptions regarding 
symmetries of $\omega$ are made at this point. The dual space of 
$V$ is denoted by $V^*$ whose elements we will denote by Greek 
letters. The set  
of linear maps $V\rightarrow V$ is denoted by $\End(V)$, called 
the endomorphisms of $V$, which forms an associative algebra over 
$\field$ (algebra multiplication being composition of maps). 
The set of invertible elements in $\End(V)$ (i.e. isomorphisms 
of $V$) will be denoted by $\group{GL}(V)$; it forms a group under 
composition. Generally, composition of maps will be denoted by 
$\circ$. 

The form $\omega$ defines an isomorphism 
\begin{equation}
\label{eq:DefOmegadown}
\omegadown:V\rightarrow V^*\,,\qquad 
\omegadown(v):=\omega(v,\cdot)\,,
\end{equation}
with inverse map being denoted by 
\begin{equation}
\label{eq:DefOmegaup} 
\omegaup:V^*\rightarrow V\,,\qquad 
\omegaup:=(\omegadown)^{-1}\,,
\end{equation}
so that 
\begin{equation}
\label{eq:OmegadownOmegaup} 
\omegaup\circ\omegadown=\mathrm{id}_{V}\qquad\text{and}\qquad
\omegadown\circ\omegaup=\mathrm{id}_{V^*}\,.
\end{equation} 

Recall that `transposition' is a map $\End(V)\rightarrow\End(V^*)$,
denoted by $A\mapsto A^\top$ and defined through 
$A^\top(\alpha):=\alpha\circ A$. This map is an anti-isomorphism of 
algebras (`anti', since it obeys $(A\circ B)^\top=B^\top\circ A^\top$). 
Different from this canonically defined notion of transposition is the 
`$\omega$-transposition', which is an isomorphism 
$\End(V)\rightarrow\End(V)$, which we denote by $A^t$ 
(the dependence on $\omega$ being implicitly understood)
and which is defined through 
\begin{equation}
\label{eq:DefOmegaTrans}
\omega(A^tu,v)=\omega(u,Av)\quad\forall u,v\in V\,.
\end{equation}
Note that the $\omega$-transposed is in $\End(V)$ whereas the 
canonical transposed is in  $\End(V^*)$. The relations between the 
two are 
\begin{equation}
\label{eq:RelTransOmegaTrans}
A^t=\omegaup\circ A^\top\circ\omegadown\quad\text{and}\quad
A^\top=\omegadown\circ A^t\circ\omegaup\,.  
\end{equation} 

\subsection{Generalized orthogonal transformations}
\label{sec:IsometriesNonDegForms}
A generalized orthogonal transformation of $(V,\omega)$ is 
any bijective map $\phi:V\rightarrow V$ such that 
$\omega\bigl(\phi(u),\phi(v)\bigr)=\omega(u,v)$ for all $u,v\in V$. 
In this subsection we shall restrict to symmetric $\omega$.  
Note that any symmetric bilinear form $\omega$ is 
uniquely determined by its quadratic form, i.e. the function 
$\hat\omega:V\rightarrow\field$, $v\mapsto\hat\omega(v):=\omega(v,v)$, 
for we have $\omega(u,v)=\frac{1}{2}\bigl(\hat\omega(u+v)-
\hat\omega(u)-\hat\omega(v)\bigr)$. It is sometimes useful to 
consider generalizations of distance measures by setting 
$d(u,v):=\sqrt{\vert\hat\omega(u-v)\vert}$. This is e.g. done in 
SR, where one speaks of timelike and spacelike distances in that 
sense. Now suppose $\varphi$ is an isometry with respect to $d$, 
i.e. $d(\varphi(u),\varphi(v))=d(u,v)$ for all $u,v$. Consider 
$\phi$ defined by $\phi(u):=\varphi(u)-\varphi(0)$. Clearly 
$\varphi$ is an isometry of $d$ if $\phi$ is a generalized 
orthogonal transformation with respect to $\omega$. 
Now, generalized orthogonal transformations are necessarily linear:
\begin{proposition}
\label{prop:IsometriesV}
Let $\omega$ be a non-degenerate symmetric bilinear form on $V$ and 
let $\phi:V\rightarrow V$ be a generalized orthogonal transformation 
with respect to $\omega$. Then $\phi$ is linear.  
\end{proposition}
\begin{proof}
Consider $I:=\omega\bigl(a\phi(u)+b\phi(v)-\phi(au+bv),w\bigr)$; 
surjectivity\footnote{Note that we only use surjectivity here, so 
that the hypotheses for this result may be slightly reduced.}
allows to write $w=\phi(z)$, so that 
$I=a\omega(u,z)+b\omega(v,z)-\omega(au+bv,z)=0$ for all $z\in V$. 
Hence the aforementioned expression for $I$ is zero for all $w\in V$, 
which by non-degeneracy of $\omega$ implies the linearity of $\phi$.
\end{proof}

Particularly simple generalized orthogonal transformations are given by 
reflections on non-degenerate hyperplanes. To explain this, let 
$v\in V$ and $v^\perp:=\{w\in V\mid\omega(v,w)=0\}\subset V$. 
$v^\perp$ is a linear subspace of co-dimension one, that is, a 
hyperplane. That it be non-degenerate means that 
$\omega\vert_{v^\perp}$ is non-degenerated, which is easily 
seen to be the case iff $\omega(v,v)\ne 0$. The reflection at the 
non-degenerate hyperplane $v^\perp$ is the map
\begin{equation}
\label{eq:Reflection1}
\rho_v(x):=x-2\,v\ \frac{v\cdot x}{v^2}\,.
\end{equation}
where for convenience we wrote $u\cdot v:=\omega(u,v)$
and $v^2:=v\cdot v$. $\rho_v$ is easily seen to be an involutive 
(i.e. $\rho_v\circ\rho_v=\text{id}_V$) generalized orthogonal 
transformation. If $\phi$ is any other generalized orthogonal 
transformation, the following equivariance property holds: 
$\phi\circ\rho_v\circ\phi^{-1}=\rho_{\phi(v)}$. An important 
result is now given by 
\begin{theorem}[Cartan, Dieudonn\'e]
Let the dimension of $V$ be $n$. Any generalized orthogonal transformation 
of $(V,\omega)$ is the composition of at most $n$ reflections. 
\end{theorem}   
\begin{proof}
Comprehensive proofs may be found in \cite{Jacobson:BasicAlgebraI}
or \cite{Berger:Geometry2}. Here we offer a proof of the weaker 
result, that any generalized orthogonal transformation is the 
composition of at most $2n-1$ reflections. So let $\phi$ be 
generalized orthogonal and $v\in V$
so that $v^2\ne 0$ (which certainly exists). Let $w=\phi(v)$, 
then $(v+w)^2+(v-w)^2=4v^2\ne 0$ so that $w+v$ and $w-v$ 
cannot simultaneously have zero squares. So let $(v\mp w)^2\ne 0$
(understood as alternatives), then $\rho_{v\mp w}(v)=\pm w$ and 
$\rho_{v\mp w}(w)=\pm v$. Hence $v$ is eigenvector with eigenvalue 
$1$ of the generalized orthogonal transformation given by   
\begin{equation}
\label{eq:proof:CartanDieu1}  
\phi'=
\begin{cases}
\rho_{v-w}\circ \phi &\text{if}\ (v-w)^2\ne 0\,,\\
\rho_v\circ\rho_{v+w}\circ \phi &\text{if}\ (v-w)^2= 0\,.
\end{cases}
\end{equation}
Consider now the generalized orthogonal transformation 
$\phi'\big\vert_{v^\perp}$
on $v^\perp$ with induced bilinear form $\omega\big\vert_{v^\perp}$,
which is non-degenerated due to $v^2\ne 0$. We now conclude by 
induction: At each dimension we need at most two reflections to 
reduce the problem by one dimension. After $n-1$ steps we have 
reduced the problem to one dimension, where we need at most 
one more reflection. Hence we need at most $2(n-1)+1=2n-1$ 
reflections which upon composition with $\phi$ produce the identity. 
Here we use that any generalized orthogonal transformation in 
$v^\perp$ can be canonically extended to $\Span\{v\}\oplus v^\perp$ by just letting 
it act trivially on $\Span\{v\}$.
\end{proof}
There are several useful applications of this result, most notably 
in the construction of the Spin groups. Other applications in SR 
are discussed in \cite{Urbantke:2002}.

\subsection{Index raising and lowering}
\label{sec:IndexMoving}
Let $\{e_a\}_{a=1,\cdots,n}$ be a basis of $V$ and  
$\{\eta^a\}_{a=1,\cdots,n}$ its \emph{(canonical) dual basis} 
of $V^*$, which is defined by $\eta^a(e_b)=\delta^a_b$.  
Using $\omegadown$ and $\omegaup$ one can define the $\omega$-duals  
of $\{e_a\}$ and $\{\eta^a\}$ respectively, given by  
\begin{subequations}
\label{eq:OmegaDualbasis12}
\begin{alignat}{4}
\label{eq:OmegaDualbasis1}
& \eta_a &&\,:=\,\omegadown(e_a) &&\in V^* &&\,,\\
\label{eq:OmegaDualbasis2}    
& e^a &&\,:=\,\omegaup(\eta^a)   && \in V &&\,,
\end{alignat}
\end{subequations}
so that, writing $\omega_{ab}:=(e_a,e_b)$ and $\omega^{ab}$ for the 
components of the inverse-transposed matrix (i.e. 
$\omega_{ac}\omega^{bc}=\omega_{ca}\omega^{cb}=\delta_a^b$), 
\begin{subequations}
\label{eq:OmegadownupBasis}
\begin{alignat}{3}
\label{eq:OmegadownBasis}
& \eta_a &&\,:=\,\omegadown(e_a) &&\,=\,\omega_{ab}\eta^b\,,\\
\label{eq:OmegaupBasis}
& e^a &&\,:=\,\omegaup(\eta^a)   &&\,=\,\omega^{ba}e_b\,.
\end{alignat}
\end{subequations}
Using components with respect to the canonical dual bases,
so that $v=v^ae_a\in V$ with $\omegadown(v)=:v_a\eta^a$ 
and $\alpha=\alpha_a\eta^a\in V^*$ with $\omegaup(\alpha)=:\alpha^ae_a$, 
one obtains the equivalent to (\ref{eq:OmegadownupBasis}) in coordinates: 
\begin{subequations}
\label{eq:OmegadownupCoord}
\begin{alignat}{2}
\label{eq:OmegadownCoord}
& v_a &&\,=\,v^b\omega_{ba}\,,\\
\label{eq:OmegaupCoord}
& \alpha^a &&\,=\,\omega^{ab}\alpha_b\,.
\end{alignat}
\end{subequations}
It should be clear from (\ref{eq:OmegadownupBasis}) and 
(\ref{eq:OmegadownupCoord}) why the maps $\omegadown$ and $\omegaup$ 
are called `index lowering' and `index raising'. Often, if there is 
no ambiguity as to what structure $\omega$ is used, the following 
notation is employed: Let $v\in V$ and $\alpha\in V^*$, then 
$\omegadown(v)=:v^\flat\in V^*$ and $\omegaup(\alpha)=:\alpha^\sharp\in V$.

Finally we remark on the choice of conventions. 
Comparing e.g. (\ref{eq:OmegadownCoord}) with (\ref{eq:OmegaupCoord})
one notices that one sums over the first index on $\omega$ for lowering
and over the second index for raising indices (on the bases 
(\ref{eq:OmegadownupBasis}) it is just the other way round). 
This is a consequence of the following requirements: 1)~raising and 
lowering of indices are mutually inverse operations, and 2)~the matrix, 
$\{\omega^{ab}\}$, used for raising indices is the 
\emph{transposed} inverse of $\{\omega_{ab}\}$. The rationale 
for the second condition is the requirement that lowering both 
indices on $\{\omega^{ab}\}$ using $\{\omega_{ab}\}$ should 
reproduce $\{\omega_{ab}\}$ and raising both indices on 
$\{\omega_{ab}\}$ using $\{\omega^{ab}\}$ should reproduce 
$\{\omega^{ab}\}$. This enforces~2). 

Note again that so far no assumptions were made concerning 
the symmetries of $\omega$. In the general case there are, 
in fact, two raising-lowering operations: One as given 
above, the other by replacing (\ref{eq:DefOmegadown}) with 
${\tilde{\omega}}^{\downarrow}(v):=\omega(\cdot,v)$, i.e.  
$v$ now being in the second rather than the first slot. 
For this second operation we have all formulae as above with 
 $\{\omega_{ab}\}$ and $\{\omega^{ab}\}$ being replaced by 
the transposed matrices. In physical applications $\omega$
is either symmetric---like in case of the Minkowski metric---or 
antisymmetric---like for the 2-spinor metric (symplectic form). 
In those cases there is---up to sign in the second case---a 
unique pair of lowering and raising operations. 

\subsection{Linear frames}
\label{sec:LinearFrames}  
A basis  $f=\{e_a\}_{a=1,\cdots,n}$ of $V$ can be viewed as 
a linear isomorphism (also denoted by $f$), $f:\field^n\rightarrow V$,
given by $f(v^1,\cdots,v^n)=v^ae_a$. With this interpretation we 
call the basis $f$ a \emph{linear frame}. Any frame $f$ induces 
an isomorphism of algebras $\End(\field^n)\rightarrow\End(V)$, given 
by $A\mapsto A^f:=f\circ A\circ f^{-1}$. If $A=\{A^b_a\}$, then 
$A^f(e_a)=A^b_ae_b$. The standard (linear) action $\phi$ of 
$\group{GL}(\field^n)$ on $\field^n$, $\phi(A,x):=Ax$, 
thereby translates in an $f$-dependent way to an action $\phi^f$ of 
$\group{GL}(\field^n)$ on $V$, defined by 
$\phi^f(A,v):=f\circ\phi(A,f^{-1}(v))$; that is, 
$(A,v)\mapsto A^fv=f(Ax)$, where $f(x)=v$.

Let $\Frames{V}$ denote the set of frames for $V$. 
The \emph{general linear group} $\group{GL}(\field^n)$ acts 
transitively and freely on $\Frames{V}$ from the right:
\begin{equation}
\label{eq:GLactionBases}
\group{GL}(V)\times\Frames{V}\rightarrow\Frames{V}\,,\qquad
(A,f)\mapsto f\cdot A:=f\circ A\,.
\end{equation}
Proper subgroups of $\group{GL}(\field^n)$ continue to act freely on $\Frames{V}$. 

\subsection{Affine spaces}
\label{sec:AffineSpaces}
An affine space over the vector space $V$ is a set $\Aff(V)$ together 
with an effective and transitive action $\phi$ of $V$, considered as 
Abelian group (group multiplication being vector addition). Since the group 
is Abelian, this suffices to imply that the action is free and simply 
transitive. One writes $\phi(m,v)=:m+v$, which defines what is meant 
by `$+$' between an element of $\Aff(V)$ and an element of $V$. 
Any ordered pair of points $(p,q)\in \Aff(V)\times\Aff(V)$ uniquely 
defines a vector $v$, namely that for which $p=q+v$. One writes 
$p-q=v$, defining what is meant by `$-$' between two elements of 
$\Aff(V)$. Considered as Abelian groups, any linear subspace 
$W\subset V$ defines a subgroup. The orbit of that subgroup in 
$\Aff(V)$ through $m\in\Aff(V)$ is an affine subspace, denoted by 
$W_m$, i.e.
\begin{equation}
\label{eq:def:AffPlane}
W_m=m+W:=\{m+w\mid w\in W\}\,,
\end{equation}
which is an affine space over $W$ in its own right of dimension 
$\text{dim}(W)$. 
One-dimensional affine subspaces are called \emph{(straight) lines}, 
two-dimensional ones \emph{planes}, and those of co-dimension one are 
called \emph{hyperplanes}. 

\subsection{Affine frames}
\label{sec:AffineFrames}
A \emph{basis} for $\Aff(V)$ is a tuple $F:=(m,f)$, where $m$ is a 
point in $\Aff(V)$ and $f$ a basis of $V$. $F$ can be considered
as a map $\field^n\rightarrow\Aff(V)$, given by $F(x):=f(x)+m$
(here $f$ is interpreted as linear frame). With this interpretation 
$F$ is called an \emph{affine frame}. We denote the set of affine 
frames by $\Frames{\Aff(V)}$. 

The \emph{general affine group} of $\field^n$ is given in the 
familiar fashion by the semi-direct product 
$\field^n\rtimes\group{GL}(\field^n)$, which acts on $\field^n$ in the 
standard way: $\phi:((a,A),x)\mapsto \phi((a,A),x):=A(x)+a$. 
Its multiplication law is given by: 
\begin{equation}
\label{eq:GeneralAffineGroup}
(a_1,A_1)(a_2,A_2)=(a_1+A_1a_2\,,\,A_1A_2)\,.
\end{equation}
Depending on the choice of a frame $F\in\Frames{\Aff(V)}$ the action 
$\phi$ of $\field^n\rtimes\group{GL}(\field^n)$ on $\field^n$ translates to 
an action $\phi^F$ of $\field^n\rtimes\group{GL}(\field^n)$ on $\Aff(V)$
as follows: $\phi^F((a,A),p):=F\circ\phi((a,A),F^{-1}(p))$;
in other words, if $F=(m,f)$ and $F(x)=p$, we have 
$\phi^F:((a,A),p)\mapsto F(Ax+a)=A^f(p-m)+m+f(a)$.    

$\field^n\rtimes\group{GL}(\field^n)$ has an obvious right action on 
$\Frames{\Aff(V)}$, given by $(g,F)\mapsto F\cdot g:=F\circ g$. 
Explicitly, for $g=(a,A)$ and $F=(m,f)$, this reads  
\begin{equation}
\label{eq:AffActionFrames}
F\cdot g=(m,f)\cdot (a,A)=(m+f(a),f\circ A)\,.
\end{equation}
%


\subsection{Lie algebras for matrix groups}
\label{sec:LieAlgebras}
\subsubsection{General considerations}
\label{sec:LieAlgebrasGenCons}
We first recall the definition of a Lie algebra: 
\begin{definition}
A \emph{Lie algebra} over $\field$ (here denoting $\reals$ or 
$\complex$) is a vector space $L$ over $\field$ endowed with 
a map (called the `Lie bracket') $L\times L\rightarrow L$,  
$(X,Y)\mapsto [X,Y]$, which for all $X,Y,Z\in L$ obeys:
\begin{subequations}
\label{eq:DefLieAlg}
\begin{alignat}{2}
\label{eq:DefLieAlg1}
&[X,Y]=-[Y,X] 
&&\qquad\text{(anti-symmetry)\,,}\\
\label{eq:DefLieAlg2}
&[aX+Y,Z]=a[X,Z]+[Y,Z]
&&\qquad\text{(linearity)\,,}\\
\label{eq:DefLieAlg3}
&[X,[Y,Z]]+[Y,[Z,X]]+[Z,[X,Y]]=0
&&\qquad\text{(Jacobi identity)\,.}
\end{alignat}
\end{subequations}
\end{definition}
A Lie \emph{subalgebra} $L'\subseteq L$ is a linear subspace which 
becomes a Lie algebra when the bracket is restricted to $L'$, i.e. 
if $[L',L']\subseteq L'$. A Lie subalgebra is called an \emph{ideal}
if the stronger condition holds that $[L',L]\subseteq L'$. It is easy 
to see that if $L'$ is an ideal the quotient $L/L'$ is again a Lie 
algebra: just define the bracket of two cosets as the coset of the 
bracket of two arbitrary representatives, which is well defined.   

In may cases of interest $L$ is already given as an associative 
algebra and the Lie bracket is then defined as commutator: 
$[X,Y]:=X\cdot Y-Y\cdot X$. This is e.g. the case if 
$L\subseteq\End(V)$ since, as already mentioned, the endomorphisms 
of a vector space $V$ form an associative algebra if the 
multiplication is taken to be the composition of maps.

Given a matrix group $\group{G}\subseteq\group{GL(n,\field)}$ we 
consider the set $C_*^1(\reals,\group{G})$ of all continuously 
differentiable curves $A:\reals\rightarrow\group{G}$ such that 
$A(0)=\mathbf{1}_n$ (unit $n\times n$-matrix). We define 
$\dot A:=\frac{d}{ds}A(s)\vert_{s=0}$, the `velocity' of the 
curve $A(s)$ at the group identity. We consider the 
set of all such velocities:  
\begin{equation}
\label{eq:DefLieAlgebra}
\Lie(\group{G}):=\bigl\{\dot A\,\mid\, A\in C^1_*(\reals,\group{G})
\bigr\}\subset \End(\reals^n)\,.
\end{equation}    
\begin{proposition}
$\Lie(\group{G})$ is a real Lie algebra.
\end{proposition}
\begin{proof}
First we prove that $\Lie(\group{G})$ is a linear space:
Let $X,Y\in \Lie(\group{G})$ and $A,B\in C^1_*(\reals,\group{G }))$ 
such that $X=\dot A$ and $Y=\dot B$. Define 
$C\in C^1_*(\reals,\group{G })$ by $C(s):=A(s)\cdot B(ks)$, where 
$k\in\reals$, then $\dot C=X+kY$, showing that $\Lie(\group{G})$
is a vector space over $\reals$. Here and below `${\,\cdot\,}$' denotes 
matrix multiplication. Now, since $\Lie(\group{G})\subseteq\End(\field^n)$, 
i.e. lies in an associative algebra, we define the Lie bracket on 
$\Lie(\group{G})$ as commutator, that is $[X,Y]:=X\cdot Y-X\cdot X$. 
This bracket clearly satisfies 
conditions (\ref{eq:DefLieAlg})). But we still have to show that 
$[X,Y]$ is in $\Lie(\group{G})$ if $X,Y$ are. That is, we have to 
show that there is a curve $C\in C^1_*(\reals,\group{G })$ such that  
$\dot C=[X,Y]$. To do this, let again $A,B\in C^1_*(\reals,\group{G })$ 
be such that $X=\dot A$ and $Y=\dot B$. Then the sought for $C$ is 
given by  
\begin{equation}
\label{eq:ProofLieAlg1}
C(s):=
\begin{cases}
A(\tau(s))\cdot B(\tau(s))\cdot A^{-1}(\tau(s))\cdot B^{-1}(\tau(s))
& \text{for}\ s\geq 0\,,\\
B(\tau(s))\cdot A(\tau(s))\cdot B^{-1}(\tau(s))\cdot A^{-1}(\tau(s))
& \text{for}\ s\leq 0\,,\\
\end{cases}
\end{equation} 
where 
\begin{equation}
\label{eq:ProofLieAlg2}
\tau(s):=
\begin{cases}
\sqrt{s} &\text{for}\ s\geq 0\,,\\
-\sqrt{-s} &\text{for}\ s\leq 0\,.
\end{cases}
\end{equation}
This curve is indeed differentiable at $s=0$ (though  
$s\mapsto A(\sqrt{s})$ and  $s\mapsto B(\sqrt{s})$ are not).
Its right derivative ($s\geq 0$) is: 
\begin{eqnarray}
\label{eq:ProofLieAlg3}
\dot C&=&\lim_{s\rightarrow 0}\frac{C(s)-\mathbf{1}_n}{s}
       =\lim_{s\rightarrow 0}\left\{\frac{\bigl[A(\tau(s)),B(\tau(s))\bigr]
         A^{-1}(\tau(s))B^{-1}(\tau(s))}{s}\right\}
\nonumber\\
       &=&\lim_{\tau\rightarrow 0}\left\{
\left [\frac{A(\tau)-\mathbf{1}_n}{\tau}\,,\,
       \frac{B(\tau)-\mathbf{1}_n}{\tau}
\right]\, A^{-1}(\tau)B^{-1}(\tau)\right\} = [X,Y]\,.
\end{eqnarray}
Its left derivative follows along the same lines, one just exchanges 
$A\leftrightarrow B$ and replaces $s$ with $-s$, leading again to 
$[X,Y]$. 
\end{proof}
\subsubsection{Some special Lie algebras}
Before we restrict attention to the Lorentz group and its 
inhomogeneous counterpart (sometimes called the Poincar\'e
group), let us describe in general the situation of which they 
are special cases. 

Consider a vector space $V$ of $n$ dimensions over the field 
$\field$ ($\reals$ or $\complex$). As before, $\End(V)$ denotes 
the associative algebra of linear maps $V\mapsto V$. 
Let $\group{GL}(V)\subset\End(V)$ denote the set of invertible 
linear maps, i.e. $\det(f)\ne 0$ (compare (\ref{eq:DetExpTrace}))
for all $f\in\group{GL}(V)$. 

Given a subgroup $\group{G}\subseteq\group{GL}(V)$, there is a 
corresponding inhomogeneous group, $\group{IG}\subseteq\group{IGL}(V)$,
given by the semi-direct product of $V$ (considered as Abelian group
under addition) with $\group{G}$, denoted by $V\rtimes\group{G}$. 
Its multiplication law is as follows: 
\begin{equation}
\label{eq:SemiDirMultLaw}
(a_1,A_1)(a_2,A_2)=(a_1+A_1(a_2)\,,\,A_1\circ A_2)\,,\qquad
a_i\in V\ A_i\in\group{G}
\end{equation} 

We endow $V$ with a non-degenerate bilinear form 
$\omega:V\times V\rightarrow\field$, which we restrict to be 
either symmetric ($\epsilon=1$) or antisymmetric ($\epsilon=-1$),
that is $\omega(v,w)=\epsilon\,\omega(w,v)$ for all $v,w\in V$. We want 
to consider the group $\group{G}\subset\group{GL}(V)$ of $\omega$
preserving maps: 
\begin{subequations}
\label{eq:DefGroupG}
\begin{equation}
\label{eq:DefGroupG1}
\group{G}\,:\,=\,
\{A\in\group{GL}(V)\mid \omega(Av,Aw)=g(v,w)\quad\forall v,w\in V\}\,.
\end{equation}   
Using the `index-lowering' map $\omegadown:V\rightarrow V^*$, 
$v\mapsto\omega(v,\cdot)$ and its inverse $\omegadown:V^*\rightarrow V$, 
the `index-raising' map (cf. Section\,\ref{sec:VectorSpaces}), this 
can also be written as  
\begin{equation}
\group{G}\,:\,=\,\{A\in\group{GL}(V)\mid 
\omegadown\circ A\circ\omegaup=(A^\top)^{-1}\}\,.
\label{eq:DefGroupG2}
\end{equation}
\end{subequations}   

The Lie algebra $\Lie({\group{G}})$ is easily obtained 
by considering curves in $\group{G}$, as explained in the 
previous subsection. Using (\ref{eq:DefGroupG}) this leads to 
\begin{subequations}
\label{eq:DefLieG}
\begin{alignat}{1}
\label{eq:DefLieG1}
\Lie(\group{G})\,:
&\,=\,\{X\in\End(V)\mid \omega(Xv,w)+\omega(v,Xw)=0\quad\forall v,w\in V\}\,,\\
\label{eq:DefLieG2}
&\,=\,\{X\in\End(V)\mid\omegadown\circ X\circ\omegaup=-X^\top\}\,.
\end{alignat}
\end{subequations}
Let us describe it more concretely in terms of components. Choose a basis 
$\{e_a\}_{a=1\cdots n}$ of $V$ and the corresponding dual basis 
$\{\eta^a\}_{a=1\cdots n}$ of $V^*$, so that $\eta^a(e_b)=\delta^a_b$. 
From~(\ref{eq:DefLieG2}) it follows that a general element 
$X^a_b\,e_a\otimes\eta^b\in\End(V)$ lies in $\Lie(\group{G})$ iff 
$X_{ab}=-\,\epsilon\,X_{ba}$, where $X_{ab}:=X^c_b\omega_{ca}$. Hence, 
writing $\eta_a:=\omega_{ab}\eta^b$ so that $\eta_a(e_b)=\omega_{ab}$
(cf. Sect.\,\ref{sec:IndexMoving}), a basis for $\Lie(\group{G})$ 
is given by the $\frac{1}{2}n(n-\epsilon)$ vectors 
\begin{equation}
\label{eq:BasisLieAlgLorentz}
M_{ab}=e_a\otimes\eta_b-\,\epsilon\,e_b\otimes\eta_a\,.
\end{equation}

The Lie algebra of the corresponding inhomogeneous group is given by 
the linear space $V\oplus\Lie(\group{G})$ and Lie bracket as follows:  
\begin{equation}
\label{eq:SemiDirektLie}
\bigl[(a_1,X_1)\,,\,(a_2,X_2)\bigr]=
\bigl(X_1(a_2)-X_2(a_1)\,,\,[X_1,X_2]\bigr)\,.
\end{equation}
Hence we obtain $\Lie(\group{IG})$ by adding to 
(\ref{eq:BasisLieAlgLorentz}) the $n$ translation generators 
\begin{equation}
\label{eq:BasisLieTransl}
T_a:=e_a\,.
\end{equation}
Together they span the $\frac {1}{2}n(n+2-\epsilon)$--dimensional Lie 
algebra $\Lie(\group{IG})$, whose commutation relations easily follow 
from (\ref{eq:BasisLieAlgLorentz},\ref{eq:SemiDirektLie},\ref{eq:BasisLieTransl}):
\begin{subequations}
\label{eq:LieAlgGenPoin}
\begin{alignat}{1}
\label{eq:LieAlgGenPoin1}
[M_{ab},M_{cd}] &\ =\ 
\omega_{ad}M_{bc}+
\omega_{bc}M_{ad}-\,\epsilon\,
\omega_{ac}M_{bd}-\,\epsilon\,
\omega_{bd}M_{ac}\,,\\
\label{eq:LieAlgGenPoin2}
 [M_{ab},T_c] &\ =\
\omega_{bc}T_a-\,\epsilon\,\omega_{ac}T_b\,,\\
\label{eq:LieAlgGenPoin3}
 [T_a,T_b] &\ =\ 0\,.
\end{alignat}
\end{subequations}

Two special cases of this general setting become relevant in 
SR:
\begin{itemize}
\item[1]
Let $V=\reals^4$ and $\omega$ symmetric with signature $(1,3)$
(one plus, three minuses). The technical name of $\group{G}$ is then 
$\group{O}(1{,}3)$. Generally, if $V=\reals^n$ and if $\omega$ is of 
signature $(p{,}q)$, where $p+q=n$, $\group{G}$ is called 
$\group{O}(p{,}q)$. $\group{O}(p{,}q)$ is isomorphic to 
$\group{O}(q{,}p)$ and $\group{O}(n{,}0)$ is just the ordinary 
orthogonal group $\group{O}(n)$ in $n$ dimensions.
\item[2]
Let $V=\complex^2$ and $\omega$ antisymmetric. In two dimensions, 
leaving an antisymmetric form invariant is equivalent to having 
unit determinant. Hence $\group{G}=\group{SL}(2{,}\complex)$, 
the group of complex $2\times 2$ matrices of unit determinant.   
The group $\group{SL}(2{,}\complex)$ is the double (and also 
universal) cover of the identity component of $\group{O}(1{,}3)$, 
often denoted by $\group{O^\uparrow_+(1{,}3)}$ or 
$\group{SO^\uparrow(1{,}3)}$. 
\end{itemize}

\subsection{Exponential map}
\label{sec:ExpMap}
Since $\End(V)$ form an associative algebra, we can form functions 
based on addition and multiplication. Writing $X^n$ for the $n$-fold 
composition $X\circ\cdots\circ X$, we can define the exponential map 
\begin{equation}
\label{eq:DefExpMap}
\exp:\End(V)\rightarrow\End(V)\,,\qquad
\exp(X):=\sum_{n=0}^\infty \frac{X^n}{n!}\,.
\end{equation} 
Note that the series converges absolutely with respect to the 
standard norms on $\End(V)$. 

Now consider `$\det$' and `$\trace$', which are the familiar 
$\field$-valued functions on $\End(V)$:
\begin{alignat}{2}
& \det(X)   &&\,:\,=\, {\det}_m\{\eta^a(Ae_b)\}\,,\\
& \trace(X) &&\,:\,=\, \eta^a(Ae_a)\,,
\end{alignat}
where $\{e_a\}$ and $\{\eta^a\}$ is any pair of dual bases (it does 
not matter which one) and where $\det_m$ is the standard determinant 
function for matrices. We have 
\begin{proposition}
\begin{equation}
\label{eq:DetExpTrace}
\det\circ\exp=\exp\circ\,\trace\,.   
\end{equation}
\end{proposition}
\begin{proof}
Assume $V$ to be complex (complexify if $V$ was
real). For $X\in\End(V)$ one can then find an eigenbasis, so that 
with respect to it $X$ is a triangular matrix, whose diagonal entries 
are its eigenvalues. Then equation (\ref{eq:DetExpTrace}) reduces 
to the statement, that the product of the exponentials of the 
eigenvalues is the exponential of their sum, which is true of course.
\end{proof}

Equation (\ref{eq:DetExpTrace}) shows that $\det(\exp(X))> 0$
for any $X\in\End(V)$. Moreover, any element $A=\exp(X)$ is connected 
to the identity by a continuous path $s\mapsto\exp(sX)$. Hence the 
image of $\End(V)$ under $\exp$ is contained in the identity 
component of $\group{GL}(V)$, which is given by the invertible 
linear maps of positive determinant, denoted by $\group{GL}^+(V)$. 

Note that the curve $s\mapsto\exp(sX)$ is a homomorphism from the 
additive group $\reals$ to $\group{GL}(V)$. Conversely, we have 
\begin{proposition}
Let $\gamma:\reals\rightarrow\group{GL}(V)$ be a homomorphism, 
i.e. a map that satisfies 
\begin{equation}
\label{eq:HomoRtoGL}
\gamma(0)=\mathbf{1}\quad\text{and}\quad
\gamma(s+t)=\gamma(s)\circ\gamma(t)\
\quad\text{for all}\ s,t\in\reals. 
\end{equation}
Then $\gamma$ must be of the form $\gamma(s)=\exp(sX)$, where 
$X=\dot\gamma(0)$. 
\end{proposition}
\begin{proof}
We consider the curve $\beta(s):=\gamma(s)\circ\exp(-sX)$, which 
satisfies $\beta(0)=\mathbf{1}$ and 
$\dot\beta(s)=\dot\gamma(s)-\gamma(s)\circ X$. But this is zero, as 
can be seen from differentiating $\gamma(s+t)=\gamma(s)\circ\gamma(t)$
with respect to $t$ at $t=0$. Hence  $\beta(s)=\mathbf{1}$ for all 
$s$, showing that $\gamma(s)=\exp(sX)$. 
\end{proof}

Let us now regard the exponential map restricted to the special 
Lie subalgebras $\Lie(\group{G})$ defined in (\ref{eq:DefLieG}). 
Since   
\begin{equation}
\label{eq:ExpEquivariance}
\omegadown\circ\exp(X)\circ\omegaup=\exp(\omegadown\circ X\circ\omegaup)
\end{equation}
for all $X\in\End(X)$, the image of $\Lie(\group{G})$ under the 
exponential map lies in $\group{G}$. More precisely, since 
$\Lie(\group{G})$ is connected and $\exp$ continuous, the image must 
also be connected. Since it also contains the identity 
($\mathbf{1}=\exp(0)$), the image of $\Lie(\group{G})$ lies in 
the identity component of $\group{G}$, denoted by 
$\group{G}_{\mathbf{1}}$. Hence we have a map 
\begin{equation}
\label{eq:ExpMap}
\exp:\Lie(\group{G})\rightarrow\group{G}_{\mathbf{1}}\,.
\end{equation}
It is clear that this map is generally not injective. Consider e.g.
the group $\group{SO}(2)$ of planar rotations, which is topologically 
a circle ($S^1$) and whose Lie algebra is the real line. $\exp$ winds 
the line infinitely often around the circle. But neither is $\exp$ 
generally onto. A relevant example is given by 
$\group{G}=\group{SL}(2,\complex)$. Its Lie algebra is given by the 
space of traceless $2\times$ matrices with complex entries. 
Now, for example, none of the matrices
\begin{equation}
\label{eq:ExpSurjExam1}
A_a:=
\begin{pmatrix}
-1&a\\
0&-1
\end{pmatrix}\quad a\ne 0
\end{equation}
can be in the image of the exponential map. To see this, first note 
that, within $\group{SL}(2,\complex)$, $A(a)$ can be continuously 
connected to the identity, e.g. by 
the path
\begin{equation}   
\label{eq:ExpSurjExam2}
A_a(s)=
\begin{pmatrix}
\exp(i\pi s)&sa\\
0&\exp(-i\pi s)
\end{pmatrix}\,.
\end{equation}
In fact, $\group{SL}(2{,}\complex)$ is connected. 
Suppose now that $\exp(X)=A_a$ for some traceless $X$. 
The eigenvalues of $X$ are $\pm\lambda\ne 0$ so that $X$ is 
diagonalizable. Let $T\in\group{GL}(2,\complex)$ such that 
$TXT^{-1}=\mathrm{diag}(\lambda,-\lambda)$, then 
$TA_aT^{-1}=\mathrm{diag}(\exp(\lambda),\exp(-\lambda))$, 
which is impossible since both eigenvalues of $A_a$ equal $-1$. 

What is however true is that the image of the exponential map 
covers a neighborhood of the group identity. This follows 
from the fact that the derivative of the smooth map 
(\ref{eq:ExpMap}) evaluated at $\mathbf{1}\in\group{G}$ is 
non-zero (it is the identity map 
$\Lie(\group{G})\rightarrow\Lie(\group{G})$). Hence, by the 
inverse-function theorem, it has a local smooth inverse.  
Moreover, we have the following
\begin{proposition}
Any $A\in\group{G}$ is the \emph{finite} product of elements in the 
image of $\exp$, that is, for any $A\in\group{G}$ there exist 
$X_i\in\Lie(\group{G})$, $i=1,\cdots,k<\infty$, such that 
\begin{equation}
\label{eq:FiniteExponentials}
A=\exp(X_1)\circ\cdots\circ\exp(X_k)\,.
\end{equation}
\end{proposition}
\begin{proof}
We first note that elements of the form 
(\ref{eq:FiniteExponentials}) obviously form a subgroup 
$\group{G}'\subset\group{G}_{\mathbf{1}}$ which contains a 
whole neighborhood $U\subset\mathbf{1}\in\group{G}_{\mathbf{1}}$,
as we have just seen. Now, for any $A\in\group{G'}$,
the map $L_A:\group{G'}\rightarrow\group{G'}$, $B\mapsto AB$,
is a smooth bijection with smooth inverse $L_{A^{-1}}$. Hence 
$L_A$ is an open map (sends open sets to open sets) so that 
$L_A(U)$ is an open neighborhood of $A\in\group{G}_{\mathbf{1}}$.
This shows that $\group{G'}\subseteq\group{G}_{\mathbf{1}}$ 
is open. Likewise one shows that all cosets of $\group{G'}$ in 
$\group{G}_{\mathbf{1}}$ are open, since they are obtained as 
images of $\group{G'}$ under $L_A$ for some
 $A\in\group{G}_{\mathbf{1}}$. But this shows that 
$\group{G'}\subseteq\group{G}_{\mathbf{1}}$ is also closed, since it 
is the complement of the union of all $\group{G'}$-cosets different 
from $\group{G'}$ itself. Being open and closed 
in the connected set $\group{G}_{\mathbf{1}}$, $\group{G'}$ is 
necessarily identical to it. [This argument shows in fact that any 
neighborhood $U$ of the identity in a topological group generates 
the identity component, in the sense that any element in 
the identity component is the finite product of elements from $U$.] 
\end{proof} 

\subsection*{Acknowledgements}
I sincerely thank J\"urgen Ehlers and Claus L\"ammerzahl for 
inviting me to the 2005 Potsdam conference on Special Relativity and 
also giving me the opportunity to present this material in 
written form. I am also grateful to Matteo Carrera and Abraham 
Ungar for remarks and suggestions which led to various improvements.

\end{document}